\documentclass[trackchanges, twocolumn]{aastex701}

\usepackage{graphicx}
\usepackage{multirow}
\usepackage{amsmath,amssymb,amsfonts}
\usepackage{amsthm}
\usepackage{mathrsfs}
\usepackage{xcolor}
\usepackage{xspace}
\usepackage[T1]{fontenc}
\usepackage{lineno}
\usepackage{comment}

\newcommand{\Me}{\ensuremath{M_{\oplus}}\xspace} 

\newcommand{\Rj}{\ensuremath{R_{\rm{Jup}}}\xspace}
\newcommand{\Mj}{\ensuremath{M_{\rm{Jup}}}\xspace}

\newcommand{\Lsun}{L_\odot}
\newcommand{\Teff}{\ensuremath{T_\mathrm{eff}}\xspace}
\newcommand{\logg}{\ensuremath{\log{g}}\xspace}
\newcommand{\lbollsun}{\ensuremath{\log(L_\mathrm{bol}/\Lsun)}}

\newcommand{\logco}{\ensuremath{\mathrm{log(^{12}CO / ^{13}CO)}\xspace}}

\newcommand{\kms}{km~s$^{-1}$\xspace}

\newcommand{\caltech}{Department of Astronomy, California Institute of Technology, Pasadena, CA 91125, USA}
\newcommand{\gps}{Division of Geological \& Planetary Sciences, California Institute of Technology, Pasadena, CA 91125, USA}
\newcommand{\ucsc}{Department of Astronomy \& Astrophysics, University of California, Santa Cruz, CA95064, USA}

\newcommand{\uclagps}{Department of Earth, Planetary, and Space Sciences, University of California, Los Angeles, CA 90095, USA}
\newcommand{\jpl}{Jet Propulsion Laboratory, California Institute of Technology, 4800 Oak Grove Dr.,Pasadena, CA 91109, USA}
\newcommand{\ucsd}{Department of Astronomy \& Astrophysics,  University of California, San Diego, La Jolla, CA 92093, USA}

\newcommand{\carnegiew}{Earth and Planets Laboratory, Carnegie Institution for Science, Washington, DC, 20015}

\newcommand{\stsci}{Space Telescope Science Institute, Baltimore, MD 21218, USA}

\newcommand{\ciera}{Center for Interdisciplinary Exploration and Research in Astrophysics, Northwestern University, 1800 Sherman Ave, Evanston, IL 60201, USA}
\newcommand{\northwestern}{Department of Physics and Astronomy, Northwestern University, 2145 Sheridan Road, Evanston, IL 60208-3112}
\newcommand{\ucsb}{Department of Physics, University of California, Santa Barbara, CA 93106, USA}

\begin{document}

\title{A Sulfur-Rich Atmosphere for the Young Jupiter Analog AF Lep b Reveals Significant Solid Accretion}

\author[0000-0002-6618-1137]{Jerry W. Xuan}
\altaffiliation{51 Pegasi b Fellow}
\affiliation{\ucsb}
\affiliation{\uclagps}
\email[show]{jxuan@ucsb.edu}

\author[0000-0001-6396-8439]{William O. Balmer}
\altaffiliation{51 Pegasi b Fellow}
\affiliation{\ciera}
\email{wbalmer@stsci.edu}

\author[0000-0003-1728-8269]{Yayaati Chachan}
\affiliation{\ucsc}
\email{ychachan@ucsc.edu}

\author[0000-0003-2233-4821]{Jean-Baptiste Ruffio}
\affiliation{\ucsd}
\email{jruffio@ucsd.edu}

\author[0000-0003-3290-6758]{Kazumasa Ohno}
\affiliation{Division of Science, National Astronomical Observatory of Japan, 2-12-1 Osawa, Mitaka, Tokyo 181-8588, Japan}
\email{ohno.k.ab.715@gmail.com}

\author[0000-0002-4918-0247]{Robert J. De Rosa}
\affiliation{European Southern Observatory, Alonso de C\'{o}rdova 3107, Vitacura, Casilla 19001, Santiago, Chile}
\email[]{rderosa@eso.org}  

\author[0000-0003-3708-241X]{Aneesh Baburaj}
\affiliation{\ciera}
\email{ababuraj@northwestern.edu}

\author[orcid=0000-0001-6975-9056]{Eric L. Nielsen}
\affiliation{Department of Astronomy, New Mexico State University, P.O. Box 30001, MSC 4500, Las Cruces, NM 88003, USA}
\email[]{nielsen@nmsu.edu}  

\author[0000-0001-5061-0462]{Ruth Murray-Clay}
\affiliation{\ucsc}
\email{rmc@ucsc.edu}

\author[orcid=0000-0002-9843-4354]{Jonathan J. Fortney}
\affiliation{\ucsc}
\email[]{}

\author[0000-0002-3191-8151]{Marshall D. Perrin}
\affiliation{\stsci}
\email{mperrin@stsci.edu}

\author[orcid=0000-0003-0774-6502]{Jason J. Wang}
\affiliation{\northwestern}
\affiliation{\ciera}
\email[]{}

\author[0000-0003-2649-2288]{Brendan P. Bowler}
\affiliation{\ucsb}
\email{bpbowler@ucsb.edu}

\author[0000-0001-7443-6550]{Alexander Madurowicz}
\affiliation{\stsci}
\email{amadurowicz@stsci.edu}

\author[0000-0003-1212-7538]{Bruce A. Macintosh}
\affiliation{Department of Astronomy and Astrophysics, UC Santa Cruz, Santa Cruz CA 95064} 
\email[]{bamacint@ucsc.edu}

\author[0000-0003-0097-4414]{Yapeng Zhang}
\altaffiliation{51 Pegasi b Fellow}
\affiliation{\caltech}
\email{yapzhang@caltech.edu}

\author[0000-0001-5578-1498]{Björn Benneke}
\affiliation{\uclagps}
\email{bbenneke@epss.ucla.edu}

\author{Alexis Bidot}
\affiliation{\stsci}
\email{abidot@stsci.edu}

\author{Geoffrey A. Blake}
\affiliation{\gps}
\email{gab@caltech.edu}

\author[0000-0003-4557-414X]{Kyle Franson}
\altaffiliation{NHFP Sagan Fellow}
\affiliation{\ucsc}
\email{kfranson@ucsc.edu}

\author[0009-0005-9021-0152]{Carrie He}
\affiliation{\uclagps}
\email{carriehe@g.ucla.edu}

\author[0000-0001-9164-7966]{Julie Inglis}
\affiliation{\ucsd}
\email{jinglis@ucsd.edu}

\author[0000-0002-5375-4725]{Heather A. Knutson}
\affiliation{\gps}
\email{hknutso2@caltech.edu}

\author{Dimitri Mawet}
\affiliation{\caltech}
\affiliation{\jpl}
\email{dmawet@astro.caltech.edu}

\author{Laurent Pueyo} 
\affiliation{\stsci}
\email{pueyo@stsci.edu}

\author[0000-0003-4203-9715]{Emily Rickman}
\affiliation{European Space Agency (ESA), ESA Office, Space Telescope Science Institute, 3700 San Martin Dr, Baltimore, MD 21218, USA}
\email[]{}

\author[0000-0002-1838-4757]{Aniket Sanghi}
\altaffiliation{NSF Graduate Research Fellow}
\affiliation{\caltech}
\email{asanghi@caltech.edu}

\author[0000-0003-0354-0187]{Nicole L. Wallack}
\affiliation{\carnegiew}
\email{nwallack@carnegiescience.edu}

\begin{abstract}
AF Lep b is one of the closest analogs to Jupiter in terms of mass ($3$--$4~\Mj$) and semi-major axis ($9$ AU) amenable to spectroscopic characterization. We present JWST/NIRSpec high-contrast spectroscopy of the planet from $2.85$--$5.3~\mu$m at $R\sim3000$, which provides detections of CO$_2$, H$_2$S, CH$_4$, $^{12}$CO (and $^{13}$CO), and H$_2$O, as well as complementary JWST/NIRCam imaging that captures the planet's continuum flux from $4.0$--$4.7~\mu$m. Combining the JWST observations with spectra from VLTI/GRAVITY and VLT/SPHERE ($1.0$--$2.5~\mu$m), we carry out atmospheric retrievals that include the effects of clouds and disequilibrium chemistry while allowing the C, O, and S abundances to vary independently. AF Lep b exhibits metal enrichment across C, O, and S with $\rm C/H=2.9\pm0.5$, $\rm O/H=3.7\pm0.6$, and $\rm S/H=4.7\pm0.7~\times$ solar (and stellar). The planet's slightly sub-solar C/O and C/S are consistent with formation near its observed location, and disfavor formation beyond the CO snowline. The sulfur enrichment in AF Lep b implies significant accretion of disk solids during formation, and we estimate the planet contains $56\pm7~\Me$ of solids. The C, O, and S enrichment levels of AF Lep b are similar to those of Jupiter, and other super-Jupiters like HR 8799 bcde. We also show that the degree of atmospheric metal enrichment of these imaged planets is similar to the bulk metal enrichment of transiting gas giants with masses greater than $\sim1~\Mj$, suggesting that the process and efficiency of metal accretion for gas giants may not be strongly dependent on orbital distance or planet mass. 

\end{abstract}

\keywords{}

\section{Introduction} 

High-contrast imaging has revealed a population of substellar companions with a stunning diversity in mass and orbital distance ($\sim0.3$--$75~\Mj$, $>3$ au). These objects are thought to originate from a continuum of formation processes, including core accretion, disk fragmentation, and cloud fragmentation (see reviews by \citealt{bowler_imaging_2016} and \citealt{Currie_review_2023}). Over the past few years, improved measurements of the atmospheric compositions, spins, orbits, and demographics of these objects, carried out on increasingly larger samples, have revealed tantalizing trends that are beginning to map the boundary between planet-like and star-like formation pathways. 

On the atmosphere side, the JWST NIRSpec and MIRI/MRS integral field units (IFU) are starting to transform our understanding of directly imaged gas giants with unprecedented high-contrast performance at moderate spectral resolutions \citep[e.g.][]{Ruffio2024, Ruffio2026, Cugno2025, Malin2025b}. JWST NIRCam and MIRI imaging are also providing complementary constraints by resolving broad molecular absorption bands and anchoring the continuum flux \citep[e.g.][]{Balmer2025b, Matthews2026, Sanghi2026}.  Compared to previous ground-based observations, JWST spectra provide a much more complete chemical inventory of planetary and brown dwarf atmospheres \citep[e.g.][]{Kiman2026, Kuhnle2026}. Recent work with NIRSpec has measured elemental abundances of sulfur and nitrogen from H$_2$S and NH$_3$ for the first time in directly imaged planets using atmospheric retrievals and chemical modeling \citep{Ruffio2026, Xuan2026}. These new elemental abundances greatly complement C/H and O/H, and can provide new insights into the planet formation process \citep[e.g.][]{Turrini2021, Schneider2021b, Chachan2023}. For instance, the enrichment in sulfur for the HR 8799 planets explicitly demonstrates a history of solid accretion from pebbles or planetesimals, whereas the C/S and N/S ratios of these planets point to additional accretion of metal-enriched gas, consistent with originating from evaporating pebbles that migrate inward across the circumstellar disk \citep{Xuan2026}.  

The metal enrichment across multiple elements in HR~8799~bcde marks a clear departure from the abundance pattern observed for more massive substellar companions (mass ratio $q\gtrsim0.01$), which show stellar-like C and O abundances as a population \citep{Hoch2023, Xuan2024b}. The observed trend in atmospheric metallicities aligns well with trends observed in the spin rates, orbital eccentricities, and demographics of directly imaged substellar companions \citep{nielsen_gemini_2019, bowler_Populationlevel_2020, Hsu2026}, which also hint at a direct fragmentation origin for widely-separated companions with $q\gtrsim0.01$.

With a dynamical mass of $3.75\pm0.50~\Mj$ ($q\approx0.003$), semi-major axis of $9$ au, and circular orbit ($e<0.07$ at $2\sigma$ confidence) \citep{Balmer2025}, AF Lep b \citep{Franson2023_AFLep, Mesa2023, DeRosa2023} is a young analog ($24\pm3$ Myr; \citealt{Bell2015}) to Jupiter, and one of the most Jupiter-like exoplanets in terms of mass and orbital distance amenable to detailed atmospheric characterization. Compared to the majority of directly imaged planets at dozens to hundreds of au, AF Lep b is located much closer to the peak of the giant planet occurrence rate inferred by Doppler surveys ($1$--$10$ au; e.g. \citealt{Fernandes2019}, \citealt{Wittenmyer2020}, \citealt{fulton_California_2021}). As such, measuring the atmospheric composition of AF Lep b provides a unique opportunity to understand the formation mechanism of the most common gas giants like Jupiter. Previous atmospheric analyses have pointed to a high metallicity for AF Lep b \citep{Zhang2023, Franson2024}, and the latest measurement using VLTI/GRAVITY $K$ band spectroscopy ($R\sim500$) yields $\rm{[C/H]}=0.75\pm0.25$ dex \citep{Balmer2025}. These ground-based observations were sensitive to CH$_4$, CO, and H$_2$O, and therefore only measured C/H and O/H.  

In this paper, we present JWST/NIRSpec IFU spectroscopy of AF Lep b from $2.8$--$5.3~\mu$m. We also present new JWST/NIRCam medium-bandpass photometry from $4.0$--$4.7~\mu$m, and new VLT/SPHERE spectroscopy ($1.0$--$1.7~\mu$m) of AF Lep b. We comprehensively model all these datasets of the planet to infer its atmospheric properties and composition. The JWST/NIRSpec data in particular allow us to constrain S/H and $^{12}$C/$^{13}$C for the first time from H$_2$S and $^{13}$CO lines. The detection of AF Lep b at $0.29\arcsec$ away from its host star with the NIRSpec IFU showcases the power of high-contrast medium-resolution spectroscopy from space. 

We organize this paper as follows: in Section~\ref{sec:obs} we describe new observations of AF Lep b and the associated data reduction methods. Section~\ref{sec:atmo} describes the modeling framework with a focus on the opacities, chemistry, clouds, and thermal structure in the atmospheric retrieval model. Section~\ref{sec:stellar_abunds} summarizes carbon, oxygen, and sulfur abundances of stars in the $\beta$ Pic moving group, and presents new measurements of these abundances for AF Lep A. In Section~\ref{sec:results}, we present the atmospheric analysis results for AF Lep b, before discussing the planet formation implications in Section~\ref{sec:discuss}, and concluding in Section~\ref{sec:conclude}.

\section{Observations and data reduction}\label{sec:obs}

In this paper, we model the JWST/NIRSpec spectra for AF Lep b in conjunction with three other datasets: 1) archival GRAVITY spectrum ($R\sim500$) in the K band ($2$--$2.5~\mu$m) from \citet{Balmer2025}, 2) new VLT/SPHERE spectrum covering $1.0$--$1.7~\mu$m, and 3) new JWST/NIRCam photometry in three medium bands (F410M, F430M, F460M) from $4.0$--$4.7~\mu$m. There is also JWST/NIRCam wide-band photometry in F444W and Keck/NIRC2 $L^\prime$ photometry for AF Lep b \citep{Franson2023_AFLep, Franson2024}. When testing retrievals that include these two additional points, we found that the posteriors are essentially unchanged. Therefore, we only use the new JWST/NIRCam medium-band observations in this paper, as they cover similar wavelengths and have higher S/N.

Together, the low-resolution spectra and photometry anchor the planet's continuum flux from $1.0$--$4.7~\mu$m, which can be complementary to continuum-subtracted high- or medium-resolution spectroscopy, as demonstrated in previous work \citep{Wang2023, Ruffio2026}. In Appendix~\ref{sec:nirspec_only}, we also compare the results with a NIRSpec-only retrieval for AF Lep b, to show the value of including the low-resolution spectra and photometry. Below we describe the new observations presented in this work. The GRAVITY spectrum is taken directly from \citet{Balmer2025}, so its data reduction is not discussed here.

\subsection{JWST/NIRSpec spectroscopy}
We observed the AF Lep system with the JWST/NIRSpec IFU in moderate-resolution spectroscopy mode ($R\sim2700$; filter F290LP; grating G395H). The observations comprise the Cycle 3 GO program 5342 (PI: Xuan), and were obtained on UT 2025 Jan 17 with a total integration time of 4.8 hr.\footnote{The JWST/NIRSpec data used in this paper can be found in MAST: \dataset[10.17909/1xbm-et73]{https://doi.org/10.17909/1xbm-et73}} The observations were conducted in a mosaic pattern with two tiles which placed the planet at different spatial locations in the IFU in order to reduce the wavelength gap between NIRSpec's two detectors (NRS1 and NRS2) around $\sim4.1\,\mu\mathrm{m}$. This observing strategy was previously recommended by \citet{Ruffio2024} following information in the NIRSpec IFU Wavelength Ranges and Gaps page\footnote{\url{https://jwst-docs.stsci.edu/jwst-near-infrared-spectrograph/nirspec-operations/nirspec-ifu-operations/nirspec-ifu-wavelength-ranges-and-gaps}}. 

Planet detection and spectral extraction are carried out using the methods described in \citet{Ruffio2026} and \citet{Xuan2026}, which build on the framework from \citet{Ruffio2024} and are implemented in the python package \texttt{BREADS}\footnote{\url{https://github.com/jruffio/breads}} \citep{breads}. Below, we briefly describe the major steps. 

\subsubsection{Pre-processing and data calibration}\label{sec:preproc}
We perform initial reductions using stage 1 of the JWST Science Calibration Pipeline v1.20.2 \citep{Bushouse2023}. In the resulting \texttt{*\_rate.fits} files, the correlated 1/f read noise and charge transfer from the saturated stellar core are fitted and subsequently subtracted in each column of the detector following \citet{Ruffio2024}.
Using stage 2 of the JWST pipeline, we then generate flux-calibrated detector images (\texttt{*\_cal.fits}). Given the presence of known wavelength-dependent coordinate offset systematics present in this and earlier versions of pipeline-processed NIRSpec IFU datacubes\footnote{These arise from chromatic dispersion in particular NIRSpec optics. This effect was initially identified in earlier high contrast IFU work by \citet{Ruffio2024}. The NIRSpec team at STScI has recently implemented a fix for this issue, which will mitigate the chromatic centroid shift in an upcoming pipeline release.}, we derive a wavelength-dependent centroid shift of the star by fitting a STPSF model to the detector images (i.e. point clouds) without reconstructing the spectral cube. Finally, we derive a continuum-normalized spectrum of the star from the speckle field by using a spline model to fit the speckle continuum. This empirical stellar model is later combined with a flexible 1D spline model to model, or subtract, the starlight row by row on the NIRSpec detectors.

\subsubsection{Planet detection} 
\begin{figure*}
    \centering
    \gridline{\fig{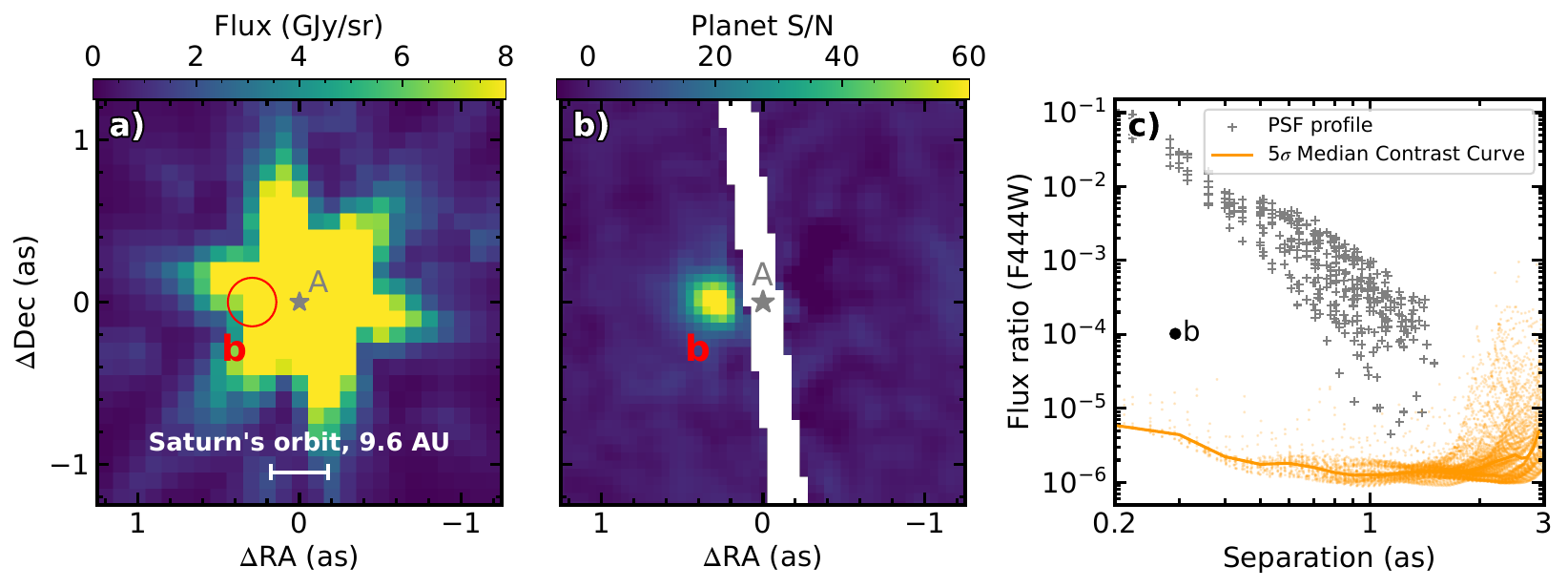}{\textwidth}{}}
    \vspace{-6mm}
    \gridline{\fig{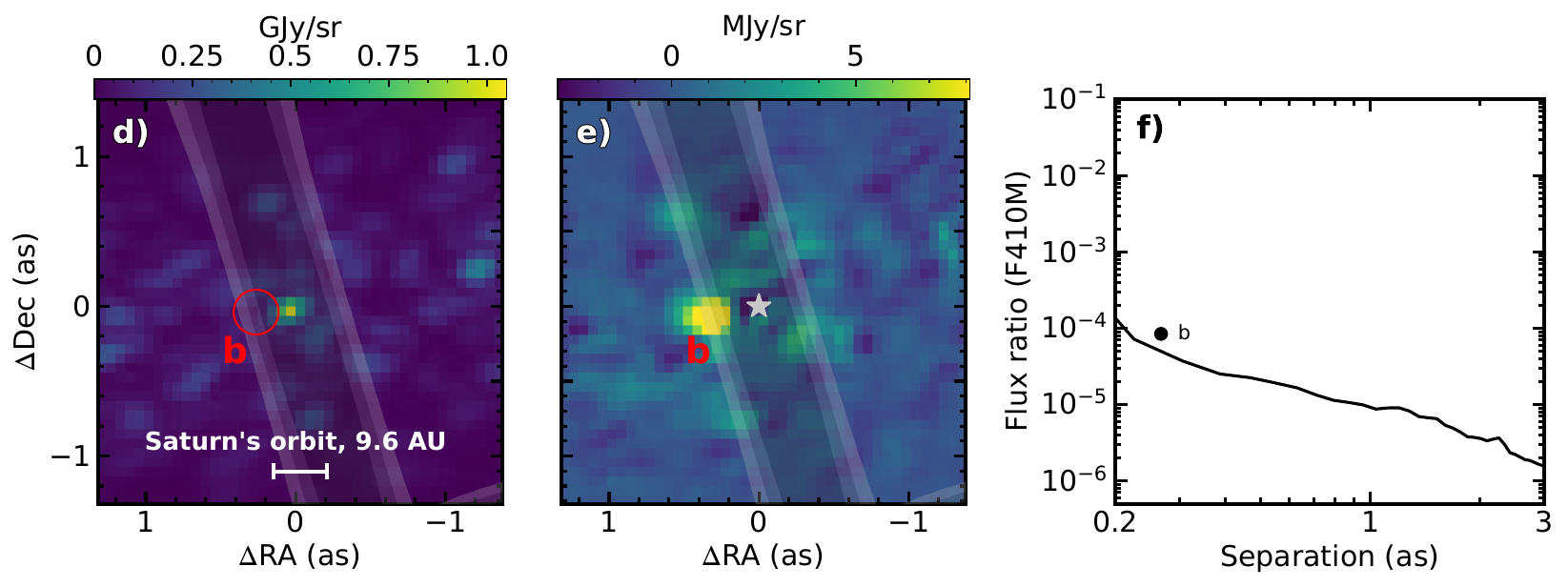}{\textwidth}{}}
    \caption{\textbf{Top:} Detection of AF Lep b with the moderate resolution mode (R$\sim$2,700) of JWST/NIRSpec IFU between $3$--$5\,\mu\mathrm{m}$.
    \textbf{a)} Median spectral cube using the standard JWST calibration pipeline and combining data from the two mosaic dither patterns. This image does not include PSF subtraction so the planet is hidden behind the starlight.
    \textbf{b)} Signal-to-noise ratio map for planet detection. AF Lep b is detected with a peak S/N of 82. The white vertical gaps arise from masking out the IFU slices containing the saturated stellar PSF core. \textbf{c)} Planet $5\sigma$ detection limits for the combined dataset with each dot representing a spatial pixel in the field of view. The flux ratio is defined in the F444W filter.
    \textbf{Bottom:} Detection of AF Lep b with the NIRCam coronagraphy (LW Bar occulter at the narrow offset position) in the F410M filter ($\lambda_{\rm eff}=4.1~\mu$m). \textbf{d)} A median image from the first roll angle, before PSF subtraction. The coronagraphic transmission function is overlaid in gray. \textbf{e)} The starlight subtracted image, showing the detection of AF Lep b. The coronagraph has been positioned to maximize the throughput at the location of AF Lep b. \textbf{f)} Planet $5\sigma$ contrast curve for the observations, calibrated using injection--recovery tests and accounting for the coronagraph throughput (perpendicular to the bar coronagraph). 
    }
    \label{fig:snrmap}
\end{figure*}

At $0.29\arcsec$, AF Lep b is $\sim200$ times fainter than the speckles at the same separation (right panel of Figure~\ref{fig:snrmap}), representing one of the most high-contrast applications to date for the JWST/NIRSpec IFU. For planet detection and S/N calculation, we fit a joint model of the planet and diffracted starlight directly to the detector images. The planet is modeled using a BT-Settl atmospheric model with $\Teff=800~K$ combined with a STPSF model \citep{Perrin2014SPIE.9143E..3XP}. For the star, we use the empirically-derived continuum normalized spectrum from Section~\ref{sec:preproc}, and modulate its continuum using a 60-node spline on each detector row, which \citet{Ruffio2024} found to be an optimal tradeoff between adequate subtraction of the starlight while preserving the planetary signal. By fitting the joint planet+star model across the field of view, the planet's detection S/N can then be calculated at each pixel. Note that previous work found the exact planet model choice does not significantly influence the detection S/N \citep{Ruffio2024, Gibbs2026}. The S/N calculation here is systematic-limited, and we normalize the S/N maps to ensure a standard deviation of unity in the absence of planet signal. After normalizing, we obtain a planet detection S/N of $\approx82$ by combining all wavelengths. The uncertainties on the planet flux are scaled to compute $5\sigma$ sensitivity curves in the NIRCam F444W filter (Figure~\ref{fig:snrmap}). To derive these sensitivity curves, we use the F444W stellar fluxes and planet-to-star flux ratios from \citet{Franson2024}.

\subsubsection{1D spectral extraction} 
While the procedure above for the S/N map calculation can in theory be used to estimate and model the planet's spectrum, it is computationally challenging in practice. Therefore, for 1D spectral extraction, we directly apply the high-pass filter to the dataset and remove the starlight. Specifically, we fit the starlight-only model everywhere on the detector with the spline node prescription from \citet{Ruffio2026}. Then, we extract the high-pass filtered planet spectrum by fitting a STPSF model at each wavelength to the combined tiles across the two mosaic pointings. The flux uncertainties and covariance matrix of the planet spectrum are estimated from an annulus of starlight residuals around the planet at a similar projected separation ($0.29\arcsec$). For details, we refer the reader to \citet{Ruffio2026}.

\subsection{JWST/NIRCam photometry}
We also observed the system using the JWST/NIRCam Long Wavelength Bar at the ``narrow'' fiducial point override offset position \citep[following the strategy outlined in][]{Perrin2018, Balmer2025b}. The observations\footnote{The JWST/NIRCam data used in this paper can be found on MAST at: \doi{10.17909/9tb6-dq26}} were collected on UT 2025 Oct 17 as part of GO 6905 (PI: Balmer), and took 1.4 h on the target star (AF Lep) split between two roll angles separated by $\sim10^\circ$, and 0.9 h on the PSF reference star (HD 33093). These observations were taken through three medium-bandpass filters covering $4.0$--$4.7~\mu$m, F410M, F430M, and F460M, which cover the $4~\mu$m continuum, $4.3~\mu$m CO$_2$ absorption feature, and the $4.6~\mu$m CO feature. 

Data reduction and photometric extraction followed \citet{Balmer2026}, and consisted of calibrating images with the \texttt{jwst} calibration pipeline \citep{Bushouse2023}, cleaning remaining artifacts and preparing images for starlight subtraction using the \texttt{spaceKLIP} package \citep{Kammerer2022, ACarter2023}, and performing starlight subtraction and photometric forward modeling using the KLIP algorithm \citep{soummer_Detection_2012b, pueyo_DETECTION_2016} via the \texttt{pyKLIP} package \citep{wang_pyKLIP_2015}. The high throughput of the \texttt{LWBAR}/narrow observing strategy resulted in a high confidence detection of the planet in all three filters. After forward modeling the companion, we subtracted the best-fit forward model from the data in order to estimate the average annular contrast \citep[computed in annuli perpendicular to the bar, as in][]{Kammerer2022}. 

\subsection{VLT/SPHERE low-resolution spectroscopy}

AF Lep was observed with the Very Large Telescope (VLT) and the Spectro-Polarimetric High-contrast Exoplanet REsearch (SPHERE; \citealp{beuzit_SPHERE_2019}) instrument on the night of 2023 January 16. The star was simultaneously observed with the Infra-Red Dual-beam Imager and Spectrograph (IRDIS; \citealp{Dohlen:2008eu}) and the Integral Field Spectrograph (IFS; \citealp{claudi_SPHERE_2008}) sub-components under ESO program ID 110.25A4.001 (PI: De Rosa). Reference star imaging was used to maximize sensitivity at the smallest inner working angles \citep[e.g.][]{Wahhaj:2021kf, Sanghi2024}. Observations of the star HD 35591, separated by 46$^\prime$, interleaved the observations of AF Lep to provide a near-simultaneous reference PSF free of any astrophysical contaminant.

A total of 39\,m of integration time was obtained on AF Lep with both IRDIS and IFS, with only 18\,m on the reference star. This difference in total integration time was compensated somewhat by the fact that the reference star is approximately 0.7\,mag brighter at both $H$ and $K$, while being the same brightness at $I$ ensuring similar AO correction. Off-axis observations of AF Lep using a neutral density filter were obtained at the start of the observing sequence for flux calibration, and centering frames where a waffle pattern on the deformable mirror introduced attenuated replicas of the central PSF were obtained for image registration. 

The IFS data for both stars were reduced and processed using the ESO SPHERE pipeline (v0.42.0) and the \texttt{VLT/SPHERE} toolkit \citep{vigan_vltsphere_2020}. In addition to performing the standard near-infrared data reduction steps such as dark subtraction, flat fielding, bad pixel fixing, the pipeline also corrected for detector anamorphism and performed the wavelength calibration and cube reconstruction. Reduced images were aligned to a common center using the centering frames taken at the start of the observing sequence; both stars were assumed to be held stationary behind the coronagraph at the same position throughout the full sequence.

\begin{figure*}
    \centering
    \includegraphics[width=0.7\linewidth]{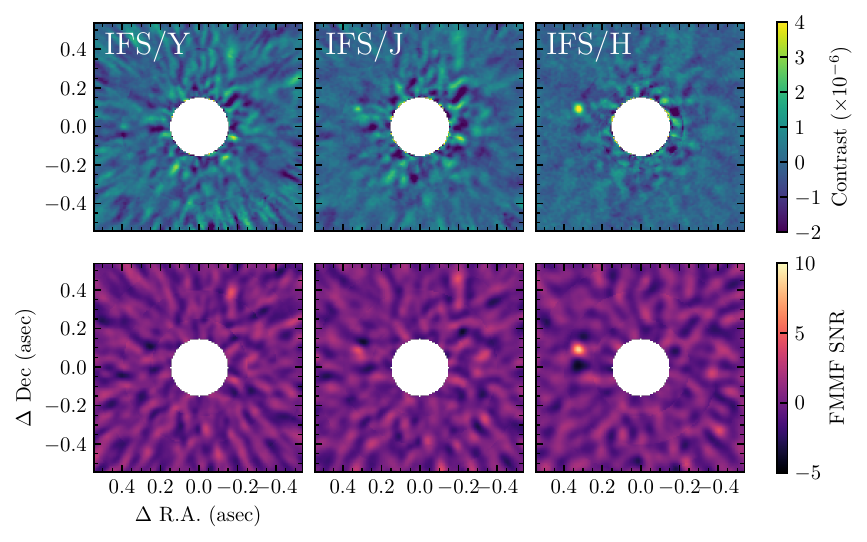}
    \caption{(top row) Residual images after PSF subtraction for the three SPHERE/IFS bands, normalized to the host star brightness. (bottom row) Detection maps computed from the same data. The planet is not detected at $Y$ at a significant level, is detected with marginal significance at $J$ ($4.4\sigma$), and is confidently detected at $H$ ($8.8\sigma$)}
    \label{fig:sphere-maps}
\end{figure*}

The post-processing of the reduced data to subtract the stellar PSF and to extract the companion spectrum follows that described in \citet{DeRosa2023}. We used \texttt{pyKLIP} \citep{wang_pyKLIP_2015} to subtract the residual stellar halo from each wavelength slice of each reduced data cube to create a detection map collapsed in wavelength. We used five annuli that were logarithmically-spaced between 15 and 100\,px radius from the central star. The model PSF was constructed from a reference library consisting of other images at the same wavelength of the target star that satisfied a movement criterion of 2\,px, and images at the same wavelength of the reference star. The model PSF was constructed using the first ten KL modes from this reference library. We did not use any spectral diversity. The quality of this dataset was significantly improved relative to the one presented in \citet{DeRosa2023}, primarily due to the greater amount of field rotation. Using the procedure described in \citet{DeRosa2023}, we estimate a contrast sensitivity at the separation of AF~Lep~b to be $3.9\times10^{-6}$, $3.6\times10^{-6}$, and $4.5\times10^{-6}$ for $Y$, $J$, and $H$, respectively, which compare to $1.1\times10^{-5}$, $7.0\times10^{-6}$, and $6.8\times10^{-6}$ in the earlier dataset. The residuals after PSF subtraction are shown in Figure~\ref{fig:sphere-maps} for each of the bands, along with the corresponding detection significance maps \citep{Ruffio2017}.

The position of the planet was measured using Bayesian KLIP-FM Astrometry (BKA; \citep{Wang:2016gl}) method, following the approach described in \citet{DeRosa2023}. We measured a separation of $\rho=336.3\pm5.0$\,mas and a position angle of $\theta=71.9\pm0.91$\,deg. The spectrum of the planet was extracted from the processed images using \texttt{extractSpec} \citep{greenbaum_GPI_2018}. A segment of the image centered on the location of the planet with a size of $\pm20$\,px and $\pm45$\,deg was defined within which a KLIP PSF subtraction was performed, using the same parameters described previously. The contrast spectrum of the planet was then extracted from the PSF-subtracted image as described in \citet{greenbaum_GPI_2018}. The uncertainties on the spectrum were estimated via injection/recovery tests. A source with the same contrast spectrum as AF~Lep~b was injected into the reduced data cube prior to PSF subtraction, which was then retrieved using the same procedure. This process was repeated at nine position angles at the same separation as the planet, and we adopted the standard deviation of the recovered contrasts at each wavelength as the uncertainty at that wavelength. The contrast spectrum and its uncertainties were converted into a spectrum in apparent flux units using the spectrum of the primary star given in \citet{DeRosa2023}.

\section{Atmospheric characterization}\label{sec:atmo}

We jointly model the high-pass filtered NIRSpec spectrum ($2.85$--$5.3\,\mu\mathrm{m}$), GRAVITY and SPHERE spectra ($1$--$2.5~\mu$m), and NIRCam photometry ($4.0$--$4.7~\mu$m) to provide a comprehensive analysis of AF Lep b's atmosphere. Specifically, we add the log likelihoods from each of these parts, and sample the posteriors using the nested sampling package \texttt{pymultinest} \citep{Buchner2014} based on \texttt{MultiNest} \citep{Feroz2009, Feroz2019}. We run \texttt{pymultinest} in constant efficiency mode with $5\%$ efficiency. We use 1000 live points and stop sampling when the estimated contribution of the remaining prior volume to the total evidence is $<1\%$. 

For the models, we use an atmospheric retrieval framework to compute synthetic spectra with the radiative-transfer code \texttt{petitRADTRANS} version 3.2.0 \citep{molliere_petitRADTRANS_2019, molliere_Retrieving_2020, Nasedkin2024}. Our retrieval approach closely follows the model for HR 8799 b in \citet{Xuan2026}, and we briefly summarize the key components in each subsection below. 

\subsection{NIRSpec forward model}
To model the NIRSpec data, we compute high-resolution line-by-line models at $R=100,000$. We first shift the model to account for the planet's radial velocity (RV), and then convolve the model with a variable Gaussian kernel to account for instrumental broadening. At each wavelength, the width of the Gaussian kernel is related to $R$. Following \citet{Xuan2026}, we fit for the NIRSpec instrumental resolution as a linear function $R_\lambda = r_0 + r\lambda$. Finally, we apply the same spline-based high-pass filter on the model that we applied on the data, and compute the log likelihood while accounting for the data covariance matrix. 

As an alternative to fitting for $R_\lambda = r_0 + r\lambda$, we also tested using the JWST User Documentation (JDox) NIRSpec resolution file (interpolated to the data wavelengths) to convolve the models.\footnote{Downloaded from \url{https://jwst-docs.stsci.edu/jwst-near-infrared-spectrograph/nirspec-instrumentation/nirspec-dispersers-and-filters}} We found that the fiducial model of fitting $R_\lambda$ is significantly favored by a log Bayes factor of ln$(B)\approx61$ compared to using the JDox resolution array. On average, the retrieved $R_\lambda$ values are $\sim20\%$ higher than those provided in JDox. On the other hand, the two models yield consistent abundances at the $<0.1$ dex level, so we follow previous work in fitting for $R_\lambda$ as part of the retrieval \citep{Xuan2024d, Ruffio2026}. We note that our inferred $R_\lambda$ values are within $5\%$ of those measured from other NIRSpec in-flight data by \citet{Shajib2025}, who also found that the JDox resolution values for the NIRSpec IFU in G395H mode are systematically low by $\approx10$--$20\%$. 

\subsection{Low-resolution spectroscopy}
At low-spectral resolutions, correlated noise arising from stellar speckles or instrumental systematics is typically the dominant noise source in the spectra of high-contrast companions \citep{Greco2016}. For the GRAVITY spectrum, we use the covariance matrix derived by \citet{Balmer2025} to account for correlated noise. The SPHERE IFS spectrum does not contain a covariance matrix, so we adopt a Gaussian process with a squared exponential kernel to empirically estimate the correlated noise. Following \citet{wang_Keck_2020} and \citet{Xuan2022}, we assume that the SPHERE error bars contain a fraction $f_{\rm amp}$ of correlated noise, and $1-f_{\rm amp}$ of white noise, and fit for $f_{\rm amp}$ and the scale of correlation $l$. 

\subsection{Opacities}\label{sec:opacities}
We include line opacities from CO and its minor isotopologue $^{13}$CO \citep{Li2015}, H$_2$O \citep{Polyansky2018}, CO$_2$ \citep{Yurchenko2026}, CH$_4$ \citep{Yurchenko2024}, NH$_3$ \citep{Coles2019}, HCN \citep{Barber2014}, H$_2$S \citep{Azzam2016}, Na \citep{Allard2019}, and K (line profiles by N. Allard, \citealt{molliere_petitRADTRANS_2019}). For continuum opacities, we include the collision induced absorption (CIA) from H$_2$-H$_2$ and H$_2$-He. In these retrievals, the reference for solar elemental abundances is \citet{asplund_Chemical_2009}. Note that compared to our previous work, we updated the opacity tables for CO$_2$ and CH$_4$ to use the latest line lists. The line list to opacity conversion was done using \texttt{line-racer} \citep{Hagele2026}. 

\subsection{Chemistry}\label{sec:chem}
We make several updates to the treatment of chemistry in this work compared to \citet{Xuan2026}. First, we construct a new chemical equilibrium table where [C/H], [O/H], and [S/H] vary independently using \texttt{easyCHEM} \citep{Lei2025}. The \texttt{easyCHEM} calculation accounts for the condensation of a dozen different species. The grid covers $[60\,\mathrm{K},\,3500\,\mathrm{K}]$ in temperature ($T$), $[10^{-7}\,\mathrm{bar},\,10^{2}\,\mathrm{bar}]$ in pressure ($P$), and $[0~\mathrm{dex},\,1.2~\mathrm{dex}]$ in [C/H], [O/H], and [S/H], with 0.06 dex spacing. We restrict the grid to super-solar abundances as our preliminary retrievals show AF Lep b's atmosphere is enriched in C, O, and S. A grid that extends to -1.2 dex requires significantly more memory, which we found to be impracticable computationally.

Second, instead of fitting for the carbon quench pressure $P_{\rm quench}$, we now fit directly for $K_{\rm zz}$ to compute the disequilibrium abundance profiles for CO, CH$_4$, H$_2$O, CO$_2$, and NH$_3$. In essence, we invert the process of computing $K_{\rm zz}$ from a retrieved $P_{\rm quench}$ \citep{Xuan2022} and find the implied $P_{\rm quench}$ given a $K_{\rm zz}$ value and the chemical reaction timescales. Specifically, the $P_{\rm quench}$ for each chemical reaction is obtained by equating the chemical reaction timescale to the mixing timescale, which is given by 

\begin{equation}
    \tau_{\rm mix} = L^2 / K_{\rm zz} 
\end{equation}

where we take $L$ to be the pressure scale height. In this paper, we assume that the effective $K_{\rm zz}$ at the quench points for CO, CO$_2$, and NH$_3$ is the same. Theoretical work on atmospheric mixing predicts that young, low-gravity imaged planets should quench in the convective zone (instead of the radiative zone), where $K_{\rm zz}$ is controlled by convective mixing \citep{Mukherjee2024} and generally described by mixing-length theory \citep{Gierasch_convect1985}. We also test a model where the $K_{\rm zz}$ at the CO$_2$ quench point is allowed to be different than that at the CO-CH$_4$ quench point (see Section~\ref{sec:kzz} for results). We take the chemical timescale equations from \citet{Zahnle_methane_2014} and \citet{Mukherjee2022}.
For the quenching of CO$_2$, we first calculate the disequilibrium H$_2$O and CO profiles, which are then used to calculate the pseudo-equilibrium CO$_2$ profile and the quenched CO$_2$ abundance, since CO$_2$ maintains equilibrium with H$_2$O and CO above their quench points \citep{Moses+11,Zahnle_methane_2014,Tsai+18,Wogan+25}.
This approach allows us to derive the disequilibrium abundance profiles for CO, CH$_4$, H$_2$O, CO$_2$, and NH$_3$ consistently, without having to fit CO$_2$ as a vertically constant abundance which was done in e.g. \citet{Xuan2026}.

\subsection{Clouds and thermal structure}\label{sec:clouds}
For the cloud model, we follow recent retrieval work by freely retrieving the mean cloud particle radius ($r_c$) and cloud base pressure ($P_{\mathrm{base}}$) \citep{Nasedkin2025, Xuan2026}. This approach provides more flexibility and relaxes assumptions of the classic EddySed model \citep{ackerman_Precipitating_2001}. The particle size distribution is assumed to be lognormal, and we retrieve its width $\sigma_g$. In this model, the cloud mass fraction is assumed to be constant with pressure above the cloud base, which corresponds to the expectation for a well-mixed atmosphere \citep{Gao2018}. Finally, $X_{\rm cloud}$ is the cloud mass fraction at the cloud base. Each cloud species has a different $r_c$, $P_{\mathrm {base}}$, and $X_{\rm cloud}$. 

We tested different cloud species (MgSiO$_3$, Mg$_2$SiO$_4$, KCl), combinations of them, as well as different cloud particle shape prescriptions between spherical (Mie scattering) and irregularly shaped (DHS or distribution of hollow spheres, \citealt{min_Modeling_2005}) particles. The cloud particles are assumed to be crystalline. Table~\ref{table:aflep_spec_results} summarizes results from different cloud models.

\begin{deluxetable*}{ll|ll}[t!]
\tablecaption{Fitted Parameters and Priors for AF Lep b Retrieval\label{tab:param_prior}}
\tabletypesize{\small}
\tablehead{
\colhead{Parameter} & \colhead{Prior} & \colhead{Parameter} & \colhead{Prior}
}
\startdata
Mass ($\Mj$)                 & $\mathcal{N}(3.75, 0.50)$ 
  & $T_{\rm ref}$ [$P=10^{-1}$] (K) & $\mathcal{U}(300, 1500)$ \\
Radius ($\Rj$)               & $\mathcal{N}(1.30, 0.05)$ 
  & $\left(d\ln{T}/d\ln{P}\right)_{1}$ [$10^2$] & $\mathcal{N}(0.15, 0.01)$ \\
$[{\rm C/H}]$                & $\mathcal{U}(0.0,1.2)$ 
  & $\left(d\ln{T}/d\ln{P}\right)_{2}$ [$10^1$] & $\mathcal{N}(0.18, 0.04)$ \\
$[{\rm O/H}]$                     & $\mathcal{U}(0.0,1.2)$ 
  & $\left(d\ln{T}/d\ln{P}\right)_{3}$ [$10^0$] & $\mathcal{N}(0.21, 0.05)$ \\
$[{\rm S/H}]$                     & $\mathcal{U}(0.0,1.2)$ 
  & $\left(d\ln{T}/d\ln{P}\right)_{4}$ [$10^{-1}$] & $\mathcal{N}(0.16, 0.06)$ \\
\logco  & $\mathcal{U}(0, 8)$  & $\left(d\ln{T}/d\ln{P}\right)_{5}$ [$10^{-2}$] & $\mathcal{N}(0.08, 0.025)$ \\
 $\log_{10}(K_{\rm zz}/\mathrm{cm^2\,s^{-1}})$ & $\mathcal{U}(2,13)$ & $\left(d\ln{T}/d\ln{P}\right)_{6}$ [$10^{-3}$] & $\mathcal{N}(0.06, 0.02)$ \\
$\log(r_{\rm cloud} / \rm {cm})$  & $\mathcal{U}(-7, 1)$ 
  & $\left(d\ln{T}/d\ln{P}\right)_{7}$ [$10^{-4}$] & $\mathcal{U}(-0.05, 0.10)$ \\
 $\log(P_{\rm cloud}/{\rm bar})$ & $\mathcal{U}(-6, 2)$  & $\left(d\ln{T}/d\ln{P}\right)_{8}$ [$10^{-5}$] & $\mathcal{U}(-0.05, 0.10)$ \\
$\sigma_{\rm g}$  & $\mathcal{U}(1.05, 3)$  & $\left(d\ln{T}/d\ln{P}\right)_{9}$ [$10^{-6}$] & $\mathcal{U}(-0.05, 0.10)$ \\
${\rm log}(X_{\rm cloud})$   & $\mathcal{U}(-8, 0)$ & $\left(d\ln{T}/d\ln{P}\right)_{10}$ [$10^{-7}$] & $\mathcal{U}(-0.05, 0.10)$ \\ \hline
Additional parameters for NIRSpec \\ \hline
RV (\kms) & $\mathcal{U}(-100, 100)$ & $r$ & $\mathcal{U}(300, 1400)$  \\ 
Error multiple$^{\rm (a)}$ & $\mathcal{U}(1 , 4)$ & $r_0$ & $\mathcal{U}(-800, 1200)$  \\ \hline
Additional parameters for SPHERE/IFS \\ \hline
${\rm log}(f_{\rm amp})$ & $\mathcal{U}(10^{-4}, 1)$ & ${\rm log}(l)$ ($\mu$m) & $\mathcal{U}(10^{-3}, 0.5)$ \\ \hline
\enddata
\tablecomments{
$\mathcal{U}$ denotes a uniform distribution with bounds in parentheses while $\mathcal{N}$ denotes a Gaussian distribution with the mean and standard deviation in parentheses. Note that the $K_{\rm zz}$ parameter here only controls the disequilibrium chemical abundances (Section~\ref{sec:chem}), and is not used in the cloud model. The mass and radius priors come from \citet{Balmer2025}. \\
$^{\rm (a)}$ Error inflation term multiplied to the NIRSpec covariance matrix.  \\
}
\end{deluxetable*}

For the thermal structure, we use the temperature gradient profile introduced by \citet{Zhang2023}, which fits for $(d\ln{T}/d\ln{P})$ values in different pressure layers, and a single reference temperature, $T_{\rm ref}$. To encompass the emission contribution function of AF Lep b, we choose 10 pressure layers spaced logarithmically between $10^{2}$ and $10^{-7}$ bars, and fit the reference temperature at $0.1$ bars. Given $T_{\rm ref}$ and quadratically interpolated $(d\ln{T}/d\ln{P})$ values, we construct the full P-T profile over 100 pressure layers for radiative transfer calculations. Following \citet{Zhang2025}, we apply Gaussian priors on the temperature gradients between $10^2$--$10^{-3}$ bars which are motivated by self-consistent atmospheric models. Outside this pressure range, we adopt wide uniform priors (see Table~\ref{tab:param_prior}). 

\subsection{Mass and radius priors}\label{sec:mr_priors}
In the atmospheric retrievals, we adopt Gaussian priors for the planet mass and radius. For the planet mass, we use the dynamical mass of $3.75\pm0.50~\Mj$ from \citet{Balmer2025}, whereas for the radius, we use a Gaussian prior of $1.30\pm0.05~\Rj$ obtained by interpolating the substellar evolutionary models from \citet{Saumon_2008} with the dynamical mass and system age \citep{Balmer2025}. These priors help prevent the retrieval from entering nonphysical parameter spaces, such as finding radii that are too small as has been observed in previous work \citep[e.g.][]{Zhang2023}.

\begin{deluxetable}{ccccc}\label{tab:stellar_ab}
\tabletypesize{\footnotesize}
\tablecaption{Elemental Abundances of Stars in $\beta$ Pic Moving Group}
\tablehead{
\colhead{Star} & \colhead{[C/H]} &  \colhead{[O/H]} & \colhead{[S/H]} &  Abundance reference
}
\startdata
HD 181327          & $-0.05 \pm 0.06$                 & $-0.10 \pm 0.06$                 & $0.08 \pm 0.20$ & 1 \\
51 Eri            & $0.03 \pm 0.08$                 & $0.04 \pm 0.08$                 & $-0.01 \pm 0.12$ & 2 \\
PZ Tel            & $-0.04 \pm 0.07$                 & $0.26 \pm 0.04$                 & $0.19 \pm 0.23$ & 3 \\
AF Lep            & $0.05\pm0.05$          & $0.15^{+0.14}_{-0.11}$          & $0.00 \pm 0.09$ & 4 \\
\hline
Weighted mean $\pm$ scatter & $0.00 \pm 0.05$ & $0.13 \pm 0.15$  & $0.02 \pm 0.09$ \\
\enddata
\tablecomments{The values in the final row are the weighted mean of all stars and the standard deviation of the different values. Since all three values in the final row are consistent with solar at the $<1\sigma$ level, we adopt the solar abundance as the reference abundance for AF Lep b in this paper.}
\tablerefs{(1) \citet{Reggiani2024}, (2) \citet{Baburaj2025}, (3) \citet{Baburaj2026}, (4) \citet{Baburaj2026c}
}
\end{deluxetable}

\section{Stellar abundances}\label{sec:stellar_abunds}
To interpret the abundances of AF Lep b, we compile C, O, and S abundance measurements for stars in the $\beta$ Pic moving group (BPMG). We choose to adopt a population-level abundance reference for the moving group instead of relying on the measurement of a single star to mitigate possible systematics, for example associated with the non-LTE correction for the oxygen abundance \citep[e.g.][]{Reggiani2024}. We first collect abundances of 51 Eri \citep{Baburaj2025}, HD 181327 \citep{Reggiani2024}, and PZ Tel \citep{Baburaj2026}, all bona fide members of BPMG \citep{Zuckerman2001}. To complement these measurements, we also present new C, O, and S abundances for AF Lep A, following the methodology of \citet{Baburaj2026}. While the observations and data analysis are detailed in \citet{Baburaj2026c}, we briefly describe the procedure here. 

The initial analysis involved determination of the stellar atmospheric parameters effective temperature ($T_{\rm eff}$), gravity ($\log{g}$), and metallicity ($\rm [M/H]$). These stellar parameters were used to generate synthetic grids using \texttt{PySME}, a python wrapper for the spectroscopic analysis code Spectroscopy Made Easy (SME; \citealt{Valenti:1996,Piskunov:2017}), the MARCS model atmospheres \citep{Marcs:2008}, and the Vienna Atomic Line Database (\texttt{VALD}; \citealt{Piskunov:2015, Ryabchikova:2015}). We use solar abundances from \citet{asplund_Chemical_2009} as our baseline. One set of grids had varying carbon and oxygen abundances, while the other set of grids had varying carbon and sulfur abundances. For each element among C, O, S, spectral features chosen had negligible contribution from atomic lines of the other two elements. Spectral fitting using the python package \texttt{SMART} \citep{Hsu:2021} was used to obtain abundances of carbon, oxygen, and sulfur. For oxygen, we subsequently apply NLTE corrections from \citet{amarsi_Carbon_2019}. This yields [C/H] = $0.05\pm0.05~\mathrm{dex}$, [O/H] = $0.15^{+0.14}_{-0.11} ~\mathrm{dex}$, and [S/H] = $0.00 \pm 0.09 ~\mathrm{dex}$.

The stellar abundances are listed in Table~\ref{tab:stellar_ab}. From these, we computed the weighted average and scatter of the measurements as proxies for the abundance of the natal molecular cloud where AF Lep b formed. We find $[\mathrm{C}/\mathrm{H}] = 0.00 \pm 0.05~\mathrm{dex}$, $[\mathrm{O}/\mathrm{H}] = 0.13 \pm 0.15~\mathrm{dex}$, and $[\mathrm{S}/\mathrm{H}] = 0.02 \pm 0.09~\mathrm{dex}$, all consistent with solar at the $<1\sigma$ level. If we instead only consider AF Lep A, this star is also solar in C, O, and S at the $<1.5\sigma$ level. Similar to \citet{Xuan2026}, we therefore adopt the solar abundance as the reference for normalizing the measurements of AF Lep b and other substellar members of BPMG. For consistency with the retrievals, we use the \citet{asplund_Chemical_2009} solar photospheric abundance. To account for the limited sample size (4) and the observed scatter in the measurements between stars, we include a conservative 0.05 dex uncertainty on the solar reference abundances for C, O, and S.

\section{Results}\label{sec:results}

\begin{figure*}[t]
\centering
\includegraphics[width=\linewidth]{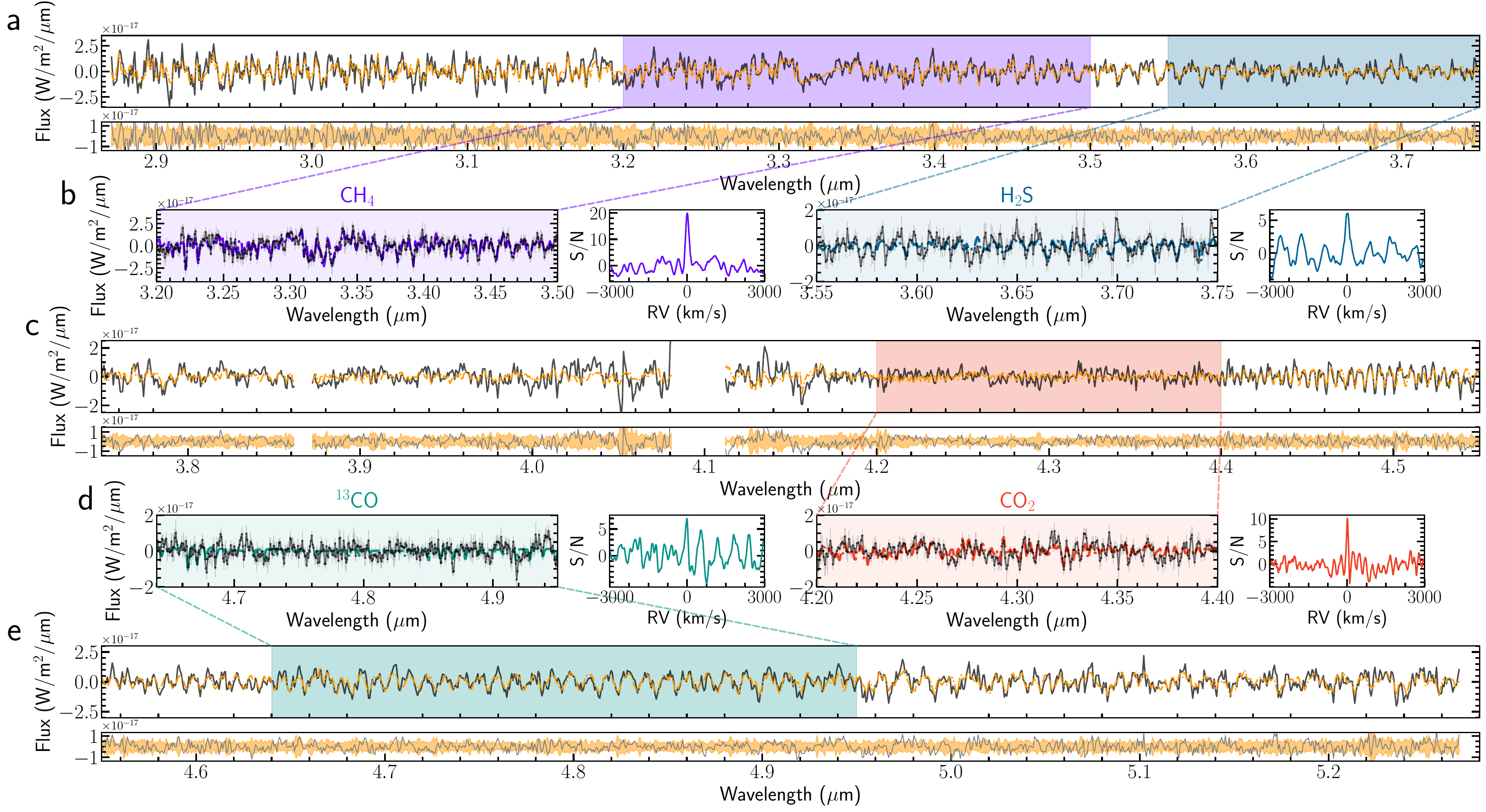}
\includegraphics[width=0.9\linewidth]{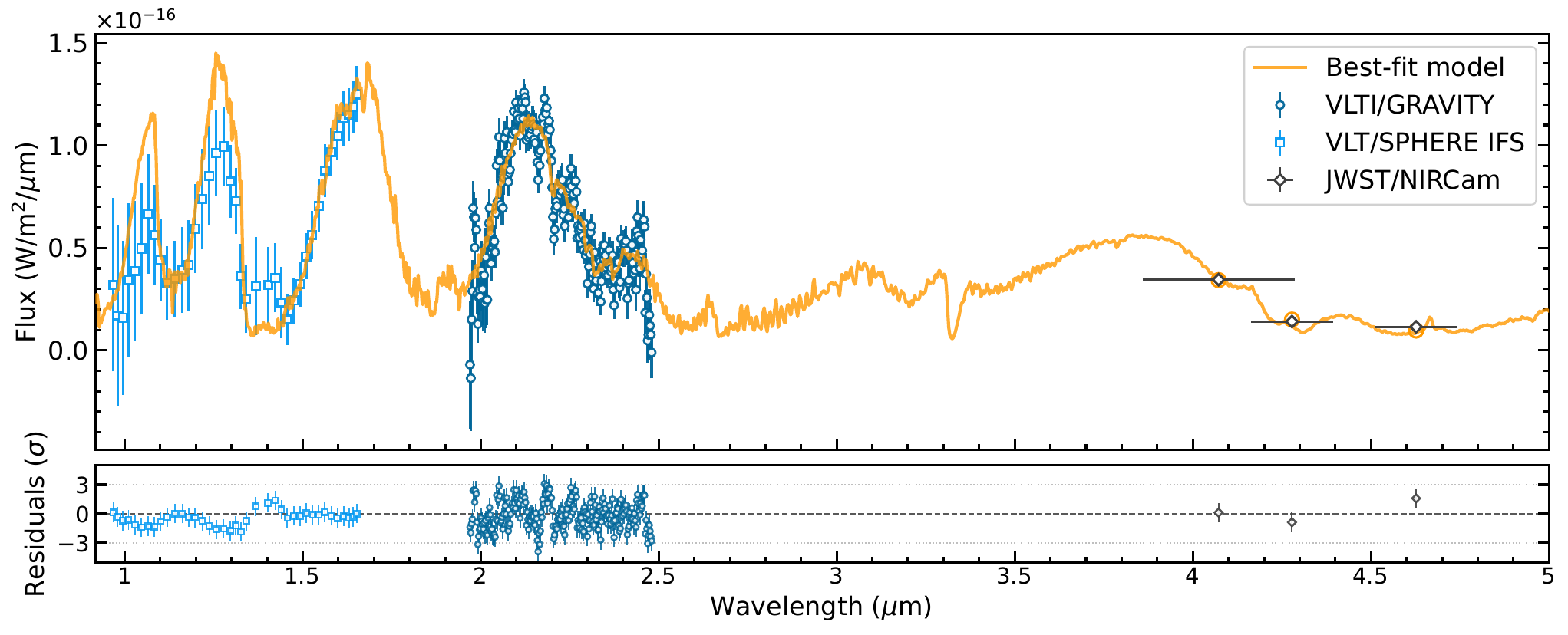}
\caption{\textbf{Top:} JWST/NIRSpec spectrum of AF Lep b. Panels a, c, e show the observed spectrum ($R\sim2,700$) in black and the best-fit \texttt{petitRADTRANS} model in orange. In the sub-panels below, the data residuals after subtracting the best-fit model are plotted as gray lines and the inflated uncertainties are shown as orange contours. Panels b and d show data residuals after fitting an atmospheric model without a given species (CH$_4$, H$_2$S, $^{13}$CO, CO$_2$) in black, and the corresponding molecular templates in color. The similarity between the data residuals and molecular templates indicates that the highlighted species contribute to the planet's spectra. On the right insets, we plot the cross-correlation functions (CCF) between the data residuals and models that are shown in the left insets, which allows us to assess the detection significance. \textbf{Bottom:} JWST/NIRCam photometry (black) and ground-based spectra of AF Lep b from VLT/SPHERE (light blue) and VLTI/GRAVITY (dark blue). The best-fit model is plotted in orange at $R=1000$. The model photometry is shown as the orange circles. These data are fitted jointly with the NIRSpec spectrum above in the retrieval. The residuals are included in the bottom panel.}\label{fig:spec}
\end{figure*}

\subsection{Molecular detections}\label{sec:free_retr}
In the JWST/NIRSpec spectra of AF Lep b, we detect CO, $^{13}$CO, H$_2$O, CH$_4$, CO$_2$, and H$_2$S. We carry out leave-one-out tests and cross-correlation analyses to assess the significance of these detections following the procedure in \citet{Xuan2026}. In short, we run a set of free retrievals where we leave one molecular species out at a time (reduced models), and a free retrieval where all molecules are included. We ran four reduced models leaving out CH$_4$, CO$_2$, H$_2$S, and $^{13}$CO opacities one at a time. Then, we plot the residuals of the reduced models against a single molecular template, and calculate the cross-correlation function (CCF) between them. As shown in Figure~\ref{fig:spec}, the data residuals after subtracting the best-fit reduced model where H$_2$S was left out show several H$_2$S lines from the planet. From the resulting CCFs, we detect H$_2$S at $6\sigma$, CH$_4$ at $20\sigma$, CO$_2$ at $10\sigma$, and $^{13}$CO at $6\sigma$.

\subsection{Atmospheric metal enrichment}
We find metal enrichment across C, O, and S for AF Lep b, with $\rm C/H=2.9\pm0.5$, $\rm O/H=3.7\pm0.6$, and $\rm S/H=4.7\pm0.7$ relative to the adopted abundance reference. As explained in Section~\ref{sec:stellar_abunds}, we find the average C, O, and S abundances of several members of the $\beta$ Pic moving group are consistent with solar, and therefore adopt solar as our abundance reference. We note that whether the planetary C/O differs from the stellar C/O does depend on the adopted abundance reference. For example, as shown in Table~\ref{tab:stellar_ab}, AF Lep A has C/O = $0.44^{+0.15}_{-0.12}$ whereas HD 181327 has C/O = $0.62\pm0.08$, even though both these stars are consistent with solar C/O at the $<1\sigma$ level. From our retrievals, the planet AF Lep b has C/O = $0.43\pm0.02$. Mitigating potential systematics in the individual stellar abundance measurements is one of the reasons we chose to use the solar abundance as the reference in this paper. Given that choice, C/O, C/S, and O/S ratios for AF Lep b are slightly sub-solar and sub-stellar, with $\mathrm{C/O}=0.78^{+0.14}_{-0.12}\times$ solar, $\mathrm{C/S}=0.61^{+0.14}_{-0.11}\times$ solar, and $\mathrm{O/S}= 0.79^{+0.17}_{-0.14}\times$ solar. We place these elemental abundances in the context of planet formation in Section~\ref{sec:discuss}.

\begin{figure*}[t!]
    \centering
    \includegraphics[width=0.75\linewidth]{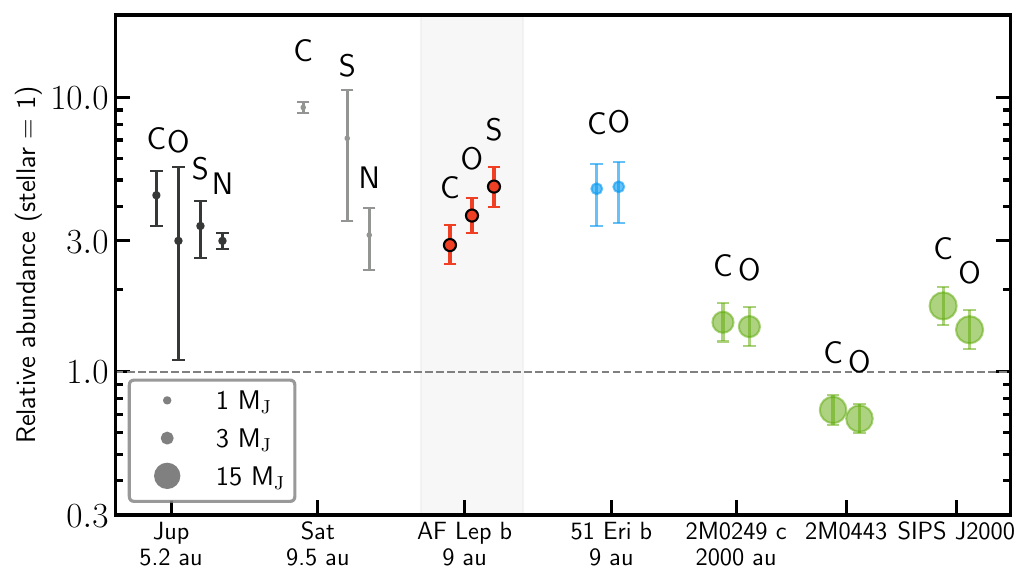}
    \caption{Elemental abundances relative to solar values for AF Lep b (red). We also plot measurements for 51 Eri b (blue) from \citet{Madurowicz2025} and three $\beta$ Pic moving group brown dwarfs in green: 2MASS J0249-0557 c, 2MASSI J0443+0002, and SIPS J2000-7523, whose abundances are from \citet{Liu2026}. The latter two brown dwarfs are isolated objects. All BPMG objects are normalized using the solar abundance, which is adopted as the reference abundance for BPMG (Section~\ref{sec:stellar_abunds}). The abundances of Jupiter and Saturn \citep{Wong2004, Li2020, Briggs1989, Fletcher2009} are shown in black and gray, respectively. The circle sizes are proportional to object mass. AF Lep b exhibits simultaneous enrichment in C, O, and S, akin to Jupiter, and is distinctly more metal-rich than the brown dwarfs.}
    \label{fig:abunds}
\end{figure*}

In Figure~\ref{fig:abunds}, we plot the measured C, O, and S abundances of AF Lep b. We compare AF Lep b to the solar system gas giants, 51 Eri b \citep{Madurowicz2025},\footnote{Since only carbon and oxygen bearing molecules have been detected in 51 Eri b, we convert the reported [M/H] and C/O from the Sonora Elf Owl fits in \citet{Madurowicz2025} to C/H and O/H. We then normalized the values to $\times$ stellar units using the adopted stellar abundances for BPMG in Section~\ref{sec:stellar_abunds}.}  and three brown dwarfs in BPMG \citep{Liu2026}. 51 Eri b is a close cousin of AF Lep b: both planets belong to the BPMG, and have similar masses ($\sim3~\Mj$ v.s. $3.75\pm0.50~\Mj$) and semi-major axes (both have $a\approx9$ au). The two planets show similar levels of carbon and oxygen enrichment. In contrast, the brown dwarfs trace the adopted composition of the moving group at the $<0.2$ dex level. The recently discovered planet $\beta$ Pic d \citep{Gibbs2026, Sutlieff2026} has a similar mass and $\Teff$ as 51 Eri b, and future atmospheric retrieval studies could also test whether this planet shares a similar degree of metal enrichment as found for AF Lep b and 51 Eri b.

Here, we adopt the \citet{Madurowicz2025} abundances for 51 Eri b, which are consistent with a recent analysis from \citet{Balmer2025b}. However, we note that some previous studies showed disagreement in 51 Eri b's abundances \citep{BrownSevilla2023, Whiteford2023}. Higher S/N and higher-resolution spectra of 51 Eri b will confirm the metal enrichment trends seen in Fig.~\ref{fig:abunds}, and test whether 51 Eri b is also enriched in sulfur like AF Lep b. Finally, we note that a recent VLT/CRIRES+ study combining 11 different nights of observations of $\beta$ Pic b ($m\approx12~\Mj$; \citealt{Lacour2021}) has found evidence of modest metal enrichment with $\rm[C/H]=0.20^{+0.16}_{-0.12}$, and a solar-like C/O with $\rm C/O=0.52\pm0.03$ \citep{GonzalesPicos2026}. Like 51 Eri b, $\beta$ Pic b has several abundance measurements in the literature \citep[e.g.][]{GRAVITY_2020, Landman2024, Ravet2025, GonzalesPicos2026}, but the more recent studies tend to find a super-solar metallicity for its atmosphere.

\renewcommand{\arraystretch}{1}
\begin{deluxetable*}{cccccccccc}[t!]
\tablecaption{\textbf{Results of Atmospheric Retrieval for AF Lep b.}\label{table:aflep_spec_results}}
\tablewidth{\textwidth}
\tablehead{
\colhead{Model} & \colhead{C/H} & \colhead{O/H} & \colhead{S/H} & \colhead{$^{12}$C/$^{13}$C} & \colhead{Radius ($\Rj$)} & \colhead{$T_\textrm{eff}$ (K)} & \lbollsun & \colhead{$\log_{\rm 10} K_{\rm zz}$} (${\rm cm^2~s^{-1}}$) & \colhead{$\ln{B}$}
}
\startdata
\textbf{MgSiO$_3$, cd (Baseline)} & $2.9\pm0.5$ & $3.7\pm0.6$ & $4.7\pm0.7$ & $121^{+37}_{-26}$ & $1.27\pm0.03$ & $765\pm10$ & $-5.28\pm0.01$ & $7.7\pm0.2$ & - \\
MgSiO$_3$, cm & $3.4\pm0.6$ & $4.3\pm0.6$ & $5.2\pm0.9$ & $120^{+36}_{-27}$ & $1.26\pm0.03$ & $773\pm10$ & $-5.27\pm0.01$ & $7.2\pm0.2$ & $-8.9$ \\
Mg$_2$SiO$_4$, cd & $3.1\pm0.6$ & $3.9\pm0.6$ & $5.6\pm0.9$ & $122^{+32}_{-24}$ & $1.41\pm0.03$ & $734\pm10$ & $-5.26\pm0.01$ & $10.7\pm0.4$ & $-18.5$ \\
MgSiO$_3$ + KCl, cd & $3.1\pm0.5$ & $4.0\pm0.6$ & $6.1\pm0.9$ & $105^{+29}_{-22}$ & $1.11\pm0.02$ & $805\pm10$ & $-5.31\pm0.01$ & $6.2\pm0.2$ & $-77.3$ \\
\enddata
\tablecomments{Central 68\% credible intervals with equal probability above and below the median. The C/H, O/H, and S/H values are quoted relative to solar \citep{asplund_Chemical_2009}, which is adopted as the abundance reference (see discussion in Section~\ref{sec:stellar_abunds}). For the models, `cd' stands for crystalline particles + DHS model, and `cm' stands for crystalline particles + Mie scattering, as described in \S~\ref{sec:clouds}.  The last column shows the log Bayes factor ln($B$) for each model, where we compute ln($B$) with respect to the baseline MgSiO$_3$, cd model.}
\end{deluxetable*}

\subsection{Carbon isotopic ratio}
We measure $\mathrm{^{12}CO / ^{13}CO}=121^{+37}_{-26}$ for AF Lep b. Other members of the $\beta$ Pic moving group provide useful benchmarks with which we can compare the isotopic ratio measured for AF Lep b. From a high-resolution retrieval study with VLT/CRIRES+, \citet{Liu2026} measured $95^{+23}_{-17}$ for the $\sim12~\Mj$ substellar companion 2MASS J0249-0557 c, and $70\pm5$ for the $\sim18~\Mj$ isolated brown dwarf 2MASSI J0443+0002. More recently, \citet{vonStauffenberg2026} measured $91^{+24}_{-17}$ for $\beta$ Pic b using VLTI/GRAVITY+ medium-resolution spectroscopy ($R\sim4000$) and \citet{GonzalesPicos2026} measured a somewhat lower ratio of $58^{+18}_{-15}$ for the same planet using VLT/CRIRES+. Compared to AF Lep b and $\beta$ Pic b, 2MASS J0249-0557 c has a very distinct system architecture: it orbits a binary pair of low-mass M dwarfs at a projected separation of $\sim2000$ AU \citep{Dupuy2018}. Before using these $\mathrm{^{12}C/^{13}C}$ ratios for formation inferences, more modeling work is required to clarify how sensitive $\mathrm{^{12}C / ^{13}C}$ may be to different formation pathways. We also note that AF Lep b's $^{13}$CO detection is only at the $6\sigma$ level in the current NIRSpec data, and higher S/N data is needed to test if there are any significant isotopic ratio differences among BPMG members.

\begin{figure}
    \centering
    \includegraphics[width=\linewidth]{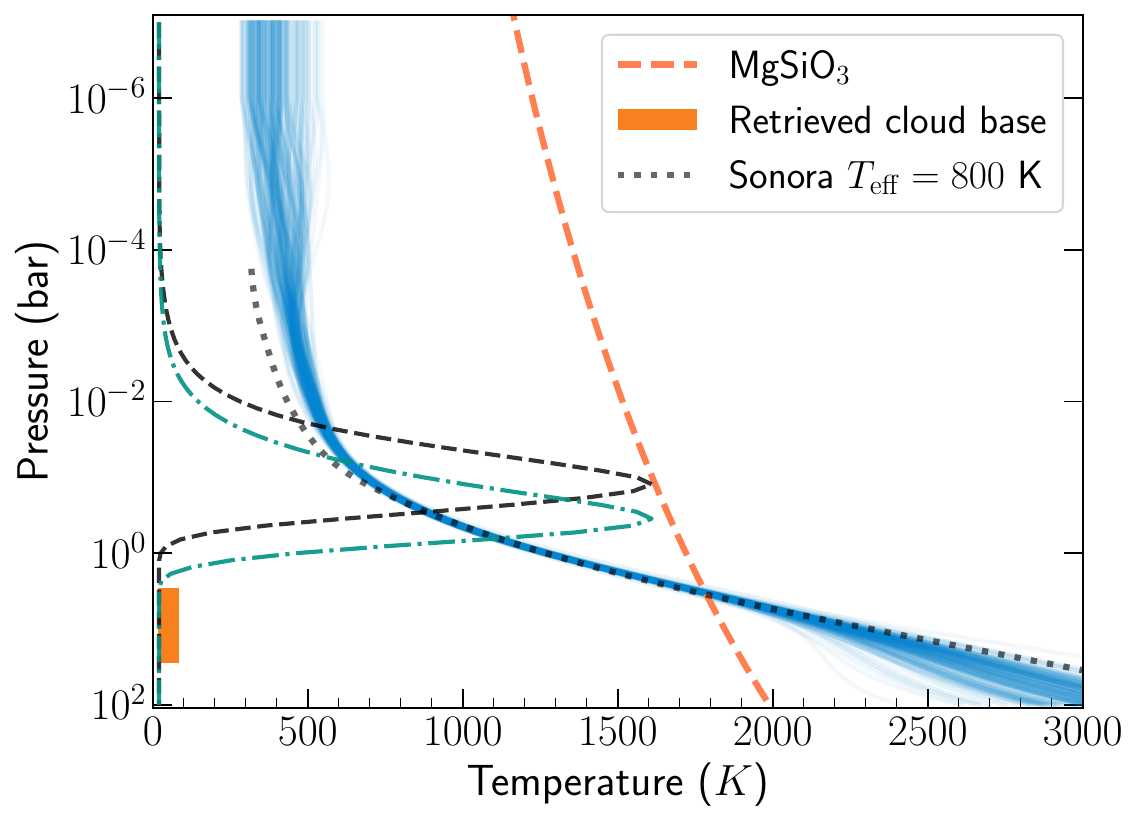}
    \caption{Random draws of the retrieved P-T profile in blue for AF Lep b. We overplot a self-consistent P-T profile from Sonora Elf Owl as the gray dotted line ($\Teff=800~$K, $\logg=3.5$). On the left side, the black dashed line indicates the weighted contribution function for the NIRSpec model ($2.85$--$5.3~\mu$m), whereas the teal dashdot line is the weighted contribution function for the $0.9$--$2.5~\mu$m model used to fit the SPHERE and GRAVITY data. The dashed orange line shows the equilibrium condensation curve for MgSiO$_3$, and the vertical orange bar indicates the $2\sigma$ retrieved cloud base pressure, which is close to the predicted cloud base from the intersection of the orange dashed line and the P-T profiles.}
    \label{fig:pt_emis}
\end{figure}

\subsection{Thermal structure, bulk parameters, and clouds}\label{sec:clouds_pt}
The retrieved PT profile for AF Lep b is shown in Figure~\ref{fig:pt_emis}, and broadly follows the shape of self-consistent PT profiles given the \citet{Zhang2023} style priors we imposed (see Table~\ref{tab:param_prior}) to prevent isothermal behavior. In the PT plot, we also show the weighted contribution functions for the NIRSpec model ($2.85$--$5.3~\mu$m) and the low-resolution model ($0.9$--$2.5~\mu$m).

In terms of bulk properties, we retrieve a planet radius of $1.27\pm0.03~\Rj$, close to the imposed Gaussian prior of $1.30\pm0.05~\Rj$. By integrating the models from $0.15$--$30~\mu$m, we obtain $\lbollsun=-5.28\pm0.01$ and $\Teff=765\pm10~$K for the planet. Given the dynamical mass and system age of $24\pm3$ Myr, evolutionary models from \citet{Saumon_2008} predict $\Teff=770\pm75~$K and $\lbollsun=-5.26\pm0.18$ \citep{Balmer2025}, which are consistent with our retrieved values. \citet{Balmer2025} noted that many of their retrieval models yielded nonphysically low radii $<1~\Rj$. In exploring different cloud and chemical prescriptions, we found that the parametrization of disequilibrium chemistry had the most noticeable impact in the retrieved radius. Specifically, parameterizing the chemistry using a carbon quench pressure with vertically constant CO$_2$ abundance converged to a smaller radius of $\approx1.18~\Rj$, despite the Gaussian radius prior. This model is also significantly disfavored in Bayesian evidence compared to the baseline model where we directly retrieve $K_{\rm zz}$ (see details in Section~\ref{sec:chem}).

Between the various cloud models we tried, the adopted baseline model with the highest Bayesian evidence has one cloud species (MgSiO$_3$) with crystalline cloud particles (DHS). As shown in Table~\ref{table:aflep_spec_results}, alternative retrievals with multiple cloud species (MgSiO$_3$ + KCl) were disfavored in Bayesian evidence compared to the baseline model. In addition, models with alternative cloud species such as Mg$_2$SiO$_4$ or KCl and different cloud particle shapes (Mie scattering instead of distribution of hollow spheres) were also disfavored. The different cloud models result in C, O, and S abundances that are consistent with the baseline model at the $<1\sigma$ level, although bulk parameters such as radius, $\Teff$, and $K_{\rm zz}$ can be more discrepant in the lower ln($B$) models.

From the baseline model, the retrieved MgSiO$_3$ cloud base pressure is $9.4^{+9.2}_{-4.7}$ bars, which is broadly consistent at the $\approx1.5~\sigma$ level with the expected MgSiO$_3$ cloud base location as determined by the intersection of the cloud condensation curve and the P-T profile ($\approx 3$ bars; see Figure~\ref{fig:pt_emis}). This agreement is consistent with silicate clouds being the dominant cloud species in AF Lep b's atmosphere. We note that the retrieved cloud base pressure does not exclusively correspond to the condensation point of MgSiO$_3$, as Mg$_2$SiO$_4$ condenses at approximately the same pressure. In addition, we note that even in the best-fit models with MgSiO$_3$ cloud particles, the $y$ and $J$ band peaks of the VLT/SPHERE data are systematically lower than the model flux at the $\approx1.5\sigma$ level. In the future, mid-infrared spectroscopy at $\sim10~\mu$m of AF Lep b is likely needed to more confidently determine the exact cloud species in its atmosphere.

Previous work on the HR 8799 planets \citep{Xuan2026} also noted close agreement between the retrieved and expected cloud base locations. From the baseline model, we retrieve a mean cloud particle radius of $r_c\approx6.6~\mu{\rm m}$, with a narrow lognormal width of $\sigma_{\rm lnorm} \approx 1.1$. This corresponds to a 68\% cloud radius range of $\approx6.0$--$7.1~\mu{\rm m}$. More recent cloud models have suggested the emergence of an elevated cloud layer at lower pressures than predicted by equilibrium cloud condensation, featuring cloud particles with smaller, sub-micron sizes (Cukier et al. in prep). Motivated by this, we tested alternative cloud models where we allow a second cloud layer with a different $r_c$ and $\sigma_{\rm lnorm}$, located with $\Delta P$ above the deeper cloud base. However, these models were not favored compared to our fiducial, single-layer cloud model (Section~\ref{sec:clouds}). In addition, we tested power-law cloud particle size distributions, as alternatives to the lognormal distribution, and also found that the data showed no preference for these. 

\begin{figure}
    \centering
    \includegraphics[width=1\linewidth]{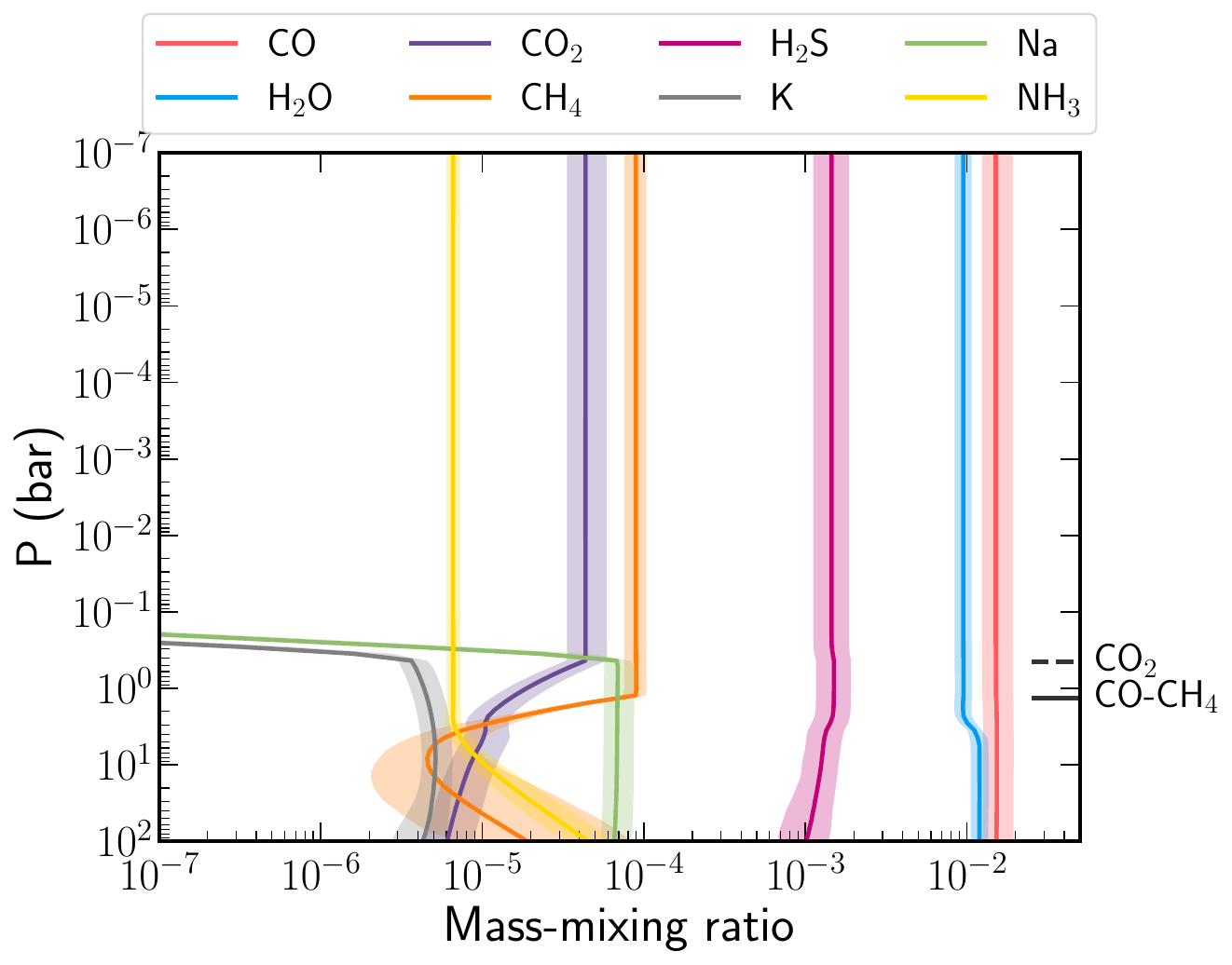}
    \caption{The retrieved mass-mixing ratios (MMR) of key gas species in AF Lep b's atmosphere. The solid line and colored ranges indicate the median and $2\sigma$ credible interval from the posteriors. The quench pressures for CO-CH$_4$ and CO$_2$ are indicated as solid and dashed black lines. These quench points correspond to the best-fit P-T profile and $K_{\rm zz}$ value.}
    \label{fig:mmr}
\end{figure}

\subsection{Disequilibrium chemistry and $K_{\rm zz}$}\label{sec:kzz}
The mass-mixing ratios of various species in the baseline disequilibrium chemistry retrieval are plotted in Figure~\ref{fig:mmr}, where we indicate the best-fit quench points of CO-CH$_4$ and CO$_2$. From this baseline retrieval, we obtain $\log_{\rm 10}(K_{\rm zz}~[{\rm cm^2~s^{-1}}])=7.7\pm0.2$ for AF Lep b. This value is quite similar to that found for HR 8799 b  ($\log_{\rm 10}(K_{\rm zz}~[{\rm cm^2~s^{-1}}])=7.8\pm0.7$; \citealt{Xuan2026}), a slightly hotter $\Teff\approx930~$K, and more massive $m\approx6~\Mj$ planet. Vertical mixing has long been known to play an important role in brown dwarf spectra, most prominently by changing the balance between CO and CH$_4$ \citep[e.g.][]{Noll1997, Oppenheimer1998, Stephens2009, Zahnle_methane_2014}. For young, imaged planets, early work from \citet{Barman2015} measured a high $\log_{\rm 10}(K_{\rm zz})\sim6$--$8$ for HR 8799 b, which is consistent with the aforementioned results from JWST/NIRSpec in \citet{Xuan2026}. \citet{Barman2011b} also suggested $\log_{\rm 10}(K_{\rm zz})\sim8$ for 2MASS 1207 b, a young, low surface gravity companion with $m\sim12$--$27 ~\Mj$ \citep{Xuan2024b}.

\begin{figure}
    \centering
    \includegraphics[width=\linewidth]{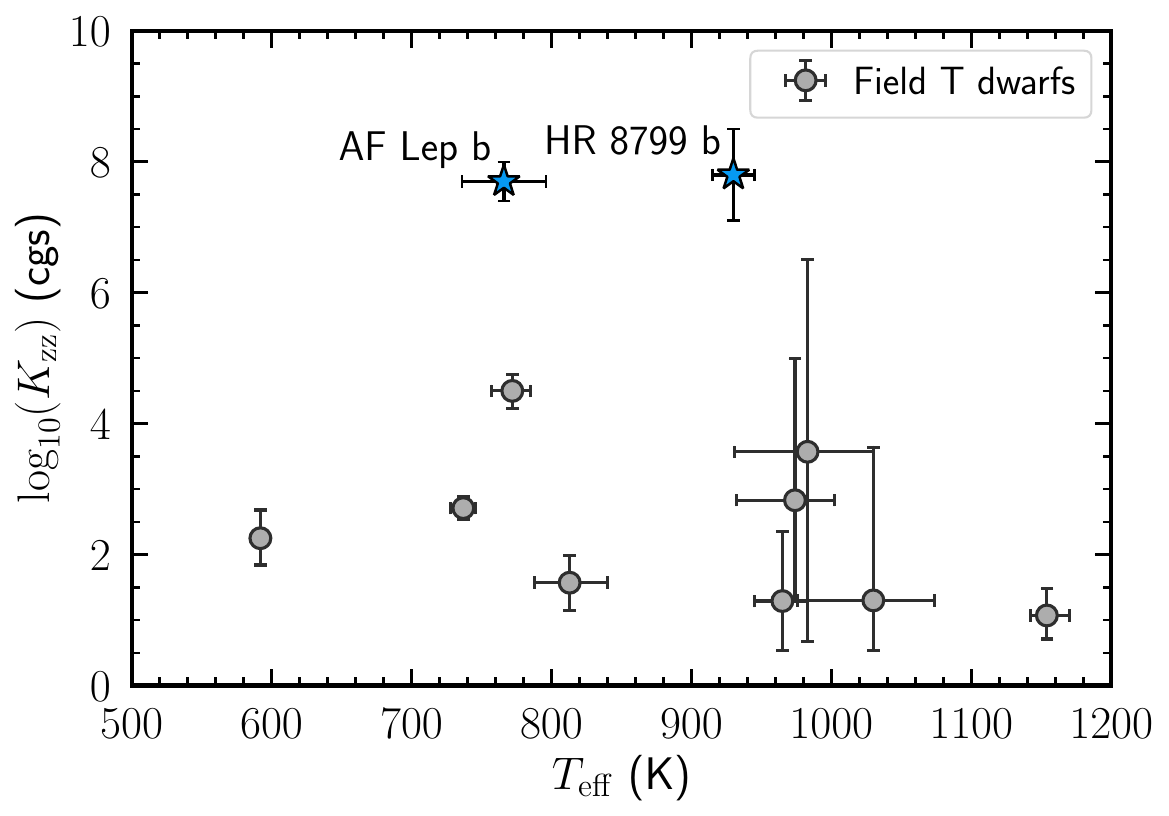}
    \caption{The $K_{\rm zz}$ values plotted against $\Teff$ for AF Lep b and HR 8799 b \citep{Xuan2026} compared to a sample of nine field T dwarfs in \citet{Mukherjee2024}. The two young planets with low surface gravities have higher inferred $K_{\rm zz}$ compared to the brown dwarfs.}
    \label{fig:kzz}
\end{figure}

In Figure~\ref{fig:kzz}, we compare the $K_{\rm zz}$ measured for AF Lep b and HR 8799 b with values measured for field T dwarfs ($\Teff$ spanning from $550$--$1150~$K) from \citet{Mukherjee2024}. The planets have more vigorous vertical mixing than the brown dwarfs, for which \citet{Mukherjee2024} measure $\log_{\rm 10}(K_{\rm zz}~[{\rm cm^2~s^{-1}}])\approx1$--$5$. \citet{Mukherjee2024} propose that the lower $K_{\rm zz}$ for the T dwarfs can be explained by quenching occurring in the radiative zones of atmospheres, due to the high gravities of these objects. In general, whether quenching occurs in the radiative or convective zone of the atmosphere is strongly dependent on surface gravity (see Figure 16 in \citealt{Mukherjee2024}). Indeed, these authors predicted that young, directly imaged planets with similar $\Teff$ but lower gravity should quench in the convective zones, which would result in higher $K_{\rm zz}$ than brown dwarfs. This prediction is supported by these new measurements for AF Lep b ($\logg\approx3.7$) and HR 8799 b ($\logg\approx4$), as shown in Figure~\ref{fig:kzz}. 

We note that our reported $K_{\rm zz}$ value mainly applies to pressures $\sim1$ bars in the atmosphere, where CO and CH$_4$ abundances are quenched. In this study we assume that $K_{\rm zz}$ is constant across the CO-CH$_4$ and CO$_2$ quench points. We also test a model which allowed two different $K_{\rm zz}$ values at the CO-CH$_4$ and CO$_2$ quench points; however, this model is strongly disfavored relative to the single $K_{\rm zz}$ model.

\subsection{Constraints on NH$_3$}
While NH$_3$ is included in our baseline disequilibrium chemistry retrieval, it is not independently detected from the free retrievals, where each species has a constant mass-mixing ratio, discussed in Section~\ref{sec:free_retr}. The baseline retrieval assumes that [N/H] follows [C/H], for which we infer $[\mathrm{C/H}]=0.46^{+0.06}_{-0.05}$. From this baseline retrieval, the expected NH$_3$ mass-mixing ratio at $\approx0.1$ bars (center of NIRSpec emission contribution function) is $\approx6\times10^{-6}$, as shown in Figure~\ref{fig:mmr}. In contrast, the free retrieval places a one-sided upper limit of $\mathrm{MMR(NH_3)}<1.1\times10^{-5}$ at the $3\sigma$ level. The expected NH$_3$ abundance from the disequilibrium chemistry retrieval is below the $3\sigma$ upper limit from the free retrieval. Therefore, the absence of an NH$_3$ detection is not in strong tension with the baseline model, but may disfavor a much more enhanced nitrogen abundance of $\rm [N/H]\gtrsim1.0$, as inferred for HR 8799 b \citep{Xuan2026}. At the NIRSpec G395H wavelengths, the NH$_3$ opacity is strongest at $\approx3~\mu$m, which is part of the NRS1 detector. We find that the NIRSpec NRS1 spectrum for AF Lep b is lower in S/N by a factor of $\approx3$--$4$ compared to that of HR 8799 b, so higher S/N observations of AF Lep b are likely needed to detect NH$_3$ and constrain N/H for this planet.  

\subsection{Planet radial velocity}\label{sec:rv}
The baseline retrieval yields a planetary RV of $14.2\pm0.8$~\kms for AF Lep b at UT 2025-01-17. Accounting for the host star's RV of $21.1\pm0.4$~\kms from \citet{GaiaDR3_2023}, this would yield a relative RV (planet-star) of $-6.9\pm0.9$~\kms at this epoch. While the NIRSpec relative RV value is in agreement with one of the two orbital solutions from \citet{Balmer2025}, it is inconsistent by $\approx15~$\kms compared to the measured RV from VLT/HiRISE and ERIS \citep{Denis2025, Hayoz2025}. Since the HiRISE measurement is from $R\sim100,000$ spectra, it is likely more reliable. Indeed, the JWST/NIRSpec wavelength calibration systematic uncertainty may be as large as $\sim20~$\kms (\url{https://jwst-docs.stsci.edu/jwst-calibration-status/nirspec-calibration-status/nirspec-ifu-calibration-status}). As in \citet{Ruffio2026}, the goal of fitting the planetary RV in this paper is to recalibrate the imperfect wavelength solution such that the models align with the data, instead of obtaining valuable constraints for the orbit (which is already well-constrained by previous work). 

\section{Discussion}\label{sec:discuss}

\begin{figure*}
    \centering
    \includegraphics[width=\linewidth]{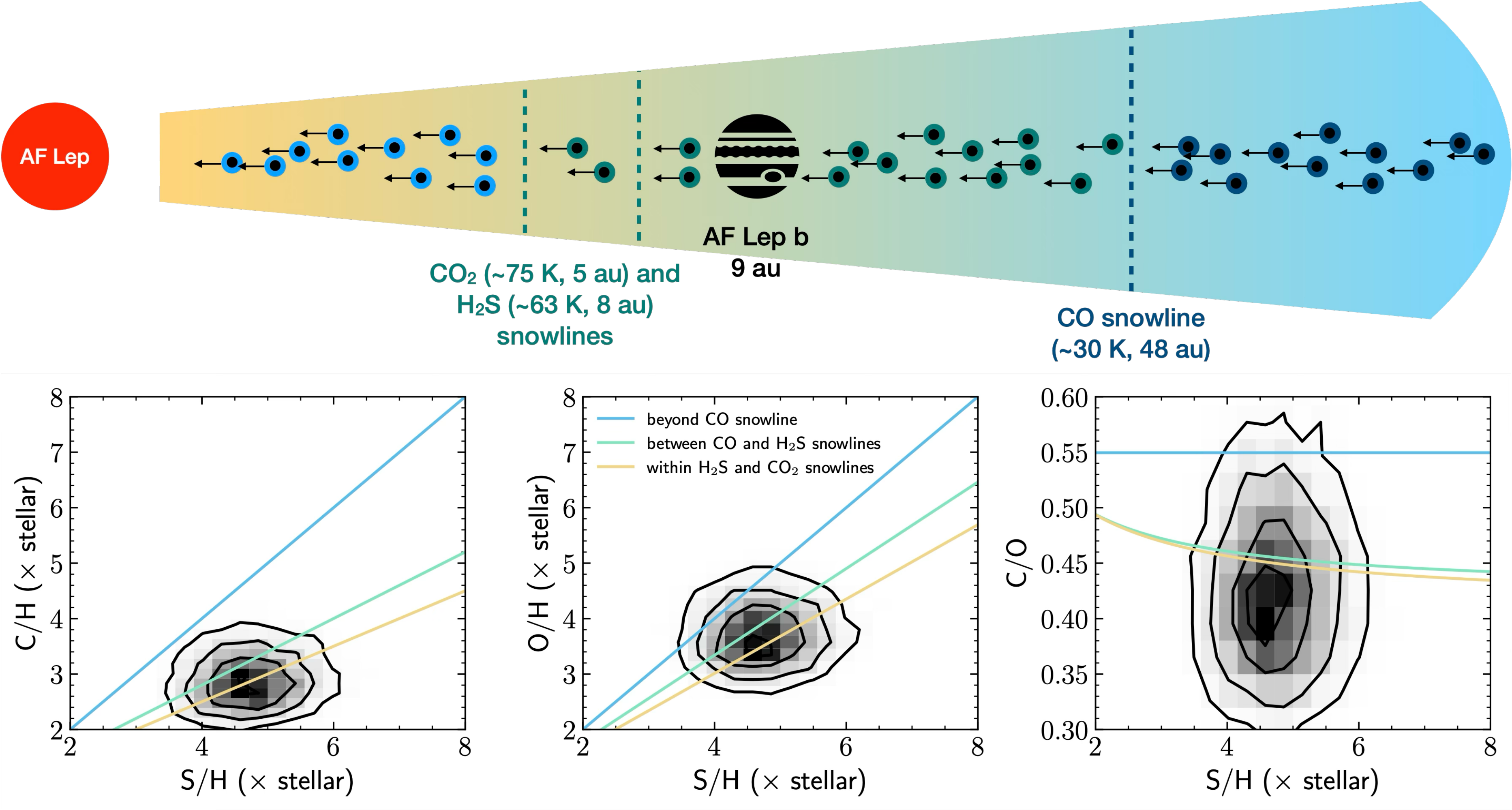}
    \caption{The measured elemental composition of AF Lep b (2D posteriors) and composition curves based on the framework in \cite{Chachan2023}. The S/H of the planet is a tracer of solid accretion as all of S is expected to be in solids at the planet's location. The y-axis shows the C/H, O/H, and C/O expected for accretion of material in different regions (see plot legend). The slightly sub-stellar C/O and C/S disfavor formation beyond the CO snowline and are compatible with accretion of material near AF Lep b's current location.}
    \label{fig:composition-curves}
\end{figure*}

\subsection{Implications for planet formation}
The elemental ratios measured in AF Lep b's atmosphere enable us to put constraints on the building blocks of the planet's envelope. We use the framework devised in \cite{Chachan2023} and Chachan et al. (in prep) to relate the measured atmospheric composition to the material accreted by the planet and to put quantitative constraints on the amount and manner of metal accretion. The different volatilities of C, O, and S allow us to probe the extent of metal accretion by solids and gas as S mostly traces refractory species while substantive fractions of C and O are sequestered in volatile species at AF Lep b's location (9 au). 

The estimated temperature of the protoplanetary disk at the planet's location is $T \sim 62$ K, based on $T_{\rm irr} = 150 \, {\rm K} \, (L_\star / L_\odot)^{2/7} (M_\star / M_\odot)^{-1/7} (r / {\rm 1 \, au})^{-3/7}$ and $L_\star = L_\odot (M_\star / M_\odot)^{1.5}$ for $M_\star = 1.2 \, M_\odot$ \citep{Ida2016}. The planet's current location is therefore comfortably within the CO snowline ($\sim 30$ K), even after accounting for the $\sim \pm 5$ K uncertainty in the sublimation temperature due to incomplete knowledge of the state of the ice \citep{Fayolle2016, Piso2016}. The CO$_2$ and H$_2$S snowlines ($\sim 75$ K and $\sim 63$ K respectively, \citealt{Fray2009, Yu2023}) are close to the planet's current location (see Figure~\ref{fig:composition-curves} for corresponding orbital distances). However, if H$_2$S and CO$_2$ are mixed with water ice, they may only sublimate much closer to the star near the water snowline \citep{Collings2004}. 
Ammonium salts, such as NH$_4$SH, have recently been proposed as an additional S reservoir in protoplanetary disks \citep{Poch2020, Altwegg2022, Nakazawa&Ohno26}; however, it should be in a condensed phase at AF Lep b's orbit, as the NH$_4$SH salt sublimes at $\gtrsim150~{\rm K}$. 
Given that H$_2$S is likely the most volatile phase of S that contains a significant fraction of the element as well as evidence supporting refractory forms of S in protoplanetary disks \citep{Kama2019,LeGal+21}, S can be assumed to be entirely in the solids at the planet's location. We also consider accretion of material inside H$_2$S and CO$_2$ snowlines and show that the small fractions of S and C contained in these molecules do not strongly affect our inferences.

Figure~\ref{fig:composition-curves} shows compositional curves for planets that accrete beyond the CO snowline, between the CO and H$_2$S snowlines, and within the CO$_2$ snowline. Each curve results from a combination of solid and gas accretion. Refractory molecules are accreted together in the form of solid pebbles or planetesimals, while volatile molecules are accreted together in the form of accumulated gas. However, solids and gas are accreted via different processes and need not be added to the planet in a known ratio. The relative contributions of solids and gas vary independently along the curves (for intuition, note that S/H is a proxy for the relative amount of mass accreted in solids and gas). We highlight that an element may be found in both volatile and refractory molecules at a particular location in the disk, so the total abundance of, for example, carbon in the planet depends on the accreted masses in both solids and gas. Outside their very inner regions, protoplanetary disks typically exhibit long chemical reaction times \citep{oberg_bergin2021}. Molecules are primarily affected by condensation and sublimation, rather than gas chemistry, and these processes control whether a particular molecule is in solid or gaseous form. Once incorporated into the planet, however, chemical reactions re-allocate the atoms into different molecules, so models need only (and can only) match the total measured atomic abundances.

In principle, the relative abundances of volatile molecules in accreted protoplanetary disk gas need not match the relative abundances averaged over the entire disk since pebble drift followed by sublimation at ice lines can enhance the gas abundance of individual species. However, we are able to obtain abundance ratios consistent with our measurements for AF Lep b without appealing to this process. We begin with a fixed set of relative molecular abundances and merely assign each molecule to be a solid or a gas depending on the disk location under consideration. The curves in Figure 8 assume that 40\% of C is in CO and 10\% is in CO$_2$ based on measurements from the interstellar medium, molecular clouds, protostellar cores, and solar system comets \citep{oberg_bergin2021}, although the exact split could be different \citep{McClure2023} and does not impact our conclusions. The remaining disk carbon is assumed to be in some refractory form that is present in solids at AF Lep b's location. In addition, 10\% of S is assumed to be in the form of volatile H$_2$S, based on \cite{Kama2019}, who estimate that $89 \pm 8$\% of S is in refractory form. 

For a planet accreting from a disk that has not experienced relative molecular redistribution due to pebble drift, super-stellar S/H indicates that a species accreted entirely with the solids would have a super-stellar abundance relative to H (matching the S enhancement), while a species accreted entirely with the gas would have a stellar abundance. The simplest formation hypothesis for AF Lep b is that it accreted its material from such a disk at its current location between the CO and H$_2$S snowlines.  AF Lep b's slightly lower volatile abundances (relative to refractory S) and sub-solar C/O are compatible with this model---some but not all of the C and O are accreted from solids, and relatively more O (compared to C) is in solid form.  
The left and right panels of Figure~\ref{fig:composition-curves} show that the planet's sub-solar C/O and C/S disfavor accretion beyond the CO snowline, where we would expect these elements to be present in stellar proportions in the disk solids. More broadly, the data are compatible with accretion between the H$_2$O and CO snowlines such that the accreted solids are relatively O and C rich. 

\begin{figure*}
    \centering
    \includegraphics[width=0.3\linewidth]{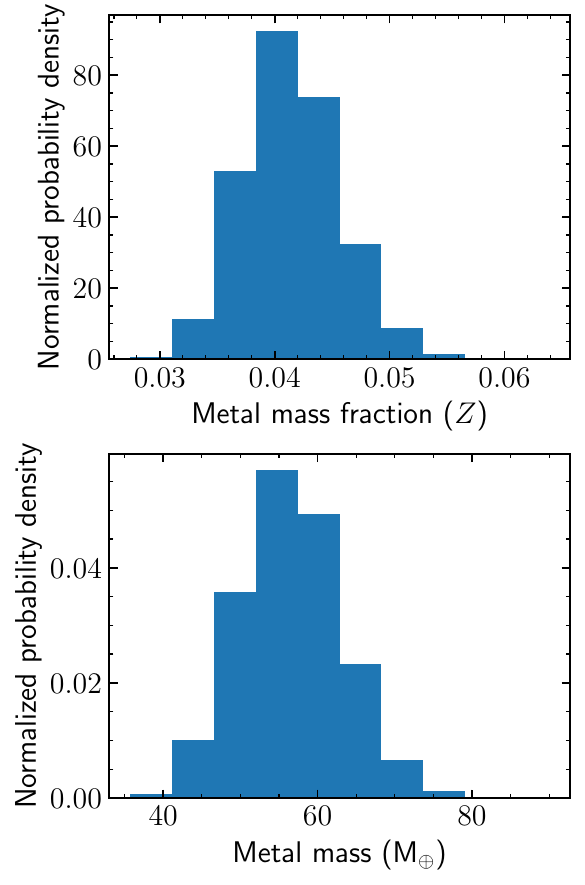}
    \includegraphics[width=0.45\linewidth]{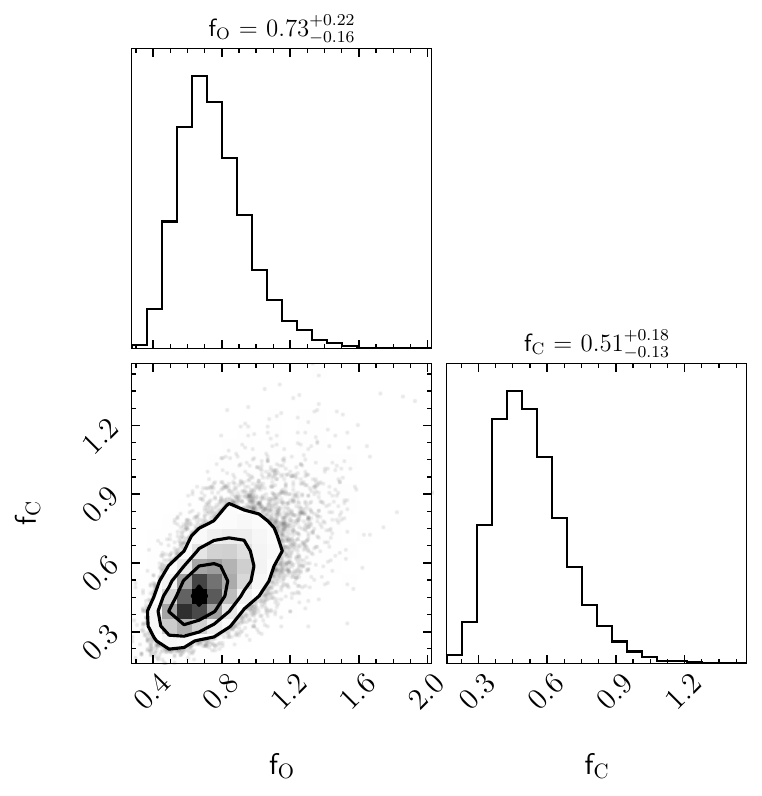}
    \caption{Left: The metal mass fraction and total metal mass of AF Lep b calculated from the measured atmospheric abundances. AF Lep b has a metal mass fraction of $0.041 \pm 0.004$ and contains $56 \pm 7 \, M_\oplus$ of metals by mass. Right: Constraints on the fraction of O and C that is needed in solids to match the observed composition with accretion of local solids and gas alone.}
    \label{fig:metal-mass}
\end{figure*}

We estimate the fraction of available carbon and oxygen ($f_{\rm C}$, $f_{\rm O}$) that is required to be in solids (as opposed to gas) for accretion of local gas and solids to produce the observed composition. This fraction is estimated using the retrieved abundance samples by calculating $f_{\rm X} = ({\rm X}/H - 1) / (\mathcal{R}/H - 1)$ for species X and refractory species $\mathcal{R}$ (sulfur in this study). Values of $f_{\rm X}$ greater than 1 imply the need for enrichment of local material (solids or gas) beyond the stellar value in species X, for example by pebble drift from more distant regions of the disk. The right panel in Figure~\ref{fig:metal-mass} shows that in the context of this model, the solids accreted by AF Lep b would be richer in C ($f_{\rm C} = 0.51^{+0.18}_{-0.13}$) and O ($f_{\rm O} = 0.73^{+0.22}_{-0.16}$) than the most pristine CI carbonaceous chondrites in the solar system, for which $f_{\rm C} \sim 0.09$ and $f_{\rm O} \sim 0.47$ \citep{Lodders2025}. This enhancement is consistent with more distant formation, beyond the CO$_2$ iceline.  CI carbonaceous chondrites either formed closer to the star, in a less ice-rich environment, or lost their volatiles over the age of the solar system \citep{Morbidelli2012, Bergin2015}. 

While the abundances of AF Lep b do not require enhancement of gas metallicity by pebble drift across the CO snowline, abundance measurements of planets in the HR 8799 system do require such enhancement \citep{Xuan2026}.  This difference may arise because the HR 8799 planets are more distant from their host star, and hence closer to the CO snowline from which additional CO gas can be sourced via pebble drift.  Alternatively, the difference may arise because AF Lep is a somewhat lower-mass star than HR 8799 or because the phase of disk evolution at which the planets formed differed.  Of course, a more complicated disk evolution model---though not required---may also be consistent with the AF Lep b measurements.

In the above, we assume that the stellar reference abundance is solar, as justified in Section~\ref{sec:stellar_abunds}. The planet's low C/S and O/S ratios, or equivalently its sulfur enrichment relative to carbon and oxygen, remain the case whether we adopt solar or the AF Lep A abundances as the reference. However, the planet's C/O is more dependent on the choice of reference abundance. In general, obtaining accurate stellar abundances is an ongoing challenge for young, rapidly rotating stars \citep[e.g.][]{Hejazi2025}, which is further complicated by systematics related to non-LTE corrections. In this paper, following \citet{Xuan2026}, we therefore adopt the solar abundances as the reference.

\subsection{Constraints on metal mass fraction of AF Lep b}
We can also use the atmospheric composition measurements of AF Lep b to put constraints on its bulk metallicity and metal mass. These constraints assume that the interior and atmosphere are fully mixed and they most likely provide lower limits on the planet's metal content as composition gradients in the planet could lead to more metal enriched interiors than what is implied by atmospheric measurements \citep{Helled2017, Thorngren2019}. This calculation includes mass contribution from elements Fe, Mg, and Si that were not measured directly. These elements constitute a significant fraction of available metal mass and it is therefore important to include their contribution. Given their refractory nature, the enrichment of Fe, Mg, and Si should be similar to S, which also traces solid accretion onto the planet. Figure~\ref{fig:metal-mass} (left panel) shows the estimated metal mass fraction ($0.041 \pm 0.004$) and metal mass ($56 \pm 7 \, M_\oplus$) of AF Lep b based on its atmospheric measurements. The vast majority of these metals would be accreted via solids based on the composition curves discussed above. Accreting this metal mass through planetesimals would require a planetesimal surface density that is $\sim 4 \times$ the density of the minimum-mass solar nebula \citep{Chiang2010}, assuming the planet accretes all the planetesimals in a feeding zone of $3.5 \times$ its Hill radius \citep{Lissauer1993}.

\begin{figure}
    \centering
    \includegraphics[width=\linewidth]{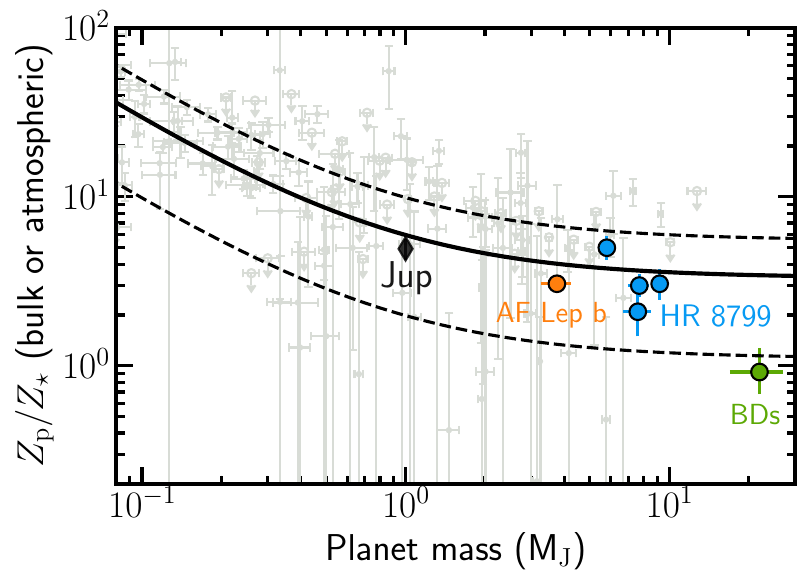}
    \caption{The metal mass fraction calculated from measured atmospheric metallicities of AF Lep b (orange; this work) and HR 8799 bcde (blue; \citealt{Xuan2026}). The bulk metallicities (from mass and radius and thermal evolution models) of close-in giant planets are overplotted in light gray \citep{Chachan2025b}. Both close-in and distant super-Jupiters remain super-stellar between $\sim1$--$10~\Mj$. Brown dwarf companions show stellar-like metallicities; the green point indicates the average and scatter of eight $10$--$30~\Mj$ companions from \citet{Xuan2024b}.}
    \label{fig:compare_transits}
\end{figure}

\subsection{Metal-rich compositions of super-Jupiters}
The conclusion that directly imaged planets need to accrete a significant solid budget to explain their atmospheric enrichment was recently inferred by \citet{Ruffio2026} and \citet{Xuan2026} for HR 8799, and in earlier work by \citet{JiWang2025}. Other metal-rich directly imaged super-Jupiters include GJ 504 b \citep{Baburaj2026b}, 51 Eri b \citep{Madurowicz2025}, HIP 99770 b \citep{Balmer2026}, and potentially, $\beta$ Pic b (\citealt{GonzalesPicos2026}). For AF Lep b and HR 8799 bcde, we note that their inferred masses of accreted solids are well above $10~\Me$, the canonical threshold for initiating runaway gas accretion \citep{pollack_formation_1996}. High metal masses have also been found for close-in ($<1$ AU) super-Jupiters based on their bulk densities and thermal evolution models \citep{Thorngren2016, Chachan2025b}. Indeed, recent work on transiting giants shows that super-Jupiters (up to $\sim10~\Mj$) continue to accrete metals as they grow in mass and the population's bulk metallicity flattens at the $3.3 \pm 0.5\times$ stellar level rather than declining to the stellar composition \citep{Chachan2025b}. Assuming that the atmospheric metallicities trace bulk metallicities, the metallicities of AF Lep b and the four HR 8799 planets strikingly agree with those of transiting super-Jupiters (see Figure~\ref{fig:compare_transits}). They are in stark contrast to the measured atmospheric composition of higher-mass substellar companions with mass ratios $\gtrsim0.01$ ($m\gtrsim10$--$20~\Mj$), which closely follow the stellar metallicity \citep[e.g.][]{Xuan2024b, GWang2025, Liu2026}, as expected for formation by direct gravitational fragmentation.

The similarly elevated metallicities of close-in ($<1$ AU) super-Jupiters, AF Lep b (9 AU) at a more solar system-like scale, and distant Jupiters like the HR 8799 planets ($15$--$70$ AU) are intriguing, as the process of metal accretion is expected to depend on orbital distance \citep[e.g.][]{Goldreich2004, Schneider2021a,Ohno2026}. It could point to either i) formation of these planets at similar orbital distances followed by substantial post-formation migration (see \citealt{Dawson2018} for a review of migration pathways for close-in giant planets) or ii) the ubiquity of metal accretion in circumstellar disks and the existence of an accretion mechanism that is not strongly dependent on orbital distance. 
In scenario ii), one possibility is late stage accretion during the disk's photoevaporation when hydrogen and helium are lost preferentially and the remaining disk material is enriched in metals \citep{Guillot2006, Desch2014}. 
Magnetically driven winds launched from disk surfaces have also been suggested to cause gradual metal enrichment \citep{Okuzumi25,Ikeda+26}.
Accretion of metal-enriched gas due to pebble drift and evaporation might also contribute a significant fraction of the observed metallicity at the population level \citep{Schneider2021a, Xuan2026, Ohno2026}. Another possibility is major mergers between planetary cores during giant planet formation, which are most efficient at boosting planet metallicity at $\sim 10$ au \citep{Ginzburg2020}. Although this mechanism might explain AF Lep b's enrichment, it cannot match the elevated median bulk metallicity of the transiting super-Jupiter population \citep{Chachan2025b}. Further measurements of the atmospheric compositions for giant planets across different orbital distances, combined with additional theoretical work, will be required to clarify how super-Jupiters acquire their significant metal budgets.

\section{Conclusion}\label{sec:conclude}
In this paper, we measured the carbon, oxygen, and sulfur abundances for AF Lep b with the goal of informing its formation and metal accretion history. We used JWST/NIRSpec IFU observations (GO 5342) which provided $3$--$5\,\mu$m spectra at $R\sim3000$, JWST/NIRCam observations (GO 6905) covering three narrow photometric bands across $4.0$--$4.7~\mu$m, and GRAVITY and SPHERE observations that extend the wavelength coverage down to $1~\mu$m. Combining all these data, we performed atmospheric retrievals using the radiative transfer code \texttt{petitRADTRANS} \citep{molliere_petitRADTRANS_2019}, and detected CO$_2$, H$_2$S, and $^{13}$CO for the first time in this planet. Besides carbon and oxygen ($\rm{C/H}=2.9\pm0.5$, $\rm{O/H}=3.7\pm0.6~\times$ solar), we find that the sulfur (from H$_2$S) abundance of AF Lep b is also enriched with $\rm{S/H}=4.7\pm0.7~\times$ solar (see Figure~\ref{fig:abunds}). By considering stellar abundances of four stars in the $\beta$ Pic moving group, including AF Lep A, we find the average stellar abundances are consistent with solar at $<1\sigma$ level. Therefore, we assume that the natal disk material that formed AF Lep b is of solar composition. 

By combining the refractory tracer of S along with the volatile abundances in C and O, we are able to place constraints on the amount and nature of material accreted onto the planet. Using the models from \citet{Chachan2023}, we infer that the compositional patterns of AF Lep b are consistent with in situ formation, and more broadly with accretion in between the H$_2$O and CO snowlines as long as nearly half of the carbon inventory is in some refractory form at AF Lep b's location. We also infer a total metal mass of $56\pm7~\Me$ based on its atmospheric measurements, which is higher than the canonical threshold for runaway gas accretion \citep{pollack_formation_1996}. Along with other imaged planets such as HR 8799 bcde, 51 Eri b, and GJ 504 b, we note an intriguing trend where directly imaged super-Jupiters have atmospheric metallicities comparable to the bulk metallicities of close-in gas giants ($<1$ AU). This trend also extends to Jupiter, which has an atmospheric metallicity of $\sim3\times$ solar and an inferred  bulk metallicity of $\sim3$--$8\times$ solar \citep[see review by][]{Helled2026}. Together, these findings could hint at a more universal metal accretion mechanism for gas giants that operates across a wide range of orbital distances. Future JWST measurements of both volatile and refractory elements as well as detailed modeling work are poised to shed more light on these trends, and consequently, how giant planet assembly unfolds.

\begin{acknowledgments}
J.W.X thanks Eugene Chiang, Rocio Kiman and Wolf Cukier for helpful discussions.
J.W.X is grateful for support from the Heising-Simons Foundation 51 Pegasi b Fellowship (grant \#2025-5887). Y.Z. is grateful for support from the Heising-Simons Foundation 51 Pegasi b Fellowship (grant \#2023-4298).
J. J. Wang acknowledges support from the Alfred P. Sloan Foundation.
This work is based on observations made with the NASA/ESA/CSA James Webb Space Telescope. The data were obtained from the Mikulski Archive for Space Telescopes at the Space Telescope Science Institute, which is operated by the Association of Universities for Research in Astronomy, Inc., under NASA contract NAS 5-03127 for JWST. These observations are associated with programs 5342 and 6905. This work is also based on observations made with ESO Telescopes at the La Silla Paranal Observatory under programme ID 110.25A4.001.

This work used the Anvil supercomputer at the Purdue Rosen Center for Advanced Computing through allocation PHY260121 from the Advanced Cyberinfrastructure Coordination Ecosystem: Services \& Support (ACCESS) program \citep{Boerner2023ACCESS}, which is supported by U.S. National Science Foundation grants \#2138259, \#2138286, \#2138307, \#2137603, and \#2138296.
 
\end{acknowledgments}

\facilities{JWST(NIRSpec)}

\software{\texttt{petitRADTRANS} \citep{molliere_Retrieving_2020}; \texttt{line-racer} \citep{Hagele2026}; \texttt{pymultinest} \citep{Buchner2014, Feroz2019} }

\newpage
\appendix

\section{JWST/NIRSpec-only retrieval}\label{sec:nirspec_only}
\begin{figure}
    \centering
    \includegraphics[width=\linewidth]{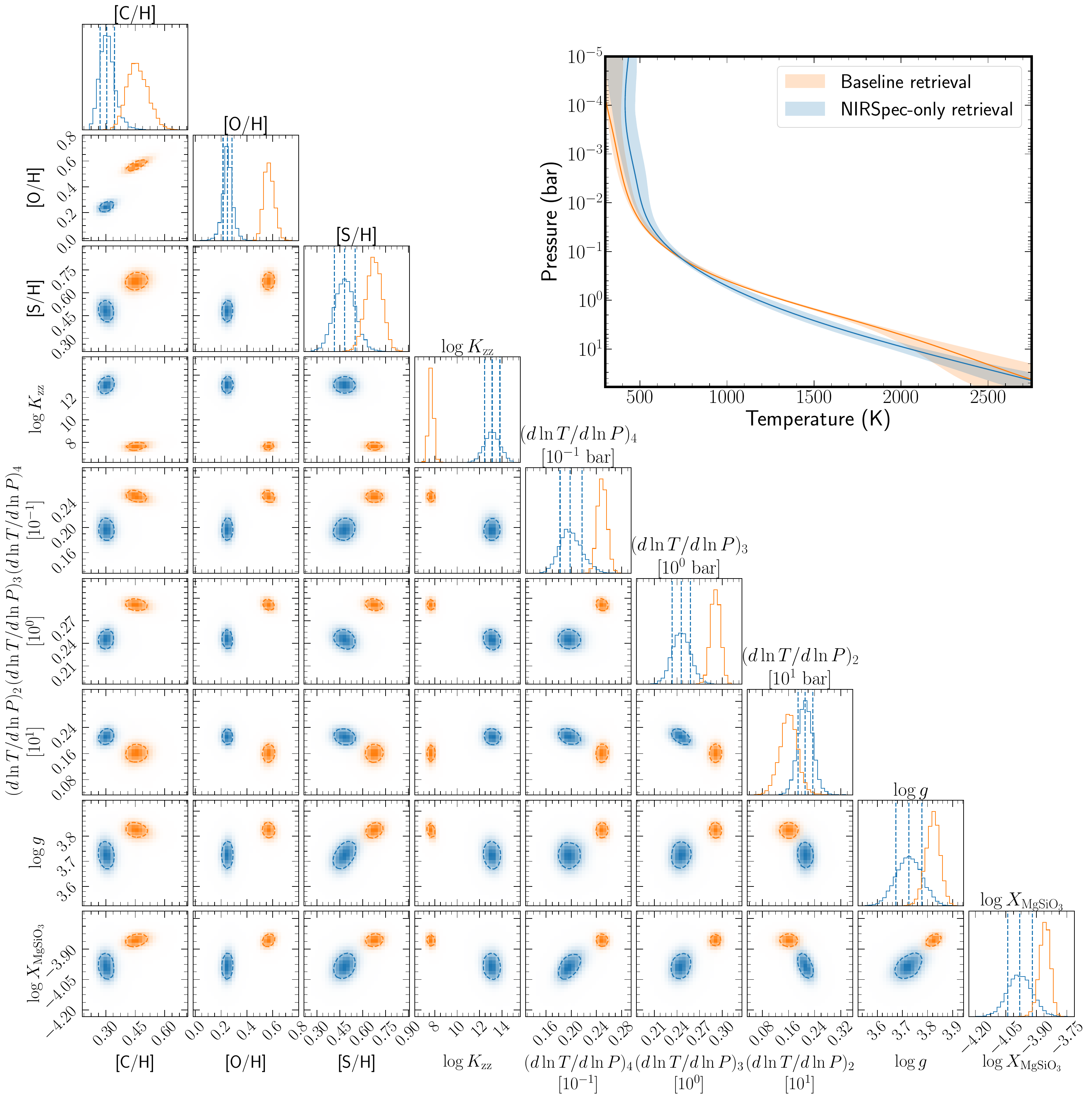}
    \caption{The joint posterior distributions of a few retrieved parameters between the baseline retrieval (orange) presented in Table~\ref{table:aflep_spec_results}, and a NIRSpec-only retrieval (blue). The top right inset shows the retrieved PT profiles: best-fit as solid lines, and $2\sigma$ contours as shaded regions. The retrieved abundances are biased in the NIRSpec-only retrieval, and the difference likely arises from the colder deep atmosphere at $\sim1$--$10$ bars found by the NIRSpec-only retrieval. The addition of low-resolution spectroscopy from $1.0$--$2.5~\mu$m probes these deeper layers, unlike the NIRSpec spectra from $3$--$5\,\mu\mathrm{m}$. In addition, the NIRSpec-only retrieval finds an implausibly large $\log K_{\rm zz}$ value. This comparison highlights the complementarity of low-resolution spectroscopy and high-pass filtered medium or high-resolution spectroscopy.}
    \label{fig:compare_corner}
\end{figure}

Here we present the results of a retrieval using only the JWST/NIRSpec spectra to illustrate the value of the low-resolution spectroscopy and NIRCam photometry. The NIRSpec-only retrieval uses the same retrieval setup as the baseline model (row 1 in Table~\ref{table:aflep_spec_results}). A few key parameters are shown in Figure~\ref{fig:compare_corner}. The joint posteriors from the NIRSpec-only run are shown in blue, while those from the baseline retrieval (including VLT/SPHERE, VLTI/GRAVITY spectra and NIRCam F410M, F430M, F460M photometry) are in orange. In the inset, we also plot the best-fit PT profiles and $2\sigma$ PT contours from the two retrievals. 

The NIRSpec-only retrieval returns [C/H] and [S/H] that are $\approx2\sigma$ different from the baseline retrieval, but the [O/H] and $\log{K_{\rm zz}}$ are discrepant by more than $3\sigma$. In addition, the NIRSpec-only retrieval finds a colder deep atmosphere between $\sim1$--$10$ bars, which is reflected in the differences in three retrieved temperature gradients at 0.1, 1, and 10 bars. The low-resolution spectra from $1.0$--$2.5~\mu$m are sensitive to deeper pressures due to their shorter wavelengths (see emission contribution functions in Figure~\ref{fig:pt_emis}), and can better constrain the thermal structure at those pressures. The $\log K_{\rm zz}\approx13$ from the NIRSpec-only retrieval is also implausibly large, as it is several orders of magnitude higher than the upper limit of $\log K_{\rm zz}\approx10$ we estimate from mixing length theory \citep{Gierasch_convect1985} for AF Lep b. The abundance shifts between the two retrievals are likely driven primarily by degeneracies between the elemental abundances, thermal structure, and $K_{\rm zz}$. Indeed, the NIRSpec-only retrieval favors a different PT profile and a much larger $K_{\rm zz}$, which changes the quench pressures and the molecular abundance profiles. By contrast, the cloud parameters are only slightly offset between the two retrievals. For example, the cloud base mass fraction ${\rm log}X_{\rm MgSiO_3}$ only changes by $\approx0.1$ dex. This comparison suggests that the high-pass filtered NIRSpec spectrum is sensitive to the cloud properties on its own, as noted in previous work (see section 5.2 of \citealt{Xuan2026}). However, caution should be taken when interpreting retrievals using only NIRSpec spectra, and the inclusion of low-resolution spectroscopy or photometry at complementary wavelengths is important.

\newpage
\bibliography{main}{}

@ARTICLE{LeGal+21,
       author = {{Le Gal}, Romane and {{\"O}berg}, Karin I. and {Teague}, Richard and {Loomis}, Ryan A. and {Law}, Charles J. and {Walsh}, Catherine and {Bergin}, Edwin A. and {M{\'e}nard}, Fran{\c{c}}ois and {Wilner}, David J. and {Andrews}, Sean M. and {Aikawa}, Yuri and {Booth}, Alice S. and {Cataldi}, Gianni and {Bergner}, Jennifer B. and {Bosman}, Arthur D. and {Cleeves}, L. Ilse and {Czekala}, Ian and {Furuya}, Kenji and {Guzm{\'a}n}, Viviana V. and {Huang}, Jane and {Ilee}, John D. and {Nomura}, Hideko and {Qi}, Chunhua and {Schwarz}, Kamber R. and {Tsukagoshi}, Takashi and {Yamato}, Yoshihide and {Zhang}, Ke},
        title = "{Molecules with ALMA at Planet-forming Scales (MAPS). XII. Inferring the C/O and S/H Ratios in Protoplanetary Disks with Sulfur Molecules}",
      journal = {\apjs},
         year = 2021,
        month = nov,
       volume = {257},
       number = {1},
          eid = {12},
        pages = {12},
          doi = {10.3847/1538-4365/ac2583},
archivePrefix = {arXiv},
       eprint = {2109.06286},
 primaryClass = {astro-ph.GA},
       adsurl = {https://ui.adsabs.harvard.edu/abs/2021ApJS..257...12L}
}

@ARTICLE{Shajib2025,
       author = {{Shajib}, Anowar J. and {Treu}, Tommaso and {Melo}, Alejandra and {Roberts-Borsani}, Guido and {Knabel}, Shawn and {Cappellari}, Michele and {Frieman}, Joshua A.},
        title = "{An accurate measurement of the spectral resolution of the JWST Near Infrared Spectrograph}",
      journal = {\aap},
         year = 2025,
        month = oct,
       volume = {702},
          eid = {L12},
        pages = {L12},
          doi = {10.1051/0004-6361/202556281},
archivePrefix = {arXiv},
       eprint = {2507.03746},
 primaryClass = {astro-ph.IM},
       adsurl = {https://ui.adsabs.harvard.edu/abs/2025A&A...702L..12S}
}

@article{Morbidelli2012,
  title = {Building {{Terrestrial Planets}}},
  author = {Morbidelli, A. and Lunine, J. I. and O'Brien, D. P. and Raymond, S. N. and Walsh, K. J.},
  year = 2012,
  month = may,
  journal = {Annual Review of Earth and Planetary Sciences, vol. 40, issue 1, pp. 251-275},
  volume = {40},
  number = {1},
  pages = {251},
  issn = {0084-6597},
  doi = {10.1146/annurev-earth-042711-105319},
  urldate = {2026-06-29},
  langid = {english}
}

@article{Bergin2015,
  title = {Tracing the Ingredients for a Habitable Earth from Interstellar Space through Planet Formation},
  author = {Bergin, Edwin A. and Blake, Geoffrey A. and Ciesla, Fred and Hirschmann, Marc M. and Li, Jie},
  year = 2015,
  month = jul,
  journal = {Proceedings of the National Academy of Sciences, vol. 112, issue 29, pp. 8965-8970},
  volume = {112},
  number = {29},
  pages = {8965},
  issn = {0027-8424},
  doi = {10.1073/pnas.1500954112},
  urldate = {2026-06-29},
  langid = {english}
}

@article{Altwegg2022,
  title = {Abundant Ammonium Hydrosulphide Embedded in Cometary Dust Grains},
  author = {Altwegg, K and Combi, M and Fuselier, S A and H{\"a}nni, N and De~Keyser, J and Mahjoub, A and M{\"u}ller, D R and Pestoni, B and Rubin, M and Wampfler, S F},
  year = 2022,
  month = nov,
  journal = {Monthly Notices of the Royal Astronomical Society},
  volume = {516},
  number = {3},
  pages = {3900--3910},
  issn = {0035-8711},
  doi = {10.1093/mnras/stac2440},
  urldate = {2026-06-11}
}

@article{Poch2020,
  title = {Ammonium Salts Are a Reservoir of Nitrogen on a Cometary Nucleus and Possibly on Some Asteroids},
  author = {Poch, Olivier and Istiqomah, Istiqomah and Quirico, Eric and Beck, Pierre and Schmitt, Bernard and Theul{\'e}, Patrice and Faure, Alexandre and {Hily-Blant}, Pierre and Bonal, Lydie and Raponi, Andrea and Ciarniello, Mauro and Rousseau, Batiste and Potin, Sandra and Brissaud, Olivier and Flandinet, Laur{\`e}ne and Filacchione, Gianrico and Pommerol, Antoine and Thomas, Nicolas and Kappel, David and Mennella, Vito and Moroz, Lyuba and Vinogradoff, Vassilissa and Arnold, Gabriele and Erard, St{\'e}phane and {Bockel{\'e}e-Morvan}, Dominique and Leyrat, C{\'e}dric and Capaccioni, Fabrizio and De Sanctis, Maria Cristina and Longobardo, Andrea and Mancarella, Francesca and Palomba, Ernesto and Tosi, Federico},
  year = 2020,
  month = mar,
  journal = {Science},
  volume = {367},
  number = {6483},
  pages = {eaaw7462},
  publisher = {American Association for the Advancement of Science},
  doi = {10.1126/science.aaw7462},
  urldate = {2026-06-11}
}

@article{Lissauer1993,
  title = {Planet Formation.},
  author = {Lissauer, Jack J.},
  year = 1993,
  journal = {Annual Review of Astronomy and Astrophysics},
  volume = {31},
  pages = {129--174},
  issn = {0066-4146},
  doi = {10.1146/annurev.aa.31.090193.001021},
  urldate = {2026-06-23},
  langid = {english}
}

@ARTICLE{Okuzumi25,
       author = {{Okuzumi}, Satoshi},
        title = "{Surface accretion as a dust retention mechanism in protoplanetary disks. I. Formulation and proof-of-concept simulations}",
      journal = {\pasj},
         year = 2025,
        month = feb,
       volume = {77},
       number = {1},
        pages = {162-177},
          doi = {10.1093/pasj/psae107},
archivePrefix = {arXiv},
       eprint = {2411.09934},
 primaryClass = {astro-ph.EP},
       adsurl = {https://ui.adsabs.harvard.edu/abs/2025PASJ...77..162O}
}

@ARTICLE{Ikeda+26,
       author = {{Ikeda}, Yoshitaka and {Ohno}, Kazumasa and {Okuzumi}, Satoshi},
        title = "{Heavy element enrichment of gas in surface-accretion disks: A possible origin of the mass─metallicity anti-correlation in exoplanets}",
      journal = {\aap},
         year = 2026,
        month = jun,
       volume = {710},
          eid = {L21},
        pages = {L21},
          doi = {10.1051/0004-6361/202660205},
archivePrefix = {arXiv},
       eprint = {2605.27289},
 primaryClass = {astro-ph.EP},
       adsurl = {https://ui.adsabs.harvard.edu/abs/2026A&A...710L..21I}
}

@ARTICLE{Nakazawa&Ohno26,
       author = {{Nakazawa}, Kanon and {Ohno}, Kazumasa},
        title = "{Sulfur Enrichment in Close-in Exoplanet Atmospheres Induced by Pebble Drift across the Salt Line}",
      journal = {\apj},
         year = 2026,
        month = mar,
       volume = {999},
       number = {1},
          eid = {130},
        pages = {130},
          doi = {10.3847/1538-4357/ae42c0},
archivePrefix = {arXiv},
       eprint = {2602.05300},
 primaryClass = {astro-ph.EP},
       adsurl = {https://ui.adsabs.harvard.edu/abs/2026ApJ...999..130N}
}

@ARTICLE{Moses+11,
       author = {{Moses}, Julianne I. and {Visscher}, C. and {Fortney}, J.~J. and {Showman}, A.~P. and {Lewis}, N.~K. and {Griffith}, C.~A. and {Klippenstein}, S.~J. and {Shabram}, M. and {Friedson}, A.~J. and {Marley}, M.~S. and {Freedman}, R.~S.},
        title = "{Disequilibrium Carbon, Oxygen, and Nitrogen Chemistry in the Atmospheres of HD 189733b and HD 209458b}",
      journal = {\apj},
         year = 2011,
        month = aug,
       volume = {737},
       number = {1},
          eid = {15},
        pages = {15},
          doi = {10.1088/0004-637X/737/1/15},
archivePrefix = {arXiv},
       eprint = {1102.0063},
 primaryClass = {astro-ph.EP},
       adsurl = {https://ui.adsabs.harvard.edu/abs/2011ApJ...737...15M}
}

@ARTICLE{Wogan+25,
       author = {{Wogan}, Nicholas F. and {Mang}, James and {Batalha}, Natasha E. and {Zahnle}, Kevin and {Mukherjee}, Sagnick and {Visscher}, Channon and {Fortney}, Jonathan J. and {Marley}, Mark S. and {Morley}, Caroline V.},
        title = "{The Sonora Substellar Atmosphere Models. V. A Correction to the Disequilibrium Abundance of CO$_{2}$ for Sonora Elf Owl}",
      journal = {Research Notes of the American Astronomical Society},
         year = 2025,
        month = may,
       volume = {9},
       number = {5},
          eid = {108},
        pages = {108},
          doi = {10.3847/2515-5172/add407},
archivePrefix = {arXiv},
       eprint = {2505.03994},
 primaryClass = {astro-ph.EP},
       adsurl = {https://ui.adsabs.harvard.edu/abs/2025RNAAS...9..108W}
}

@ARTICLE{Tsai+18,
       author = {{Tsai}, Shang-Min and {Kitzmann}, Daniel and {Lyons}, James R. and {Mendon{\c{c}}a}, Jo{\~a}o and {Grimm}, Simon L. and {Heng}, Kevin},
        title = "{Toward Consistent Modeling of Atmospheric Chemistry and Dynamics in Exoplanets: Validation and Generalization of the Chemical Relaxation Method}",
      journal = {\apj},
         year = 2018,
        month = jul,
       volume = {862},
       number = {1},
          eid = {31},
        pages = {31},
          doi = {10.3847/1538-4357/aac834},
archivePrefix = {arXiv},
       eprint = {1711.08492},
 primaryClass = {astro-ph.EP},
       adsurl = {https://ui.adsabs.harvard.edu/abs/2018ApJ...862...31T}
}

@article{Dawson2018,
  title = {Origins of {{Hot Jupiters}}},
  author = {Dawson, Rebekah I. and Johnson, John Asher},
  year = 2018,
  month = sep,
  journal = {Annual Review of Astronomy and Astrophysics, vol. 56, p.175-221},
  volume = {56},
  pages = {175},
  issn = {0066-4146},
  doi = {10.1146/annurev-astro-081817-051853},
  urldate = {2026-06-22},
  langid = {english}
}

@article{Fayolle2016,
  title = {N2 and {{CO Desorption Energies}} from {{Water Ice}}},
  author = {Fayolle, Edith C. and Balfe, Jodi and Loomis, Ryan and Bergner, Jennifer and Graninger, Dawn and Rajappan, Mahesh and {\"O}berg, Karin I.},
  year = 2016,
  month = jan,
  journal = {\apj},
  volume = {816},
  pages = {L28},
  publisher = {IOP},
  issn = {0004-637X},
  doi = {10.3847/2041-8205/816/2/L28},
  urldate = {2025-02-11}
}

@article{Piso2016,
  title = {{{THE ROLE OF ICE COMPOSITIONS FOR SNOWLINES AND THE C}}/{{N}}/{{O RATIOS IN ACTIVE DISKS}}},
  author = {Piso, Ana-Maria A. and Pegues, Jamila and {\"O}berg, Karin I.},
  year = 2016,
  month = dec,
  journal = {\apj},
  volume = {833},
  number = {2},
  pages = {203},
  publisher = {The American Astronomical Society},
  issn = {0004-637X},
  doi = {10.3847/1538-4357/833/2/203},
  urldate = {2025-05-06},
  langid = {english}
}

@article{Collings2004,
  title = {A Laboratory Survey of the Thermal Desorption of Astrophysically Relevant Molecules},
  author = {Collings, M. P. and Anderson, M. A. and Chen, R. and Dever, J. W. and Viti, S. and Williams, D. A. and McCoustra, M. R. S.},
  year = 2004,
  month = nov,
  journal = {MNRAS},
  volume = {354},
  number = {4},
  pages = {1133--1140},
  issn = {0035-8711, 1365-2966},
  doi = {10.1111/j.1365-2966.2004.08272.x},
  urldate = {2026-06-04},
  langid = {english}
}

@article{Ohno2026,
  title = {A Dichotomy of the Mass-metallicity Relation of Exoplanetary Atmospheres Demarcated by Their Birthplace},
  author = {Ohno, Kazumasa and Ikoma, Masahiro and Okuzumi, Satoshi and Kimura, Tadahiro},
  year = 2026,
  month = apr,
  journal = {Publications of the Astronomical Society of Japan},
  volume = {78},
  pages = {493--523},
  publisher = {OUP},
  issn = {0004-6264},
  doi = {10.1093/pasj/psaf157},
  urldate = {2026-06-04}
}

@article{Guillot2006,
  title = {The Composition of {{Jupiter}}: Sign of a (Relatively) Late Formation in a Chemically Evolved Protosolar Disc},
  shorttitle = {The Composition of {{Jupiter}}},
  author = {Guillot, Tristan and Hueso, Ricardo},
  year = 2006,
  month = mar,
  journal = {MNRAS},
  volume = {367},
  number = {1},
  pages = {L47-L51},
  issn = {1745-3933, 1745-3925},
  doi = {10.1111/j.1745-3933.2006.00137.x},
  urldate = {2025-02-13},
  langid = {english}
}

@article{Desch2014,
  title = {Jupiter's {{Noble Gas Abundances May Require External UV Irradiation}} of the {{Solar Nebula}}},
  author = {Desch, S. J. and Monga, N.},
  year = 2014,
  month = mar,
  pages = {1725},
  urldate = {2025-02-13}
}

@ARTICLE{Lodders2025,
       author = {{Lodders}, K. and {Bergemann}, M. and {Palme}, H.},
        title = "{Solar System Elemental Abundances from the Solar Photosphere and CI-Chondrites}",
      journal = {\ssr},
         year = 2025,
        month = mar,
       volume = {221},
       number = {2},
          eid = {23},
        pages = {23},
          doi = {10.1007/s11214-025-01146-w},
archivePrefix = {arXiv},
       eprint = {2502.10575},
 primaryClass = {astro-ph.SR},
       adsurl = {https://ui.adsabs.harvard.edu/abs/2025SSRv..221...23L}
}

@article{Fray2009,
  title = {Sublimation of Ices of Astrophysical Interest: {{A}} Bibliographic Review},
  shorttitle = {Sublimation of Ices of Astrophysical Interest},
  author = {Fray, N. and Schmitt, B.},
  year = 2009,
  month = dec,
  journal = {Planetary and Space Science},
  volume = {57},
  number = {14-15},
  pages = {2053--2080},
  issn = {00320633},
  doi = {10.1016/j.pss.2009.09.011},
  urldate = {2026-06-04},
  langid = {english}
}

@article{Yu2023,
  title = {Material {{Properties}} of {{Organic Liquids}}, {{Ices}}, and {{Hazes}} on {{Titan}}},
  author = {Yu, Xinting and Yu, Yue and Garver, Julia and Li, Jialin and Hawthorn, Abigale and {Sciamma-O'Brien}, Ella and Zhang, Xi and Barth, Erika},
  year = 2023,
  month = jun,
  journal = {\apjs},
  volume = {266},
  number = {2},
  pages = {30},
  publisher = {The American Astronomical Society},
  issn = {0067-0049},
  doi = {10.3847/1538-4365/acc6cf},
  urldate = {2026-06-05},
  langid = {english}
}

@ARTICLE{Baburaj2026,
       author = {{Baburaj}, Aneesh and {Konopacky}, Quinn M. and {Theissen}, Christopher A. and {Gerasimov}, Roman and {Hoch}, Kielan K.~W.},
        title = "{A High-resolution Spectroscopic Survey of Directly Imaged Companion Hosts. II. Diversity in C/O Ratios among Host Stars}",
      journal = {\aj},
         year = 2026,
        month = jan,
       volume = {171},
       number = {1},
          eid = {21},
        pages = {21},
          doi = {10.3847/1538-3881/ae1a6b},
archivePrefix = {arXiv},
       eprint = {2510.17774},
 primaryClass = {astro-ph.EP},
       adsurl = {https://ui.adsabs.harvard.edu/abs/2026AJ....171...21B}
}

@ARTICLE{Zuckerman2001,
       author = {{Zuckerman}, B. and {Song}, Inseok and {Bessell}, M.~S. and {Webb}, R.~A.},
        title = "{The {\ensuremath{\beta}} Pictoris Moving Group}",
      journal = {\apjl},
         year = 2001,
        month = nov,
       volume = {562},
       number = {1},
        pages = {L87-L90},
          doi = {10.1086/337968},
       adsurl = {https://ui.adsabs.harvard.edu/abs/2001ApJ...562L..87Z}
}

@ARTICLE{Kuhnle2026,
       author = {{K{\"u}hnle}, H. and {Matthews}, E.~C. and {Molli{\`e}re}, P. and {Patapis}, P. and {Zhang}, Z. and {Nasedkin}, E. and {Gasman}, D. and {Whiteford}, N. and {Wang}, H.~S. and {Ravet}, M. and {Chauvin}, G. and {Bonnefoy}, M. and {Barrado}, D. and {Glauser}, A.~M. and {Quanz}, S.~P.},
        title = "{Oxygen and nitrogen isotopologs on cold COCONUTS-2b observed with MIRI/MRS}",
      journal = {\aap},
         year = 2026,
        month = jun,
       volume = {710},
          eid = {A355},
        pages = {A355},
          doi = {10.1051/0004-6361/202659604},
archivePrefix = {arXiv},
       eprint = {2604.26850},
 primaryClass = {astro-ph.EP},
       adsurl = {https://ui.adsabs.harvard.edu/abs/2026A&A...710A.355K}
}

@inproceedings{Boerner2023ACCESS,
  author    = {Boerner, Timothy J. and Deems, Stephen and Furlani, Thomas R. and Knuth, Shelley L. and Towns, John},
  title     = {{ACCESS}: Advancing Innovation: {NSF}'s Advanced Cyberinfrastructure Coordination Ecosystem: Services \& Support},
  booktitle = {Practice and Experience in Advanced Research Computing},
  series    = {PEARC '23},
  year      = {2023},
  month     = jul,
  pages     = {173--176},
  publisher = {ACM},
  doi       = {10.1145/3569951.3597559},
  url       = {https://doi.org/10.1145/3569951.3597559}
}

@ARTICLE{Baburaj2026b,
    doi = {10.3847/1538-3881/ae6919},
    url = {https://doi.org/10.3847/1538-3881/ae6919},
    year = {2026},
    month = {jun},
    publisher = {The American Astronomical Society},
    volume = {172},
    number = {1},
    pages = {28},
    author = {Baburaj, Aneesh and Ruffio, Jean-Baptiste and Perrin, Marshall and Xuan, Jerry W. and Balmer, William O. and Chachan, Yayaati and Konopacky, Quinn M. and Barman, Travis S. and Mâlin, Mathilde and Hoch, Kielan K. W. and Rickman, Emily and Ward-Duong, Kimberly and Pueyo, Laurent and Girard, Julien H. and Rebollido, Isabel and Bidot, Alexis and Chen, Christine and Worthen, Kadin and Lu, Cicero and Kammerer, Jens and van der Marel, Roeland P. and Lewis, Nikole K. and Valenti, Jeff and Seager, Sara and Stark, Chris and Soummer, Rémi and Anderson, Jay and Lajoie, Charles-Philippe and Clampin, Mark and Mountain, C. Matt},
    title = {JWST-TST High Contrast: First Direct Spectroscopy of GJ 504 b Reveals Clouds and Possible Metal Enrichment},
    journal = {AJ}
}

@ARTICLE{Kiman2026,
       author = {{Kiman}, Rocio and {Beichman}, Charles A. and {Ruiz Diaz}, Azul and {Faherty}, Jacqueline K. and {Lacy}, Brianna and {Su{\'a}rez}, Genaro and {Marocco}, Federico and {Kirkpatrick}, J. Davy and {Gagn{\'e}}, Jonathan and {Copeland}, Jessica and {Burningham}, Ben and {Whiteford}, Niall and {Rowland}, Melanie J. and {Bardalez Gagliuffi}, Daniella C. and {Vos}, Johanna M. and {Schneider}, Adam C. and {Gonzales}, Eileen C. and {Alejandro Merchan}, Sherelyn and {Rothermich}, Austin and {Smart}, Richard and {Costa}, Edgardo and {Mendez}, Rene A.},
        title = "{The Diversity of Cold Worlds: Age and Characterization of the Exoplanet COCONUTS-2 b}",
      journal = {\aj},
         year = 2026,
        month = feb,
       volume = {171},
       number = {2},
          eid = {60},
        pages = {60},
          doi = {10.3847/1538-3881/ae230f},
archivePrefix = {arXiv},
       eprint = {2511.20923},
 primaryClass = {astro-ph.EP},
       adsurl = {https://ui.adsabs.harvard.edu/abs/2026AJ....171...60K}
}

@ARTICLE{Matthews2026,
       author = {{Matthews}, Elisabeth C. and {Mang}, James and {Carter}, Aarynn L. and {M{\^a}lin}, Mathlide and {Morley}, Caroline V. and {Rajpoot}, Bhavesh and {Boogaard}, Leindert A. and {Burt}, Jennifer A. and {Crossfield}, Ian J.~M. and {Feng}, Fabo and {Marie Lagrange}, Anne- and {Phillips}, Mark W.},
        title = "{A Second Visit to Eps Ind Ab with JWST: New Photometry Confirms Ammonia and Suggests Thick Clouds in the Exoplanet Atmosphere of the Closest Super-Jupiter}",
      journal = {\apjl},
         year = 2026,
        month = may,
       volume = {1002},
       number = {1},
          eid = {L5},
        pages = {L5},
          doi = {10.3847/2041-8213/ae5823},
       adsurl = {https://ui.adsabs.harvard.edu/abs/2026ApJ..1002L...5M}
}

@ARTICLE{Sanghi2026,
       author = {{Sanghi}, Aniket and {Thompson}, William and {Mang}, James and {Xuan}, Jerry W. and {Mawet}, Dimitri and {Ruffio}, Jean-Baptiste and {Zhang}, Yapeng and {Wang}, Jason J. and {Morley}, Caroline V. and {Nielsen}, Eric and {Roberson}, William and {Matthews}, Elisabeth and {Carter}, Aarynn L. and {Crossfield}, Ian J.~M. and {M{\^a}lin}, Mathilde and {Benneke}, Bj{\"o}rn and {Bidot}, Alexis and {G{\'a}sp{\'a}r}, Andr{\'a}s and {He}, Carrie and {Horstman}, Katelyn and {Madurowicz}, Alexander and {Marois}, Christian and {Oppenheimer}, Rebecca and {Perrin}, Marshall},
        title = "{Worlds Next Door. IV. Mapping the Late Stages of Giant Planet Evolution with a Precise Dynamical Mass and Luminosity for ϵ Ind Ab}",
      journal = {\aj},
         year = 2026,
        month = aug,
       volume = {172},
       number = {2},
          eid = {115},
        pages = {115},
          doi = {10.3847/1538-3881/ae74c8},
archivePrefix = {arXiv},
       eprint = {2603.08787},
 primaryClass = {astro-ph.EP},
       adsurl = {https://ui.adsabs.harvard.edu/abs/2026AJ....172..115S}
}

@ARTICLE{Malin2025b,
       author = {{M{\^a}lin}, Mathilde and {Ward-Duong}, Kimberly and {Grant}, Sierra L. and {Arulanantham}, Nicole and {Tabone}, Beno{\^\i}t and {Pueyo}, Laurent and {Perrin}, Marshall and {Balmer}, William O. and {Betti}, Sarah and {Chen}, Christine H. and {Debes}, John H. and {Girard}, Julien H. and {Hoch}, Kielan K.~W. and {Kammerer}, Jens and {Lu}, Cicero and {Rebollido}, Isabel and {Rickman}, Emily and {Robinson}, Connor and {Worthen}, Kadin and {van der Marel}, Roeland P. and {Lewis}, Nikole K. and {Seager}, Sara and {Valenti}, Jeff A. and {Soummer}, Remi},
        title = "{JWST-TST High Contrast: Medium-resolution spectroscopy reveals a carbon-rich circumplanetary disk around the young accreting exoplanet Delorme 1 AB b}",
      journal = {\aap},
         year = 2025,
        month = dec,
       volume = {704},
          eid = {A181},
        pages = {A181},
          doi = {10.1051/0004-6361/202556792},
archivePrefix = {arXiv},
       eprint = {2510.07253},
 primaryClass = {astro-ph.EP},
       adsurl = {https://ui.adsabs.harvard.edu/abs/2025A&A...704A.181M}
}

@ARTICLE{Li2015,
       author = {{Li}, Gang and {Gordon}, Iouli E. and {Rothman}, Laurence S. and {Tan}, Yan and {Hu}, Shui-Ming and {Kassi}, Samir and {Campargue}, Alain and {Medvedev}, Emile S.},
        title = "{Rovibrational Line Lists for Nine Isotopologues of the CO Molecule in the X $^{1}${\ensuremath{\Sigma}}$^{+}$ Ground Electronic State}",
      journal = {\apjs},
         year = 2015,
        month = jan,
       volume = {216},
       number = {1},
          eid = {15},
        pages = {15},
          doi = {10.1088/0067-0049/216/1/15},
       adsurl = {https://ui.adsabs.harvard.edu/abs/2015ApJS..216...15L}
}

@ARTICLE{Franson2024,
       author = {{Franson}, Kyle and {Balmer}, William O. and {Bowler}, Brendan P. and {Pueyo}, Laurent and {Zhou}, Yifan and {Rickman}, Emily and {Zhang}, Zhoujian and {Mukherjee}, Sagnick and {Pearce}, Tim D. and {Bardalez Gagliuffi}, Daniella C. and {Biddle}, Lauren I. and {Brandt}, Timothy D. and {Bowens-Rubin}, Rachel and {Crepp}, Justin R. and {Davidson}, James W. and {Faherty}, Jacqueline and {Ginski}, Christian and {Horch}, Elliott P. and {Morgan}, Marvin and {Morley}, Caroline V. and {Perrin}, Marshall D. and {Sanghi}, Aniket and {Salama}, Ma{\"\i}ssa and {Theissen}, Christopher A. and {Tran}, Quang H. and {Wolf}, Trevor N.},
        title = "{JWST/NIRCam 4─5 {\ensuremath{\mu}}m Imaging of the Giant Planet AF Lep b}",
      journal = {\apjl},
         year = 2024,
        month = oct,
       volume = {974},
       number = {1},
          eid = {L11},
        pages = {L11},
          doi = {10.3847/2041-8213/ad736a},
archivePrefix = {arXiv},
       eprint = {2406.09528},
 primaryClass = {astro-ph.EP},
       adsurl = {https://ui.adsabs.harvard.edu/abs/2024ApJ...974L..11F}
}

@ARTICLE{Xuan2026,
       author = {{Xuan}, Jerry W. and {Ruffio}, Jean-Baptiste and {Chachan}, Yayaati and {Ohno}, Kazumasa and {Kesseli}, Aurora and {Murray-Clay}, Ruth and {Lee}, Eve J. and {Moses}, Julianne I. and {Balmer}, William O. and {Baburaj}, Aneesh and {Blake}, Geoffrey A. and {Johnstone}, Doug and {Zhang}, Yapeng and {Knutson}, Heather A. and {Mawet}, Dimitri and {Beichman}, Charles and {Hodapp}, Klaus and {Perrin}, Marshall D. and {Konopacky}, Quinn and {Meyer}, Michael and {Bryden}, Geoffrey and {Greene}, Thomas P. and {Leisenring}, Jarron and {Ygouf}, Marie and {Benneke}, Bj{\"o}rn and {Inglis}, Julie and {Wallack}, Nicole L.},
        title = "{The Compositions of the HR 8799 Planets Reflect Accretion of Both Solids and Metal-enriched Gas}",
      journal = {\apj},
         year = 2026,
        month = mar,
       volume = {1000},
       number = {1},
          eid = {27},
        pages = {27},
          doi = {10.3847/1538-4357/ae448f},
archivePrefix = {arXiv},
       eprint = {2602.09422},
 primaryClass = {astro-ph.EP},
       adsurl = {https://ui.adsabs.harvard.edu/abs/2026ApJ..1000...27X}
}

@ARTICLE{Stephens2009,
       author = {{Stephens}, D.~C. and {Leggett}, S.~K. and {Cushing}, Michael C. and {Marley}, Mark S. and {Saumon}, D. and {Geballe}, T.~R. and {Golimowski}, David A. and {Fan}, Xiaohui and {Noll}, K.~S.},
        title = "{The 0.8-14.5 {\ensuremath{\mu}}m Spectra of Mid-L to Mid-T Dwarfs: Diagnostics of Effective Temperature, Grain Sedimentation, Gas Transport, and Surface Gravity}",
      journal = {\apj},
         year = 2009,
        month = sep,
       volume = {702},
       number = {1},
        pages = {154-170},
          doi = {10.1088/0004-637X/702/1/154},
archivePrefix = {arXiv},
       eprint = {0906.2991},
 primaryClass = {astro-ph.SR},
       adsurl = {https://ui.adsabs.harvard.edu/abs/2009ApJ...702..154S}
}

@ARTICLE{Ruffio2026,
       author = {{Ruffio}, Jean-Baptiste and {Xuan}, Jerry W. and {Chachan}, Yayaati and {Kesseli}, Aurora and {Lee}, Eve J. and {Beichman}, Charles and {Hodapp}, Klaus and {Balmer}, William O. and {Konopacky}, Quinn and {Perrin}, Marshall D. and {Mawet}, Dimitri and {Knutson}, Heather A. and {Bryden}, Geoffrey and {Greene}, Thomas P. and {Johnstone}, Doug and {Leisenring}, Jarron and {Meyer}, Michael and {Ygouf}, Marie},
        title = "{Jupiter-like uniform metal enrichment in a system of multiple giant exoplanets}",
      journal = {Nature Astronomy},
         year = 2026,
        month = apr,
       volume = {10},
        pages = {511-521},
          doi = {10.1038/s41550-026-02783-z},
archivePrefix = {arXiv},
       eprint = {2601.08227},
 primaryClass = {astro-ph.EP},
       adsurl = {https://ui.adsabs.harvard.edu/abs/2026NatAs..10..511R}
}

@ARTICLE{Yurchenko2026,
       author = {{Yurchenko}, Sergei N. and {Barnfield}, Marco G. and {Bowesman}, Charles A. and {Brady}, Ryan P. and {Guest}, Elizabeth R. and {Kefala}, Kyriaki and {Ni}, Qing-He and {Perri}, Armando N. and {Smola}, Oleksiy A. and {Solokov}, Andrei and {Tao}, Chenyi and {Tennyson}, Jonathan},
        title = "{ExoMol line lists ─ LXIII. ExoMol line lists for 12 isotopologues of CO$_{2}$}",
      journal = {\mnras},
         year = 2026,
        month = jan,
       volume = {545},
       number = {3},
          eid = {staf2135},
        pages = {staf2135},
          doi = {10.1093/mnras/staf2135},
archivePrefix = {arXiv},
       eprint = {2512.13889},
 primaryClass = {astro-ph.EP},
       adsurl = {https://ui.adsabs.harvard.edu/abs/2026MNRAS.545f2135Y}
}

@ARTICLE{Gao2018,
       author = {{Gao}, Peter and {Marley}, Mark S. and {Ackerman}, Andrew S.},
        title = "{Sedimentation Efficiency of Condensation Clouds in Substellar Atmospheres}",
      journal = {\apj},
         year = 2018,
        month = mar,
       volume = {855},
       number = {2},
          eid = {86},
        pages = {86},
          doi = {10.3847/1538-4357/aab0a1},
archivePrefix = {arXiv},
       eprint = {1802.06241},
 primaryClass = {astro-ph.EP},
       adsurl = {https://ui.adsabs.harvard.edu/abs/2018ApJ...855...86G}
}

@ARTICLE{Chiang2010,
       author = {{Chiang}, E. and {Youdin}, A.~N.},
        title = "{Forming Planetesimals in Solar and Extrasolar Nebulae}",
      journal = {Annu. Rev. Earth Planet. Sci.},
         year = 2010,
        month = may,
       volume = {38},
        pages = {493-522},
          doi = {10.1146/annurev-earth-040809-152513},
archivePrefix = {arXiv},
       eprint = {0909.2652},
 primaryClass = {astro-ph.EP},
       adsurl = {https://ui.adsabs.harvard.edu/abs/2010AREPS..38..493C}
}

@ARTICLE{Ida2016,
       author = {{Ida}, S. and {Guillot}, T. and {Morbidelli}, A.},
        title = "{The radial dependence of pebble accretion rates: A source of diversity in planetary systems. I. Analytical formulation}",
      journal = {\aap},
         year = 2016,
        month = jun,
       volume = {591},
          eid = {A72},
        pages = {A72},
          doi = {10.1051/0004-6361/201628099},
archivePrefix = {arXiv},
       eprint = {1604.01291},
 primaryClass = {astro-ph.EP},
       adsurl = {https://ui.adsabs.harvard.edu/abs/2016A&A...591A..72I}
}

@ARTICLE{DeRosa2023,
       author = {{De Rosa}, Robert J. and {Nielsen}, Eric L. and {Wahhaj}, Zahed and {Ruffio}, Jean-Baptiste and {Kalas}, Paul G. and {Peck}, Anne E. and {Hirsch}, Lea A. and {Roberson}, William},
        title = "{Direct imaging discovery of a super-Jovian around the young Sun-like star AF Leporis}",
      journal = {\aap},
         year = 2023,
        month = apr,
       volume = {672},
          eid = {A94},
        pages = {A94},
          doi = {10.1051/0004-6361/202345877},
archivePrefix = {arXiv},
       eprint = {2302.06332},
 primaryClass = {astro-ph.EP},
       adsurl = {https://ui.adsabs.harvard.edu/abs/2023A&A...672A..94D}
}

@ARTICLE{Mesa2023,
       author = {{Mesa}, D. and {Gratton}, R. and {Kervella}, P. and {Bonavita}, M. and {Desidera}, S. and {D'Orazi}, V. and {Marino}, S. and {Zurlo}, A. and {Rigliaco}, E.},
        title = "{AF Lep b: The lowest-mass planet detected by coupling astrometric and direct imaging data}",
      journal = {\aap},
         year = 2023,
        month = apr,
       volume = {672},
          eid = {A93},
        pages = {A93},
          doi = {10.1051/0004-6361/202345865},
archivePrefix = {arXiv},
       eprint = {2302.06213},
 primaryClass = {astro-ph.EP},
       adsurl = {https://ui.adsabs.harvard.edu/abs/2023A&A...672A..93M}
}

@ARTICLE{Cugno2025,
       author = {{Cugno}, Gabriele and {Grant}, Sierra L.},
        title = "{A Carbon-rich Disk Surrounding a Planetary-mass Companion}",
      journal = {\apjl},
         year = 2025,
        month = oct,
       volume = {991},
       number = {2},
          eid = {L46},
        pages = {L46},
          doi = {10.3847/2041-8213/ae0290},
archivePrefix = {arXiv},
       eprint = {2509.15209},
 primaryClass = {astro-ph.EP},
       adsurl = {https://ui.adsabs.harvard.edu/abs/2025ApJ...991L..46C}
}

@ARTICLE{Madurowicz2025,
       author = {{Madurowicz}, Alexander and {Ruffio}, Jean-Baptiste and {Macintosh}, Bruce and {Perrin}, Marshall and {Konopacky}, Quinn M. and {Baburaj}, Aneesh and {Hoch}, Kielan},
        title = "{Direct Spectroscopy of 51 Eridani b with JWST NIRSpec}",
      journal = {\aj},
         year = 2025,
        month = dec,
       volume = {170},
       number = {6},
          eid = {326},
        pages = {326},
          doi = {10.3847/1538-3881/ae1028},
archivePrefix = {arXiv},
       eprint = {2510.08327},
 primaryClass = {astro-ph.EP},
       adsurl = {https://ui.adsabs.harvard.edu/abs/2025AJ....170..326M}
}

@ARTICLE{Yurchenko2024,
       author = {{Yurchenko}, Sergei N. and {Owens}, Alec and {Kefala}, Kyriaki and {Tennyson}, Jonathan},
        title = "{ExoMol line lists - LVII. High accuracy ro-vibrational line list for methane (CH$_{4}$)}",
      journal = {\mnras},
         year = 2024,
        month = feb,
       volume = {528},
       number = {2},
        pages = {3719-3729},
          doi = {10.1093/mnras/stae148},
       adsurl = {https://ui.adsabs.harvard.edu/abs/2024MNRAS.528.3719Y}
}

@ARTICLE{Franson2023_AFLep,
       author = {{Franson}, Kyle and {Bowler}, Brendan P. and {Zhou}, Yifan and {Pearce}, Tim D. and {Bardalez Gagliuffi}, Daniella C. and {Biddle}, Lauren I. and {Brandt}, Timothy D. and {Crepp}, Justin R. and {Dupuy}, Trent J. and {Faherty}, Jacqueline and {Jensen-Clem}, Rebecca and {Morgan}, Marvin and {Sanghi}, Aniket and {Theissen}, Christopher A. and {Tran}, Quang H. and {Wolf}, Trevor N.},
        title = "{Astrometric Accelerations as Dynamical Beacons: A Giant Planet Imaged inside the Debris Disk of the Young Star AF Lep}",
      journal = {\apjl},
         year = 2023,
        month = jun,
       volume = {950},
       number = {2},
          eid = {L19},
        pages = {L19},
          doi = {10.3847/2041-8213/acd6f6},
archivePrefix = {arXiv},
       eprint = {2302.05420},
 primaryClass = {astro-ph.EP},
       adsurl = {https://ui.adsabs.harvard.edu/abs/2023ApJ...950L..19F}
}

@article{Goldreich2004,
  title = {Final {{Stages}} of {{Planet Formation}}},
  author = {Goldreich, Peter and Lithwick, Yoram and Sari, Re'em},
  year = {2004},
  month = oct,
  journal = {\apj},
  volume = {614},
  number = {1},
  pages = {497--507},
  issn = {0004-637X, 1538-4357},
  doi = {10.1086/423612},
  urldate = {2025-02-11},
  langid = {english}
}

@ARTICLE{Ginzburg2020,
       author = {{Ginzburg}, Sivan and {Chiang}, Eugene},
        title = "{Heavy-metal Jupiters by major mergers: metallicity versus mass for giant planets}",
      journal = {\mnras},
         year = 2020,
        month = oct,
       volume = {498},
       number = {1},
        pages = {680-688},
          doi = {10.1093/mnras/staa2500},
archivePrefix = {arXiv},
       eprint = {2006.12500},
 primaryClass = {astro-ph.EP},
       adsurl = {https://ui.adsabs.harvard.edu/abs/2020MNRAS.498..680G}
}

@ARTICLE{GWang2025,
       author = {{Wang}, Gavin and {Xuan}, Jerry W. and {Gonz{\'a}lez Picos}, Dar{\'\i}o and {Zhang}, Zhoujian and {Zhang}, Yapeng and {Mawet}, Dimitri and {Hsu}, Chih-Chun and {Wang}, Jason J. and {Blake}, Geoffrey A. and {Ruffio}, Jean-Baptiste and {Horstman}, Katelyn and {Sappey}, Ben and {Xin}, Yinzi and {Finnerty}, Luke and {Echeverri}, Daniel and {Jovanovic}, Nemanja and {Baker}, Ashley and {Bartos}, Randall and {Calvin}, Benjamin and {Cetre}, Sylvain and {Delorme}, Jacques-Robert and {Doppmann}, Gregory W. and {Fitzgerald}, Michael P. and {Liberman}, Joshua and {L{\'o}pez}, Ronald A. and {Morris}, Evan and {Pezzato-Rovner}, Jacklyn and {Phillips}, Caprice L. and {Schofield}, Tobias and {Skemer}, Andrew and {Wallace}, J. Kent and {Wang}, Ji},
        title = "{Chemical and Isotopic Homogeneity between the L Dwarf CD-35 2722 B and Its Early M Host Star}",
      journal = {\apj},
         year = 2026,
        month = feb,
       volume = {997},
       number = {2},
          eid = {195},
        pages = {195},
          doi = {10.3847/1538-4357/ae232f},
archivePrefix = {arXiv},
       eprint = {2511.19588},
 primaryClass = {astro-ph.EP},
       adsurl = {https://ui.adsabs.harvard.edu/abs/2026ApJ...997..195W}
}

@ARTICLE{Chachan2025b,
       author = {{Chachan}, Yayaati and {Fortney}, Jonathan J. and {Ohno}, Kazumasa and {Thorngren}, Daniel and {Murray-Clay}, Ruth},
        title = "{Revising the Giant Planet Mass─Metallicity Relation: Deciphering the Formation Sequence of Giant Planets}",
      journal = {\apj},
         year = 2025,
        month = nov,
       volume = {994},
       number = {1},
          eid = {43},
        pages = {43},
          doi = {10.3847/1538-4357/ae0cbf},
archivePrefix = {arXiv},
       eprint = {2509.20428},
 primaryClass = {astro-ph.EP},
       adsurl = {https://ui.adsabs.harvard.edu/abs/2025ApJ...994...43C}
}

@ARTICLE{Mukherjee2024,
       author = {{Mukherjee}, Sagnick and {Fortney}, Jonathan J. and {Morley}, Caroline V. and {Batalha}, Natasha E. and {Marley}, Mark S. and {Karalidi}, Theodora and {Visscher}, Channon and {Lupu}, Roxana and {Freedman}, Richard and {Gharib-Nezhad}, Ehsan},
        title = "{The Sonora Substellar Atmosphere Models. IV. Elf Owl: Atmospheric Mixing and Chemical Disequilibrium with Varying Metallicity and C/O Ratios}",
      journal = {\apj},
         year = 2024,
        month = mar,
       volume = {963},
       number = {1},
          eid = {73},
        pages = {73},
          doi = {10.3847/1538-4357/ad18c2},
archivePrefix = {arXiv},
       eprint = {2402.00756},
 primaryClass = {astro-ph.EP},
       adsurl = {https://ui.adsabs.harvard.edu/abs/2024ApJ...963...73M}
}

@ARTICLE{Hsu2026,
       author = {{Hsu}, Chih-Chun and {Wang}, Jason J. and {Xuan}, Jerry W. and {Zhang}, Yapeng and {Ruffio}, Jean-Baptiste and {Mawet}, Dimitri and {Finnerty}, Luke and {Horstman}, Katelyn and {Cronin}, Julianne and {Xin}, Yinzi and {Sappey}, Ben and {Echeverri}, Daniel and {Jovanovic}, Nemanja and {Baker}, Ashley and {Bartos}, Randall and {Blake}, Geoffrey A. and {Calvin}, Benjamin and {Cetre}, Sylvain and {Delorme}, Jacques-Robert and {Doppmann}, Gregory W. and {Fitzgerald}, Michael P. and {Konopacky}, Quinn M. and {Liberman}, Joshua and {L{\'o}pez}, Ronald A. and {Morris}, Evan and {Pezzato}, Jacklyn and {Schofield}, Tobias and {Skemer}, Andrew and {Wallace}, J. Kent and {Wang}, Ji},
        title = "{Distinct Rotational Evolution of Giant Planets and Brown Dwarf Companions}",
      journal = {\aj},
         year = 2026,
        month = apr,
       volume = {171},
       number = {4},
          eid = {224},
        pages = {224},
          doi = {10.3847/1538-3881/ae434b},
archivePrefix = {arXiv},
       eprint = {2601.05976},
 primaryClass = {astro-ph.EP},
       adsurl = {https://ui.adsabs.harvard.edu/abs/2026AJ....171..224H}
}

@article{Xuan2022,
   author = {Jerry W Xuan and Jason Wang and Jean-Baptiste Ruffio and Heather Knutson and Dimitri Mawet and Paul Mollière and Jared Kolecki and Arthur Vigan and Sagnick Mukherjee and Nicole Wallack and Ji Wang and Ashley Baker and Randall Bartos and Geoffrey A Blake and Charlotte Z Bond and Marta Bryan and Benjamin Calvin and Sylvain Cetre and Mark Chun and Jacques-Robert Delorme and Greg Doppmann and Daniel Echeverri and Luke Finnerty and Michael P Fitzgerald and Katelyn Horstman and Julie Inglis and Nemanja Jovanovic and Ronald López and Emily C Martin and Evan Morris and Jacklyn Pezzato and Sam Ragland and Bin Ren and Garreth Ruane and Ben Sappey and Tobias Schofield and Andrew Skemer and Taylor Venenciano and J Kent Wallace and Peter Wizinowich},
   doi = {10.3847/1538-4357/ac8673},
   issn = {0004-637X},
   issue = {2},
   journal = {\apj},
   pages = {54},
   title = {A Clear View of a Cloudy Brown Dwarf Companion from High-resolution Spectroscopy},
   volume = {937},
   url = {https://iopscience.iop.org/article/10.3847/1538-4357/ac8673/meta https://iopscience.iop.org/article/10.3847/1538-4357/ac8673/pdf https://iopscience.iop.org/article/10.3847/1538-4357/ac8673},
   year = {2022},
}

@article{Wang2023,
   author = {Ji Wang and Jason J Wang and Jean-Baptiste Ruffio and Geoffrey A Blake and Dimitri Mawet and Ashley Baker and Randall Bartos and Charlotte Z Bond and Benjamin Calvin and Sylvain Cetre and Jacques-Robert Delorme and Greg Doppmann and Daniel Echeverri and Luke Finnerty and Michael P Fitzgerald and Nemanja Jovanovic and Ronald Lopez and Emily C Martin and Evan Morris and Jacklyn Pezzato and Sam Ragland and Garreth Ruane and Ben Sappey and Tobias Schofield and Andrew Skemer and Taylor Venenciano and J Kent Wallace and Peter Wizinowich and Jerry W Xuan and Marta L Bryan and Arpita Roy and Nicole L Wallack},
   doi = {10.3847/1538-3881/ac9f19},
   issn = {0004-6256},
   journal = {\aj},
   pages = {4},
   title = {Retrieving C and O Abundance of HR 8799 c by Combining High- and Low-resolution Data},
   volume = {165},
   url = {https://ui.adsabs.harvard.edu/abs/2023AJ....165....4W https://ui.adsabs.harvard.edu/link_gateway/2023AJ....165....4W/ARTICLE},
   year = {2023},
}

@article{bowler_imaging_2016,
	title = {Imaging {Extrasolar} {Giant} {Planets}},
	volume = {128},
	issn = {1538-3873},
	url = {http://stacks.iop.org/1538-3873/128/i=968/a=102001},
	doi = {10.1088/1538-3873/128/968/102001},
	language = {en},
	number = {968},
	urldate = {2018-09-28},
	journal = {\pasp},
	author = {Bowler, Brendan P.},
	year = {2016},
	pages = {102001}
}

@article{nielsen_gemini_2019,
	doi = {10.3847/1538-3881/ab16e9},
	url = {https://doi.org/10.3847%2F1538-3881%2Fab16e9},
	year = 2019,
	month = {jun},
	publisher = {American Astronomical Society},
	volume = {158},
	number = {1},
	pages = {13},
	author = {Eric L. Nielsen and Robert J. De Rosa and Bruce Macintosh and Jason J. Wang and Jean-Baptiste Ruffio and Eugene Chiang and Mark S. Marley and Didier Saumon and Dmitry Savransky and S. Mark Ammons and Vanessa P. Bailey and Travis Barman and C{\'{e}}lia Blain and Joanna Bulger and Adam Burrows and Jeffrey Chilcote and Tara Cotten and Ian Czekala and Rene Doyon and Gaspard Duch{\^{e}}ne and Thomas M. Esposito and Daniel Fabrycky and Michael P. Fitzgerald and Katherine B. Follette and Jonathan J. Fortney and Benjamin L. Gerard and Stephen J. Goodsell and James R. Graham and Alexandra Z. Greenbaum and Pascale Hibon and Sasha Hinkley and Lea A. Hirsch and Justin Hom and Li-Wei Hung and Rebekah Ilene Dawson and Patrick Ingraham and Paul Kalas and Quinn Konopacky and James E. Larkin and Eve J. Lee and Jonathan W. Lin and J{\'{e}}r{\^{o}}me Maire and Franck Marchis and Christian Marois and Stanimir Metchev and Maxwell A. Millar-Blanchaer and Katie M. Morzinski and Rebecca Oppenheimer and David Palmer and Jennifer Patience and Marshall Perrin and Lisa Poyneer and Laurent Pueyo and Roman R. Rafikov and Abhijith Rajan and Julien Rameau and Fredrik T. Rantakyrö and Bin Ren and Adam C. Schneider and Anand Sivaramakrishnan and Inseok Song and Remi Soummer and Melisa Tallis and Sandrine Thomas and Kimberly Ward-Duong and Schuyler Wolff},
	title = {The Gemini Planet Imager Exoplanet Survey: Giant Planet and Brown Dwarf Demographics from 10 to 100 au},
	journal = {\aj},
}

@ARTICLE{Kama2019,
       author = {{Kama}, Mihkel and {Shorttle}, Oliver and {Jermyn}, Adam S. and {Folsom}, Colin P. and {Furuya}, Kenji and {Bergin}, Edwin A. and {Walsh}, Catherine and {Keller}, Lindsay},
        title = "{Abundant Refractory Sulfur in Protoplanetary Disks}",
      journal = {ApJ},
         year = 2019,
        month = nov,
       volume = {885},
       number = {2},
          eid = {114},
        pages = {114},
          doi = {10.3847/1538-4357/ab45f8},
archivePrefix = {arXiv},
       eprint = {1908.05169},
 primaryClass = {astro-ph.EP},
       adsurl = {https://ui.adsabs.harvard.edu/abs/2019ApJ...885..114K}
}

@ARTICLE{JiWang2025,
       author = {{Wang}, Ji},
        title = "{Early Accretion of Large Amounts of Solids for Directly Imaged Exoplanets}",
      journal = {ApJ},
         year = 2025,
        month = mar,
       volume = {981},
       number = {2},
          eid = {138},
        pages = {138},
          doi = {10.3847/1538-4357/adb42c},
archivePrefix = {arXiv},
       eprint = {2310.00088},
 primaryClass = {astro-ph.EP},
       adsurl = {https://ui.adsabs.harvard.edu/abs/2025ApJ...981..138W}
}

@ARTICLE{Barman2015,
       author = {{Barman}, Travis S. and {Konopacky}, Quinn M. and {Macintosh}, Bruce and {Marois}, Christian},
        title = "{Simultaneous Detection of Water, Methane, and Carbon Monoxide in the Atmosphere of Exoplanet HR8799b}",
      journal = {\apj},
         year = 2015,
        month = may,
       volume = {804},
       number = {1},
          eid = {61},
        pages = {61},
          doi = {10.1088/0004-637X/804/1/61},
archivePrefix = {arXiv},
       eprint = {1503.03539},
 primaryClass = {astro-ph.EP},
       adsurl = {https://ui.adsabs.harvard.edu/abs/2015ApJ...804...61B}
}

@ARTICLE{Denis2025,
       author = {{Denis}, A. and {Vigan}, A. and {Costes}, J. and {Chauvin}, G. and {Radcliffe}, A. and {Ravet}, M. and {Balmer}, W. and {Palma-Bifani}, P. and {Petrus}, S. and {Parmentier}, V. and {Martos}, S. and {Simonnin}, A. and {Bonnefoy}, M. and {Cadet}, R. and {Forveille}, T. and {Charnay}, B. and {Kiefer}, F. and {Lagrange}, A.-M. and {Chiavassa}, A. and {Stolker}, T. and {Lavail}, A. and {Godoy}, N. and {Janson}, M. and {Pourcelot}, R. and {Delorme}, P. and {Rickman}, E. and {Cont}, D. and {Reiners}, A. and {De Rosa}, R. and {Anwand-Heerwart}, H. and {Charles}, Y. and {Costille}, A. and {El Morsy}, M. and {Garcia}, J. and {Houll{\'e}}, M. and {Lopez}, M. and {Murray}, G. and {Muslimov}, E. and {Otten}, G.~P.~P.~L. and {Paufique}, J. and {Phillips}, M. and {Seemann}, U. and {Viret}, A. and {Zins}, G.},
        title = "{Characterization of AF Lep b at high spectral resolution with VLT/HiRISE}",
      journal = {\aap},
         year = 2025,
        month = apr,
       volume = {696},
          eid = {A6},
        pages = {A6},
          doi = {10.1051/0004-6361/202453108},
archivePrefix = {arXiv},
       eprint = {2502.19558},
 primaryClass = {astro-ph.EP},
       adsurl = {https://ui.adsabs.harvard.edu/abs/2025A&A...696A...6D}
}

@ARTICLE{Hayoz2025,
       author = {{Hayoz}, J. and {Bonse}, M.~J. and {Dannert}, F. and {Garvin}, E.~O. and {Cugno}, G. and {Patapis}, P. and {Gebhard}, T.~D. and {Balmer}, W.~O. and {De Rosa}, R.~J. and {Agudo Berbel}, A. and {Cao}, Y. and {Orban de Xivry}, G. and {Stolker}, T. and {Davies}, R. and {Absil}, O. and {Schmid}, H.~M. and {Quanz}, S.~P. and {Agapito}, G. and {Baruffolo}, A. and {Black}, M. and {Bonaglia}, M. and {Briguglio}, R. and {Carbonaro}, L. and {Cresci}, G. and {Dallilar}, Y. and {Deysenroth}, M. and {Di Antonio}, I. and {Di Cianno}, A. and {Di Rico}, G. and {Doelman}, D. and {Dolci}, M. and {Eisenhauer}, F. and {Esposito}, S. and {Fantinel}, D. and {Ferruzzi}, D. and {Feuchtgruber}, H. and {F{\"o}rster Schreiber}, N.~M. and {Gao}, X. and {Genzel}, R. and {Gillessen}, S. and {Glauser}, A.~M. and {Grani}, P. and {Hartl}, M. and {Henry}, D. and {Huber}, H. and {Keller}, C. and {Kenworthy}, M. and {Kravchenko}, K. and {Lightfoot}, J. and {Lunney}, D. and {Lutz}, D. and {MacIntosh}, M. and {Mannucci}, F. and {Ott}, T. and {Pearson}, D. and {Puglisi}, A. and {Rabien}, S. and {Rau}, C. and {Riccardi}, A. and {Salasnich}, B. and {Shimizu}, T. and {Snik}, F. and {Sturm}, E. and {Tacconi}, L. and {Taylor}, W. and {Valentini}, A. and {Waring}, C. and {Wiezorrek}, E. and {Xompero}, M.},
        title = "{High-contrast spectroscopy with the new VLT/ERIS instrument: Molecular maps and radial velocity of the gas giant AF Lep b}",
      journal = {\aap},
         year = 2025,
        month = jun,
       volume = {698},
          eid = {A87},
        pages = {A87},
          doi = {10.1051/0004-6361/202453297},
archivePrefix = {arXiv},
       eprint = {2502.19961},
 primaryClass = {astro-ph.EP},
       adsurl = {https://ui.adsabs.harvard.edu/abs/2025A&A...698A..87H}
}

@ARTICLE{vonStauffenberg2026,
       author = {{von Stauffenberg}, A. and {Sauter}, J. and {Molli{\`e}re}, P. and {Ravet}, M. and {Trevascus}, D. and {Brandner}, W. and {Berdeu}, A. and {Bonnefoy}, M. and {Bourdarot}, G. and {Le Bouquin}, J.-B. and {Chauvin}, G. and {Eisenhauer}, F. and {Houll{\'e}}, M. and {Kreidberg}, L. and {Matthews}, E. and {Millour}, F. and {Scigliuto}, J. and {Wang}, J. and {Xuan}, J.~W. and {Zhang}, Y. and {Gravity + Collaboration}},
        title = "{$^{13}$CO and potential variability in {\ensuremath{\beta}} Pictoris b with GRAVITY+}",
      journal = {\aap},
         year = 2026,
        month = jul,
       volume = {711},
          eid = {L2},
        pages = {L2},
          doi = {10.1051/0004-6361/202660275},
archivePrefix = {arXiv},
       eprint = {2606.11972},
 primaryClass = {astro-ph.EP},
       adsurl = {https://ui.adsabs.harvard.edu/abs/2026A&A...711L...2V}
}

@ARTICLE{Helled2017,
       author = {{Helled}, Ravit and {Stevenson}, David},
        title = "{The Fuzziness of Giant Planets{\textquoteright} Cores}",
      journal = {\apjl},
         year = 2017,
        month = may,
       volume = {840},
       number = {1},
          eid = {L4},
        pages = {L4},
          doi = {10.3847/2041-8213/aa6d08},
archivePrefix = {arXiv},
       eprint = {1704.01299},
 primaryClass = {astro-ph.EP},
       adsurl = {https://ui.adsabs.harvard.edu/abs/2017ApJ...840L...4H}
}

@ARTICLE{Schneider2021a,
       author = {{Schneider}, Aaron David and {Bitsch}, Bertram},
        title = "{How drifting and evaporating pebbles shape giant planets. I. Heavy element content and atmospheric C/O}",
      journal = {\aap},
         year = 2021,
        month = oct,
       volume = {654},
          eid = {A71},
        pages = {A71},
          doi = {10.1051/0004-6361/202039640},
archivePrefix = {arXiv},
       eprint = {2105.13267},
 primaryClass = {astro-ph.EP},
       adsurl = {https://ui.adsabs.harvard.edu/abs/2021A&A...654A..71S}
}

@article{oberg_bergin2021,
   author = {Karin I. Öberg and Edwin A. Bergin},
   doi = {10.1016/j.physrep.2020.09.004},
   issn = {03701573},
   journal = {Physics Reports},
   month = {1},
   pages = {1-48},
   publisher = {Elsevier B.V.},
   title = {Astrochemistry and compositions of planetary systems},
   volume = {893},
   year = {2021},
}

@ARTICLE{Schneider2021b,
       author = {{Schneider}, Aaron David and {Bitsch}, Bertram},
        title = "{How drifting and evaporating pebbles shape giant planets. II. Volatiles and refractories in atmospheres}",
      journal = {\aap},
         year = 2021,
        month = oct,
       volume = {654},
          eid = {A72},
        pages = {A72},
          doi = {10.1051/0004-6361/202141096},
archivePrefix = {arXiv},
       eprint = {2109.03589},
 primaryClass = {astro-ph.EP},
       adsurl = {https://ui.adsabs.harvard.edu/abs/2021A&A...654A..72S}
}

@article{pollack_formation_1996,
	title = {Formation of the {Giant} {Planets} by {Concurrent} {Accretion} of {Solids} and {Gas}},
	volume = {124},
	issn = {0019-1035},
	url = {http://www.sciencedirect.com/science/article/pii/S0019103596901906},
	doi = {10.1006/icar.1996.0190},
	number = {1},
	urldate = {2019-03-26},
	journal = {Icarus},
	author = {Pollack, James B. and Hubickyj, Olenka and Bodenheimer, Peter and Lissauer, Jack J. and Podolak, Morris and Greenzweig, Yuval},
	month = nov,
	year = {1996},
	pages = {62--85},
}

@ARTICLE{Nasedkin2024,
       author = {{Nasedkin}, E. and {Molli{\`e}re}, P. and {Lacour}, S. and {Nowak}, M. and {Kreidberg}, L. and {Stolker}, T. and {Wang}, J.~J. and {Balmer}, W.~O. and {Kammerer}, J. and {Shangguan}, J. and {Abuter}, R. and {Amorim}, A. and {Asensio-Torres}, R. and {Benisty}, M. and {Berger}, J.-P. and {Beust}, H. and {Blunt}, S. and {Boccaletti}, A. and {Bonnefoy}, M. and {Bonnet}, H. and {Bordoni}, M.~S. and {Bourdarot}, G. and {Brandner}, W. and {Cantalloube}, F. and {Caselli}, P. and {Charnay}, B. and {Chauvin}, G. and {Chavez}, A. and {Choquet}, E. and {Christiaens}, V. and {Cl{\'e}net}, Y. and {Coud{\'e} Du Foresto}, V. and {Cridland}, A. and {Davies}, R. and {Dembet}, R. and {Dexter}, J. and {Drescher}, A. and {Duvert}, G. and {Eckart}, A. and {Eisenhauer}, F. and {F{\"o}rster Schreiber}, N.~M. and {Garcia}, P. and {Garcia Lopez}, R. and {Gendron}, E. and {Genzel}, R. and {Gillessen}, S. and {Girard}, J.~H. and {Grant}, S. and {Haubois}, X. and {Hei{\ss}el}, G. and {Henning}, Th. and {Hinkley}, S. and {Hippler}, S. and {Houll{\'e}}, M. and {Hubert}, Z. and {Jocou}, L. and {Keppler}, M. and {Kervella}, P. and {Kurtovic}, N.~T. and {Lagrange}, A.-M. and {Lapeyr{\`e}re}, V. and {Le Bouquin}, J.-B. and {Lutz}, D. and {Maire}, A.-L. and {Mang}, F. and {Marleau}, G.-D. and {M{\'e}rand}, A. and {Monnier}, J.~D. and {Mordasini}, C. and {Ott}, T. and {Otten}, G.~P.~P.~L. and {Paladini}, C. and {Paumard}, T. and {Perraut}, K. and {Perrin}, G. and {Pfuhl}, O. and {Pourr{\'e}}, N. and {Pueyo}, L. and {Ribeiro}, D.~C. and {Rickman}, E. and {Ruffio}, J.~B. and {Rustamkulov}, Z. and {Shimizu}, T. and {Sing}, D. and {Stadler}, J. and {Straub}, O. and {Straubmeier}, C. and {Sturm}, E. and {Tacconi}, L.~J. and {van Dishoeck}, E.~F. and {Vigan}, A. and {Vincent}, F. and {von Fellenberg}, S.~D. and {Widmann}, F. and {Winterhalder}, T.~O. and {Woillez}, J. and {Yazici}, {\c{S}}. and {Gravity Collaboration}},
        title = "{Four-of-a-kind? Comprehensive atmospheric characterisation of the HR 8799 planets with VLTI/GRAVITY}",
      journal = {\aap},
         year = 2024,
        month = jul,
       volume = {687},
          eid = {A298},
        pages = {A298},
          doi = {10.1051/0004-6361/202449328},
archivePrefix = {arXiv},
       eprint = {2404.03776},
 primaryClass = {astro-ph.EP},
       adsurl = {https://ui.adsabs.harvard.edu/abs/2024A&A...687A.298N}
}

@article{bowler_Populationlevel_2020,
  title = {Population-Level {{Eccentricity Distributions}} of {{Imaged Exoplanets}} and {{Brown Dwarf Companions}}: {{Dynamical Evidence}} for {{Distinct Formation Channels}}},
  shorttitle = {Population-Level {{Eccentricity Distributions}} of {{Imaged Exoplanets}} and {{Brown Dwarf Companions}}},
  author = {Bowler, Brendan P. and Blunt, Sarah C. and Nielsen, Eric L.},
  year = {2020},
  month = feb,
  journal = {\aj},
  volume = {159},
  pages = {63},
  issn = {0004-6256},
  doi = {10.3847/1538-3881/ab5b11}
}

@article{Saumon_2008,
	doi = {10.1086/592734},
	url = {https://doi.org/10.1086%2F592734},
	year = 2008,
	month = {dec},
	publisher = {{IOP} Publishing},
	volume = {689},
	number = {2},
	pages = {1327--1344},
	author = {D. Saumon and Mark S. Marley},
	title = {The Evolution of L and T Dwarfs in Color-Magnitude Diagrams},
	journal = {\apj},
}

@ARTICLE{BrownSevilla2023,
       author = {{Brown-Sevilla}, S.~B. and {Maire}, A. -L. and {Molli{\`e}re}, P. and {Samland}, M. and {Feldt}, M. and {Brandner}, W. and {Henning}, Th. and {Gratton}, R. and {Janson}, M. and {Stolker}, T. and {Hagelberg}, J. and {Zurlo}, A. and {Cantalloube}, F. and {Boccaletti}, A. and {Bonnefoy}, M. and {Chauvin}, G. and {Desidera}, S. and {D'Orazi}, V. and {Lagrange}, A. -M. and {Langlois}, M. and {Menard}, F. and {Mesa}, D. and {Meyer}, M. and {Pavlov}, A. and {Petit}, C. and {Rochat}, S. and {Rouan}, D. and {Schmidt}, T. and {Vigan}, A. and {Weber}, L.},
        title = "{Revisiting the atmosphere of the exoplanet 51 Eridani b with VLT/SPHERE}",
      journal = {\aap},
         year = 2023,
        month = may,
       volume = {673},
          eid = {A98},
        pages = {A98},
          doi = {10.1051/0004-6361/202244826},
archivePrefix = {arXiv},
       eprint = {2211.14330},
 primaryClass = {astro-ph.EP},
       adsurl = {https://ui.adsabs.harvard.edu/abs/2023A&A...673A..98B}
}

@article{Whiteford2023,
   author = {Niall Whiteford and Alistair Glasse and Katy L. Chubb and Daniel Kitzmann and Shrishmoy Ray and Mark W. Phillips and Beth A. Biller and Paul I. Palmer and Ken Rice and Ingo P. Waldmann and Quentin Changeat and Nour Skaf and Jason Wang and Billy Edwards and Ahmed Al-Refaie},
   doi = {10.1093/mnras/stad670},
   issn = {13652966},
   issue = {1},
   journal = {MNRAS},
   month = {10},
   pages = {1375-1400},
   publisher = {Oxford University Press},
   title = {Retrieval study of cool directly imaged exoplanet 51 Eri b},
   volume = {525},
   year = {2023},
}

@article{Coles2019,
   author = {Phillip A. Coles and Sergei N. Yurchenko and Jonathan Tennyson},
   doi = {10.1093/mnras/stz2778},
   issn = {13652966},
   issue = {4},
   journal = {MNRAS},
   month = {12},
   pages = {4481-4488},
   publisher = {Oxford University Press},
   title = {ExoMol molecular line lists - XXXV. A rotation-vibration line list for hot ammonia},
   volume = {490},
   year = {2019},
}

@article{Polyansky2018,
   author = {Oleg L Polyansky and Aleksandra A Kyuberis and Nikolai F Zobov and Jonathan Tennyson and Sergei N Yurchenko and Lorenzo Lodi},
   doi = {10.1093/mnras/sty1877},
   issn = {0035-8711},
   issue = {2},
   journal = {MNRAS},
   pages = {2597-2608},
   title = {ExoMol molecular line lists XXX: a complete high-accuracy line list for water},
   volume = {480},
   url = {https://doi.org/10.1093/mnras/sty1877 https://academic.oup.com/mnras/article-pdf/480/2/2597/28250193/sty1877.pdf https://academic.oup.com/mnras/article/480/2/2597/5054049},
   year = {2018},
}

@article{Azzam2016,
   author = {Ala'a A.A. Azzam and Jonathan Tennyson and Sergei N. Yurchenko and Olga V. Naumenko},
   doi = {10.1093/mnras/stw1133},
   issn = {13652966},
   issue = {4},
   journal = {MNRAS},
   month = {8},
   pages = {4063-4074},
   publisher = {Oxford University Press},
   title = {ExoMol molecular line lists - XVI. The rotation-vibration spectrum of hot H2S},
   volume = {460},
   year = {2016},
}

@ARTICLE{Dupuy2018,
       author = {{Dupuy}, Trent J. and {Liu}, Michael C. and {Allers}, Katelyn N. and {Biller}, Beth A. and {Kratter}, Kaitlin M. and {Mann}, Andrew W. and {Shkolnik}, Evgenya L. and {Kraus}, Adam L. and {Best}, William M.~J.},
        title = "{The Hawaii Infrared Parallax Program. III. 2MASS J0249-0557 c: A Wide Planetary-mass Companion to a Low-mass Binary in the {\ensuremath{\beta}} Pic Moving Group}",
      journal = {\aj},
         year = 2018,
        month = aug,
       volume = {156},
       number = {2},
          eid = {57},
        pages = {57},
          doi = {10.3847/1538-3881/aacbc2},
archivePrefix = {arXiv},
       eprint = {1807.05235},
 primaryClass = {astro-ph.EP},
       adsurl = {https://ui.adsabs.harvard.edu/abs/2018AJ....156...57D}
}

@ARTICLE{Wittenmyer2020,
       author = {{Wittenmyer}, Robert A. and {Wang}, Songhu and {Horner}, Jonathan and {Butler}, R.~P. and {Tinney}, C.~G. and {Carter}, B.~D. and {Wright}, D.~J. and {Jones}, H.~R.~A. and {Bailey}, J. and {O'Toole}, S.~J. and {Johns}, Daniel},
        title = "{Cool Jupiters greatly outnumber their toasty siblings: occurrence rates from the Anglo-Australian Planet Search}",
      journal = {\mnras},
         year = 2020,
        month = feb,
       volume = {492},
       number = {1},
        pages = {377-383},
          doi = {10.1093/mnras/stz3436},
archivePrefix = {arXiv},
       eprint = {1912.01821},
 primaryClass = {astro-ph.EP},
       adsurl = {https://ui.adsabs.harvard.edu/abs/2020MNRAS.492..377W}
}

@article{ackerman_Precipitating_2001,
  title = {Precipitating {{Condensation Clouds}} in {{Substellar Atmospheres}}},
  author = {Ackerman, Andrew S. and Marley, Mark S.},
  year = {2001},
  month = aug,
  journal = {\apj},
  volume = {556},
  pages = {872--884},
  issn = {0004-637X},
  doi = {10.1086/321540}
}

@article{asplund_Chemical_2009,
  title = {The {{Chemical Composition}} of the {{Sun}}},
  author = {Asplund, Martin and Grevesse, Nicolas and Sauval, A. Jacques and Scott, Pat},
  year = {2009},
  month = sep,
  journal = {\araa},
  volume = {47},
  number = {1},
  pages = {481},
  issn = {0066-4146},
  doi = {10.1146/annurev.astro.46.060407.145222},
  language = {en}
}

@ARTICLE{Sutlieff2026,
       author = {{Sutlieff}, Ben J. and {Bonse}, Markus J. and {Christiaens}, Valentin and {Fontanive}, Cl{\'e}mence and {Matthews}, Elisabeth C. and {Parker}, Luke T. and {Pearce}, Tim D. and {Birkby}, Jayne L. and {Biller}, Beth A. and {Dupuy}, Trent J. and {Garvin}, Emily O. and {Iskandarli}, Leyla and {Kammerer}, Jens and {Zhou}, Yifan and {De Rosa}, Robert J. and {Carter}, Aarynn L. and {Hinkley}, Sasha and {Kenworthy}, Matthew A. and {Balmer}, William O. and {Hammond}, Iain and {Mang}, James and {Morley}, Caroline V. and {Neeser}, Mark J. and {Absil}, Olivier and {Boccaletti}, Anthony and {Bonavita}, Mariangela and {Bowler}, Brendan P. and {Chen}, Xueqing and {Dannert}, Felix A. and {Girard}, Julien H. and {Kasper}, Markus and {Lagrange}, Anne-Marie and {Liu}, Pengyu and {Orban de Xivry}, Gilles and {Poon}, Michael and {Quanz}, Sascha P. and {Serra}, Beno{\^\i}t and {Vos}, Johanna M. and {Wagner}, Kevin and {Wang}, Jason and {Sch{\"o}lkopf}, Bernhard and {Agapito}, Guido and {Agudo Berbel}, Alex and {Apai}, D{\'a}niel and {Baruffolo}, Andrea and {Black}, Martin and {Bonaglia}, Marco and {Briguglio}, Runa and {Cao}, Yixian and {Carbonaro}, Luca and {Chapman}, Lee and {Cresci}, Giovanni and {Dallilar}, Yigit and {Davies}, Richard and {Deysenroth}, Matthias and {Di Antonio}, Ivan and {Di Cianno}, Amico and {Di Rico}, Gianluca and {Doelman}, David and {Dolci}, Mauro and {Eisenhauer}, Frank and {Esposito}, Simone and {Ferruzzi}, Debora and {Feuchtgruber}, Helmut and {F{\"o}rster-Schreiber}, Natascha and {Franson}, Kyle and {Genzel}, Reinhard and {Gillessen}, Stefan and {Gonzales}, Eileen C. and {Hartl}, Michael and {Hayoz}, Jean and {Huber}, Heinrich and {Keller}, Christoph and {Kravchenko}, Kateryna and {Leisenring}, Jarron and {Lightfoot}, John and {Lunney}, David and {Lutz}, Dieter and {Macintosh}, Mike and {Mannucci}, Filippo and {Metchev}, Stanimir and {Ott}, Thomas and {Pearson}, David and {Puglisi}, Alfio and {Rabien}, Sebastian and {Rau}, Christian and {Riccardi}, Armando and {Salasnich}, Bernardo and {Shimizu}, Taro and {Snik}, Frans and {Sturm}, Eckhard and {Su{\'a}rez}, Genaro and {Tacconi}, Linda and {Tan}, Xianyu and {Taylor}, William and {Waring}, Christopher and {Xompero}, Marco},
        title = "{Direct Imaging Discovery of Giant Exoplanet {\ensuremath{\beta}} Pictoris d: A Decade-long Game of Hide-and-seek}",
      journal = {\apjl},
         year = 2026,
        month = jul,
       volume = {1006},
       number = {1},
          eid = {L10},
        pages = {L10},
          doi = {10.3847/2041-8213/ae80a0},
archivePrefix = {arXiv},
       eprint = {2606.23801},
 primaryClass = {astro-ph.EP},
       adsurl = {https://ui.adsabs.harvard.edu/abs/2026ApJ..1006L..10S}
}

@article{fulton_California_2021,
     author = {{Fulton}, Benjamin J. and {Rosenthal}, Lee J. and {Hirsch}, Lea A. and {Isaacson}, Howard and {Howard}, Andrew W. and {Dedrick}, Cayla M. and {Sherstyuk}, Ilya A. and {Blunt}, Sarah C. and {Petigura}, Erik A. and {Knutson}, Heather A. and {Behmard}, Aida and {Chontos}, Ashley and {Crepp}, Justin R. and {Crossfield}, Ian J.~M. and {Dalba}, Paul A. and {Fischer}, Debra A. and {Henry}, Gregory W. and {Kane}, Stephen R. and {Kosiarek}, Molly and {Marcy}, Geoffrey W. and {Rubenzahl}, Ryan A. and {Weiss}, Lauren M. and {Wright}, Jason T.},
        title = "{California Legacy Survey. II. Occurrence of Giant Planets beyond the Ice Line}",
      journal = {\apjs},
         year = 2021,
        month = jul,
       volume = {255},
       number = {1},
          eid = {14},
        pages = {14},
          doi = {10.3847/1538-4365/abfcc1},
archivePrefix = {arXiv},
       eprint = {2105.11584},
 primaryClass = {astro-ph.EP},
       adsurl = {https://ui.adsabs.harvard.edu/abs/2021ApJS..255...14F}
}

@article{Zhang2023,
doi = {10.3847/1538-3881/acf768},
url = {https://dx.doi.org/10.3847/1538-3881/acf768},
year = {2023},
month = {oct},
publisher = {The American Astronomical Society},
volume = {166},
number = {5},
pages = {198},
author = {Zhoujian Zhang and Paul Mollière and Keith Hawkins and Catherine Manea and Jonathan J. Fortney and Caroline V. Morley and Andrew Skemer and Mark S. Marley and Brendan P. Bowler and Aarynn L. Carter and Kyle Franson and Zachary G. Maas and Christopher Sneden},
title = {ELemental abundances of Planets and brown dwarfs Imaged around Stars (ELPIS). I. Potential Metal Enrichment of the Exoplanet AF Lep b and a Novel Retrieval Approach for Cloudy Self-luminous Atmospheres},
journal = {AJ},
}

@article{greenbaum_GPI_2018,
  title = {{{GPI Spectra}} of {{HR}} 8799 c, d, and e from 1.5 to 2.4 {$M$}m with {{KLIP Forward Modeling}}},
  author = {Greenbaum, Alexandra Z. and Pueyo, Laurent and Ruffio, Jean-Baptiste and Wang, Jason J. and De Rosa, Robert J. and Aguilar, Jonathan and Rameau, Julien and Barman, Travis and Marois, Christian and Marley, Mark S. and Konopacky, Quinn and Rajan, Abhijith and Macintosh, Bruce and Ansdell, Megan and Arriaga, Pauline and Bailey, Vanessa P. and Bulger, Joanna and Burrows, Adam S. and Chilcote, Jeffrey and Cotten, Tara and Doyon, Rene and Duch{\^e}ne, Gaspard and Fitzgerald, Michael P. and Follette, Katherine B. and Gerard, Benjamin and Goodsell, Stephen J. and Graham, James R. and Hibon, Pascale and Hung, Li-Wei and Ingraham, Patrick and Kalas, Paul and Larkin, James E. and Maire, J{\'e}r{\^o}me and Marchis, Franck and Metchev, Stanimir and {Millar-Blanchaer}, Maxwell A. and Nielsen, Eric L. and Norton, Andrew and Oppenheimer, Rebecca and Palmer, David and Patience, Jennifer and Perrin, Marshall D. and Poyneer, Lisa and Rantakyr{\"o}, Fredrik T. and Savransky, Dmitry and Schneider, Adam C. and Sivaramakrishnan, Anand and Song, Inseok and Soummer, R{\'e}mi and Thomas, Sandrine and Wallace, J. Kent and {Ward-Duong}, Kimberly and Wiktorowicz, Sloane and Wolff, Schuyler},
  year = {2018},
  month = jun,
  journal = {\aj},
  volume = {155},
  pages = {226},
  issn = {0004-6256},
  doi = {10.3847/1538-3881/aabcb8}
}

@article{pueyo_DETECTION_2016,
  title = {{{DETECTION AND CHARACTERIZATION OF EXOPLANETS USING PROJECTIONS ON KARHUNEN}}\textendash{{LOEVE EIGENIMAGES}}: {{FORWARD MODELING}}},
  shorttitle = {{{DETECTION AND CHARACTERIZATION OF EXOPLANETS USING PROJECTIONS ON KARHUNEN}}\textendash{{LOEVE EIGENIMAGES}}},
  author = {Pueyo, Laurent},
  year = {2016},
  month = jun,
  journal = {ApJ},
  volume = {824},
  number = {2},
  pages = {117},
  publisher = {{American Astronomical Society}},
  issn = {0004-637X},
  doi = {10.3847/0004-637X/824/2/117},
  language = {en}
}

@article{wang_pyKLIP_2015,
  title = {{{pyKLIP}}: {{PSF Subtraction}} for {{Exoplanets}} and {{Disks}}},
  shorttitle = {{{pyKLIP}}},
  author = {Wang, Jason J. and Ruffio, Jean-Baptise and De Rosa, Robert J. and Aguilar, Jonathan and Wolff, Schuyler G. and Pueyo, Laurent},
  year = {2015},
  month = jun,
  journal = {Astrophysics Source Code Library},
  pages = {ascl:1506.001}
}

@article{molliere_petitRADTRANS_2019,
  title = {{{petitRADTRANS}}. {{A Python}} Radiative Transfer Package for Exoplanet Characterization and Retrieval},
  author = {Molli{\`e}re, P. and Wardenier, J. P. and {van Boekel}, R. and Henning, Th. and Molaverdikhani, K. and Snellen, I. A. G.},
  year = {2019},
  month = jul,
  journal = {\aap},
  volume = {627},
  pages = {A67},
  issn = {0004-6361},
  doi = {10.1051/0004-6361/201935470}
}

@article{molliere_Retrieving_2020,
  title = {Retrieving Scattering Clouds and Disequilibrium Chemistry in the Atmosphere of {{HR}} 8799e},
  author = {Molli{\`e}re, P. and Stolker, T. and Lacour, S. and Otten, G. P. P. L. and Shangguan, J. and Charnay, B. and Molyarova, T. and Nowak, M. and Henning, Th. and Marleau, G.-D. and Semenov, D. A. and {van Dishoeck}, E. and Eisenhauer, F. and Garcia, P. and Garcia Lopez, R. and Girard, J. H. and Greenbaum, A. Z. and Hinkley, S. and Kervella, P. and Kreidberg, L. and Maire, A.-L. and Nasedkin, E. and Pueyo, L. and Snellen, I. A. G. and Vigan, A. and Wang, J. and {de Zeeuw}, P. T. and Zurlo, A.},
  year = {2020},
  month = aug,
  journal = {\aap},
  volume = {640},
  pages = {A131},
  issn = {0004-6361},
  doi = {10.1051/0004-6361/202038325}
}

@article{Zahnle_methane_2014,
  title = {Methane, {{Carbon Monoxide}}, and {{Ammonia}} in {{Brown Dwarfs}} and {{Self}}-{{Luminous Giant Planets}}},
  author = {Zahnle, Kevin J. and Marley, Mark S.},
  year = {2014},
  month = dec,
  journal = {\apj},
  volume = {797},
  pages = {41},
  issn = {0004-637X},
  doi = {10.1088/0004-637X/797/1/41}
}

@inproceedings{claudi_SPHERE_2008,
  title = {{{SPHERE IFS}}: The Spectro Differential Imager of the {{VLT}} for Exoplanets Search},
  shorttitle = {{{SPHERE IFS}}},
  booktitle = {{{SPIE Astronomical Telescopes}} + {{Instrumentation}}},
  author = {Claudi, R. U. and Turatto, M. and Gratton, R. G. and Antichi, J. and Bonavita, M. and Bruno, P. and Cascone, E. and De Caprio, V. and Desidera, S. and Giro, E. and Mesa, D. and Scuderi, S. and Dohlen, K. and Beuzit, J. L. and Puget, P.},
  editor = {McLean, Ian S. and Casali, Mark M.},
  year = {2008},
  month = jul,
  pages = {70143E},
  address = {{Marseille, France}},
  doi = {10.1117/12.788366}
}

@article{vigan_vltsphere_2020,
  title = {Vlt-Sphere: Automatic {{VLT}}/{{SPHERE}} Data Reduction and Analysis},
  shorttitle = {Vlt-Sphere},
  author = {Vigan, Arthur},
  year = {2020},
  month = sep,
  journal = {Astrophysics Source Code Library},
  pages = {ascl:2009.002}
}

@article{beuzit_SPHERE_2019,
  title = {{{SPHERE}}: The Exoplanet Imager for the {{Very Large Telescope}}},
  shorttitle = {{{SPHERE}}},
  author = {Beuzit, J.-L. and Vigan, A. and Mouillet, D. and Dohlen, K. and Gratton, R. and Boccaletti, A. and Sauvage, J.-F. and Schmid, H. M. and Langlois, M. and Petit, C. and Baruffolo, A. and Feldt, M. and Milli, J. and Wahhaj, Z. and Abe, L. and Anselmi, U. and Antichi, J. and Barette, R. and Baudrand, J. and Baudoz, P. and Bazzon, A. and Bernardi, P. and Blanchard, P. and Brast, R. and Bruno, P. and Buey, T. and Carbillet, M. and Carle, M. and Cascone, E. and Chapron, F. and Charton, J. and Chauvin, G. and Claudi, R. and Costille, A. and De Caprio, V. and {de Boer}, J. and Delboulb{\'e}, A. and Desidera, S. and Dominik, C. and Downing, M. and Dupuis, O. and Fabron, C. and Fantinel, D. and Farisato, G. and Feautrier, P. and Fedrigo, E. and Fusco, T. and Gigan, P. and Ginski, C. and Girard, J. and Giro, E. and Gisler, D. and Gluck, L. and Gry, C. and Henning, T. and Hubin, N. and Hugot, E. and Incorvaia, S. and Jaquet, M. and Kasper, M. and Lagadec, E. and Lagrange, A.-M. and Le Coroller, H. and Le Mignant, D. and Le Ruyet, B. and Lessio, G. and Lizon, J.-L. and Llored, M. and Lundin, L. and Madec, F. and Magnard, Y. and Marteaud, M. and Martinez, P. and Maurel, D. and M{\'e}nard, F. and Mesa, D. and {M{\"o}ller-Nilsson}, O. and Moulin, T. and Moutou, C. and Orign{\'e}, A. and Parisot, J. and Pavlov, A. and Perret, D. and Pragt, J. and Puget, P. and Rabou, P. and Ramos, J. and Reess, J.-M. and Rigal, F. and Rochat, S. and Roelfsema, R. and Rousset, G. and Roux, A. and Saisse, M. and Salasnich, B. and Santambrogio, E. and Scuderi, S. and Segransan, D. and Sevin, A. and Siebenmorgen, R. and Soenke, C. and Stadler, E. and Suarez, M. and Tiph{\`e}ne, D. and Turatto, M. and Udry, S. and Vakili, F. and Waters, L. B. F. M. and Weber, L. and Wildi, F. and Zins, G. and Zurlo, A.},
  year = {2019},
  month = nov,
  journal = {\aap},
  volume = {631},
  pages = {A155},
  issn = {0004-6361},
  doi = {10.1051/0004-6361/201935251},
  langid = {english}
}

@article{amarsi_Carbon_2019,
  title = {Carbon, Oxygen, and Iron Abundances in Disk and Halo Stars. {{Implications}} of {{3D}} Non-{{LTE}} Spectral Line Formation},
  author = {Amarsi, A. M. and Nissen, P. E. and Sk{\'u}lad{\'o}ttir, {\'A}},
  year = {2019},
  month = oct,
  journal = {\aap},
  volume = {630},
  pages = {A104},
  issn = {0004-6361},
  doi = {10.1051/0004-6361/201936265},
  langid = {english}
}

@article{wang_Keck_2020,
  title = {Keck/{{NIRC2 L}}'-Band {{Imaging}} of {{Jovian-mass Accreting Protoplanets}} around {{PDS}} 70},
  author = {Wang, Jason J. and Ginzburg, Sivan and Ren, Bin and Wallack, Nicole and Gao, Peter and Mawet, Dimitri and Bond, Charlotte Z. and Cetre, Sylvain and Wizinowich, Peter and De Rosa, Robert J. and Ruane, Garreth and Liu, Michael C. and Absil, Olivier and Alvarez, Carlos and Baranec, Christoph and Choquet, {\'E}lodie and Chun, Mark and Defr{\`e}re, Denis and Delorme, Jacques-Robert and Duch{\^e}ne, Gaspard and Forsberg, Pontus and Ghez, Andrea and Guyon, Olivier and Hall, Donald N. B. and Huby, Elsa and Jolivet, A{\"i}ssa and {Jensen-Clem}, Rebecca and Jovanovic, Nemanja and Karlsson, Mikael and Lilley, Scott and Matthews, Keith and M{\'e}nard, Fran{\c c}ois and Meshkat, Tiffany and {Millar-Blanchaer}, Maxwell and Ngo, Henry and {Orban de Xivry}, Gilles and Pinte, Christophe and Ragland, Sam and Serabyn, Eugene and Catal{\'a}n, Ernesto Vargas and Wang, Ji and Wetherell, Ed and Williams, Jonathan P. and Ygouf, Marie and Zuckerman, Ben},
  year = {2020},
  month = jun,
  journal = {\aj},
  volume = {159},
  pages = {263},
  doi = {10.3847/1538-3881/ab8aef}
}

@ARTICLE{Liu2026,
       author = {{Liu}, Yurou and {Zhang}, Yapeng and {Xuan}, Jerry W. and {Mawet}, Dimitri and {Snellen}, Ignas and {Landman}, Rico and {Stolker}, Tomas and {de Regt}, Sam and {Kesseli}, Aurora and {Rice}, Malena},
        title = "{Chemistry and Isotope Ratios of Substellar Atmospheres in the {\ensuremath{\beta}} Pictoris Young Moving Group and Vicinity}",
      journal = {\apj},
         year = 2026,
        month = jun,
       volume = {1004},
       number = {2},
          eid = {172},
        pages = {172},
          doi = {10.3847/1538-4357/ae6806},
archivePrefix = {arXiv},
       eprint = {2605.01012},
 primaryClass = {astro-ph.EP},
       adsurl = {https://ui.adsabs.harvard.edu/abs/2026ApJ..1004..172L}
}

@INPROCEEDINGS{Currie_review_2023,
       author = {{Currie}, T. and {Biller}, B. and {Lagrange}, A. and {Marois}, C. and {Guyon}, O. and {Nielsen}, E.~L. and {Bonnefoy}, M. and {De Rosa}, R.~J.},
        title = "{Direct Imaging and Spectroscopy of Extrasolar Planets}",
    booktitle = {Protostars and Planets VII},
         year = 2023,
       editor = {{Inutsuka}, S. and {Aikawa}, Y. and {Muto}, T. and {Tomida}, K. and {Tamura}, M.},
       series = {Astronomical Society of the Pacific Conference Series},
       volume = {534},
        month = jul,
        pages = {799},
          doi = {10.48550/arXiv.2205.05696},
archivePrefix = {arXiv},
       eprint = {2205.05696},
 primaryClass = {astro-ph.EP},
       adsurl = {https://ui.adsabs.harvard.edu/abs/2023ASPC..534..799C}
}

@INCOLLECTION{Gierasch_convect1985,
       author = {{Gierasch}, P.~J. and {Conrath}, B.~J.},
        title = "{Energy conversion processes in the outer planets.}",
    booktitle = {Recent Advances in Planetary Meteorology},
         year = 1985,
       editor = {{Hunt}, G.~E.},
        pages = {121-146},
        publisher = {Cambridge University Press},
       adsurl = {https://ui.adsabs.harvard.edu/abs/1985rapm.book..121G},
}

@ARTICLE{Ruffio2024,
       author = {{Ruffio}, Jean-Baptiste and {Perrin}, Marshall D. and {Hoch}, Kielan K.~W. and {Kammerer}, Jens and {Konopacky}, Quinn M. and {Pueyo}, Laurent and {Madurowicz}, Alex and {Rickman}, Emily and {Theissen}, Christopher A. and {Agrawal}, Shubh and {Greenbaum}, Alexandra Z. and {Miles}, Brittany E. and {Barman}, Travis S. and {Balmer}, William O. and {Llop-Sayson}, Jorge and {Girard}, Julien H. and {Rebollido}, Isabel and {Soummer}, R{\'e}mi and {Allen}, Natalie H. and {Anderson}, Jay and {Beichman}, Charles A. and {Bellini}, Andrea and {Bryden}, Geoffrey and {Espinoza}, N{\'e}stor and {Glidden}, Ana and {Huang}, Jingcheng and {Lewis}, Nikole K. and {Libralato}, Mattia and {Louie}, Dana R. and {Sohn}, Sangmo Tony and {Seager}, Sara and {van der Marel}, Roeland P. and {Wakeford}, Hannah R. and {Watkins}, Laura L. and {Ygouf}, Marie and {Mountain}, C. Matt},        title = "{JWST-TST High Contrast: Achieving Direct Spectroscopy of Faint Substellar Companions Next to Bright Stars with the NIRSpec Integral Field Unit}",
      journal = {AJ},
         year = 2024,
        month = aug,
       volume = {168},
       number = {2},
          eid = {73},
        pages = {73},
          doi = {10.3847/1538-3881/ad5281},
archivePrefix = {arXiv},
       eprint = {2310.09902},
 primaryClass = {astro-ph.EP},
       adsurl = {https://ui.adsabs.harvard.edu/abs/2024AJ....168...73R},
}

@ARTICLE{Mukherjee2022,
       author = {{Mukherjee}, Sagnick and {Fortney}, Jonathan J. and {Batalha}, Natasha E. and {Karalidi}, Theodora and {Marley}, Mark S. and {Visscher}, Channon and {Miles}, Brittany E. and {Skemer}, Andrew J.~I.},
        title = "{Probing the Extent of Vertical Mixing in Brown Dwarf Atmospheres with Disequilibrium Chemistry}",
      journal = {\apj},
         year = 2022,
        month = oct,
       volume = {938},
       number = {2},
          eid = {107},
        pages = {107},
          doi = {10.3847/1538-4357/ac8dfb},
archivePrefix = {arXiv},
       eprint = {2208.14317},
 primaryClass = {astro-ph.EP},
       adsurl = {https://ui.adsabs.harvard.edu/abs/2022ApJ...938..107M}
}

@ARTICLE{Xuan2024d,
       author = {{Xuan}, Jerry W. and {Perrin}, Marshall D. and {Mawet}, Dimitri and {Knutson}, Heather A. and {Mukherjee}, Sagnick and {Zhang}, Yapeng and {Hoch}, Kielan K.~W. and {Wang}, Jason J. and {Inglis}, Julie and {Wallack}, Nicole L. and {Ruffio}, Jean-Baptiste},
        title = "{Atmospheric Abundances and Bulk Properties of the Binary Brown Dwarf Gliese 229Bab from JWST/MIRI Spectroscopy}",
      journal = {\apjl},
         year = 2024,
        month = dec,
       volume = {977},
       number = {2},
          eid = {L32},
        pages = {L32},
          doi = {10.3847/2041-8213/ad92f9},
archivePrefix = {arXiv},
       eprint = {2411.10571},
 primaryClass = {astro-ph.SR},
       adsurl = {https://ui.adsabs.harvard.edu/abs/2024ApJ...977L..32X}
}

@article{min_Modeling_2005,
  title = {Modeling Optical Properties of Cosmic Dust Grains Using a Distribution of Hollow Spheres},
  author = {Min, M. and Hovenier, J. W. and {de Koter}, A.},
  year = {2005},
  month = mar,
  journal = {\aap},
  volume = {432},
  number = {3},
  pages = {909},
  issn = {0004-6361},
  doi = {10.1051/0004-6361:20041920},
  langid = {english}
}

@ARTICLE{GRAVITY_2020,
       author = {{GRAVITY Collaboration} and {Nowak}, M. and {Lacour}, S. and {Molli{\`e}re}, P. and {Wang}, J. and {Charnay}, B. and {van Dishoeck}, E.~F. and {Abuter}, R. and {Amorim}, A. and {Berger}, J.~P. and {Beust}, H. and {Bonnefoy}, M. and {Bonnet}, H. and {Brandner}, W. and {Buron}, A. and {Cantalloube}, F. and {Collin}, C. and {Chapron}, F. and {Cl{\'e}net}, Y. and {Coud{\'e} Du Foresto}, V. and {de Zeeuw}, P.~T. and {Dembet}, R. and {Dexter}, J. and {Duvert}, G. and {Eckart}, A. and {Eisenhauer}, F. and {F{\"o}rster Schreiber}, N.~M. and {F{\'e}dou}, P. and {Garcia Lopez}, R. and {Gao}, F. and {Gendron}, E. and {Genzel}, R. and {Gillessen}, S. and {Hau{\ss}mann}, F. and {Henning}, T. and {Hippler}, S. and {Hubert}, Z. and {Jocou}, L. and {Kervella}, P. and {Lagrange}, A.-M. and {Lapeyr{\`e}re}, V. and {Le Bouquin}, J.-B. and {L{\'e}na}, P. and {Maire}, A.-L. and {Ott}, T. and {Paumard}, T. and {Paladini}, C. and {Perraut}, K. and {Perrin}, G. and {Pueyo}, L. and {Pfuhl}, O. and {Rabien}, S. and {Rau}, C. and {Rodr{\'\i}guez-Coira}, G. and {Rousset}, G. and {Scheithauer}, S. and {Shangguan}, J. and {Straub}, O. and {Straubmeier}, C. and {Sturm}, E. and {Tacconi}, L.~J. and {Vincent}, F. and {Widmann}, F. and {Wieprecht}, E. and {Wiezorrek}, E. and {Woillez}, J. and {Yazici}, S. and {Ziegler}, D.},
        title = "{Peering into the formation history of {\ensuremath{\beta}} Pictoris b with VLTI/GRAVITY long-baseline interferometry}",
      journal = {\aap},
         year = 2020,
        month = jan,
       volume = {633},
          eid = {A110},
        pages = {A110},
          doi = {10.1051/0004-6361/201936898},
archivePrefix = {arXiv},
       eprint = {1912.04651},
 primaryClass = {astro-ph.EP},
       adsurl = {https://ui.adsabs.harvard.edu/abs/2020A&A...633A.110G}
}

@ARTICLE{Feroz2019,
       author = {{Feroz}, Farhan and {Hobson}, Michael P. and {Cameron}, Ewan and {Pettitt}, Anthony N.},
        title = "{Importance Nested Sampling and the MultiNest Algorithm}",
      journal = {The Open Journal of Astrophysics},
         year = 2019,
        month = nov,
       volume = {2},
       number = {1},
          eid = {10},
        pages = {10},
          doi = {10.21105/astro.1306.2144},
archivePrefix = {arXiv},
       eprint = {1306.2144},
 primaryClass = {astro-ph.IM},
       adsurl = {https://ui.adsabs.harvard.edu/abs/2019OJAp....2E..10F}
}

@ARTICLE{Feroz2009,
       author = {{Feroz}, F. and {Hobson}, M.~P. and {Bridges}, M.},
        title = "{MULTINEST: an efficient and robust Bayesian inference tool for cosmology and particle physics}",
      journal = {Mon. Not. R. Astron. Soc.},
         year = 2009,
        month = oct,
       volume = {398},
       number = {4},
        pages = {1601-1614},
          doi = {10.1111/j.1365-2966.2009.14548.x},
archivePrefix = {arXiv},
       eprint = {0809.3437},
 primaryClass = {astro-ph},
       adsurl = {https://ui.adsabs.harvard.edu/abs/2009MNRAS.398.1601F}
}

@ARTICLE{Buchner2014,
       author = {{Buchner}, J. and {Georgakakis}, A. and {Nandra}, K. and {Hsu}, L. and {Rangel}, C. and {Brightman}, M. and {Merloni}, A. and {Salvato}, M. and {Donley}, J. and {Kocevski}, D.},
        title = "{X-ray spectral modelling of the AGN obscuring region in the CDFS: Bayesian model selection and catalogue}",
      journal = {A\&A},
         year = 2014,
        month = apr,
       volume = {564},
          eid = {A125},
        pages = {A125},
          doi = {10.1051/0004-6361/201322971},
archivePrefix = {arXiv},
       eprint = {1402.0004},
 primaryClass = {astro-ph.HE},
       adsurl = {https://ui.adsabs.harvard.edu/abs/2014A&A...564A.125B}
}

@ARTICLE{Thorngren2019,
       author = {{Thorngren}, Daniel and {Fortney}, Jonathan J.},
        title = "{Connecting Giant Planet Atmosphere and Interior Modeling: Constraints on Atmospheric Metal Enrichment}",
      journal = {\apjl},
         year = 2019,
        month = apr,
       volume = {874},
       number = {2},
          eid = {L31},
        pages = {L31},
          doi = {10.3847/2041-8213/ab1137},
archivePrefix = {arXiv},
       eprint = {1811.11859},
 primaryClass = {astro-ph.EP},
       adsurl = {https://ui.adsabs.harvard.edu/abs/2019ApJ...874L..31T}
}

@article{soummer_Detection_2012b,
  title = {Detection and {{Characterization}} of {{Exoplanets}} and {{Disks Using Projections}} on {{Karhunen-Lo\`eve Eigenimages}}},
  author = {Soummer, R{\'e}mi and Pueyo, Laurent and Larkin, James},
  year = {2012},
  month = aug,
  journal = {\apj},
  volume = {755},
  pages = {L28},
  issn = {0004-637X},
  doi = {10.1088/2041-8205/755/2/L28}
}

@ARTICLE{Bell2015,
       author = {{Bell}, Cameron P.~M. and {Mamajek}, Eric E. and {Naylor}, Tim},
        title = "{A self-consistent, absolute isochronal age scale for young moving groups in the solar neighbourhood}",
      journal = {\mnras},
         year = 2015,
        month = nov,
       volume = {454},
       number = {1},
        pages = {593-614},
          doi = {10.1093/mnras/stv1981},
archivePrefix = {arXiv},
       eprint = {1508.05955},
 primaryClass = {astro-ph.SR},
       adsurl = {https://ui.adsabs.harvard.edu/abs/2015MNRAS.454..593B}
}

@ARTICLE{Balmer2026,
       author = {{Balmer}, William O. and {Pueyo}, Laurent and {Messier}, Ashley and {Bruinsma}, Evelyn and {Jones}, Jeremy and {Matuszewska}, Klara and {Perrin}, Marshall D. and {Girard}, Julien H. and {Leisenring}, Jarron M. and {Lawson}, Kellen and {van der Marel}, Roeland P. and {Kammerer}, Jens and {Carter}, Aarynn and {M{\^a}lin}, Mathilde and {Ward-Duong}, Kimberly and {Hoch}, Kielan K.~W. and {Rickman}, Emily and {Seager}, Sara},
        title = "{Direct Images of CO$_{2}$ Absorption in the Atmosphere of a Super-Jupiter: Enhanced Metallicity Suggestive of Formation in a Disk}",
      journal = {\apjl},
         year = 2026,
        month = apr,
       volume = {1001},
       number = {2},
          eid = {L26},
        pages = {L26},
          doi = {10.3847/2041-8213/ae374a},
archivePrefix = {arXiv},
       eprint = {2604.09785},
 primaryClass = {astro-ph.EP},
       adsurl = {https://ui.adsabs.harvard.edu/abs/2026ApJ..1001L..26B}
}

@ARTICLE{Landman2024,
       author = {{Landman}, R. and {Stolker}, T. and {Snellen}, I.~A.~G. and {Costes}, J. and {de Regt}, S. and {Zhang}, Y. and {Gandhi}, S. and {Molliere}, P. and {Kesseli}, A. and {Vigan}, A. and {Sanchez-L{\'o}pez}, A.},
        title = "{{\ensuremath{\beta}} Pictoris b through the eyes of the upgraded CRIRES+. Atmospheric composition, spin rotation, and radial velocity}",
      journal = {\aap},
         year = 2024,
        month = feb,
       volume = {682},
          eid = {A48},
        pages = {A48},
          doi = {10.1051/0004-6361/202347846},
archivePrefix = {arXiv},
       eprint = {2311.13527},
 primaryClass = {astro-ph.EP},
       adsurl = {https://ui.adsabs.harvard.edu/abs/2024A&A...682A..48L}
}

@ARTICLE{Greco2016,
       author = {{Greco}, Johnny P. and {Brandt}, Timothy D.},
        title = "{The Measurement, Treatment, and Impact of Spectral Covariance and Bayesian Priors in Integral-field Spectroscopy of Exoplanets}",
      journal = {\apj},
         year = 2016,
        month = dec,
       volume = {833},
       number = {2},
          eid = {134},
        pages = {134},
          doi = {10.3847/1538-4357/833/2/134},
archivePrefix = {arXiv},
       eprint = {1602.00691},
 primaryClass = {astro-ph.EP},
       adsurl = {https://ui.adsabs.harvard.edu/abs/2016ApJ...833..134G}
}

@ARTICLE{GaiaDR3_2023,
       author = {{Gaia Collaboration} and {Vallenari}, A. and {Brown}, A.~G.~A. and {Prusti}, T. and {de Bruijne}, J.~H.~J. and {Arenou}, F. and {Babusiaux}, C. and {Biermann}, M. and {Creevey}, O.~L. and {Ducourant}, C. and {Evans}, D.~W. and {Eyer}, L. and {Guerra}, R. and {Hutton}, A. and {Jordi}, C. and {Klioner}, S.~A. and {Lammers}, U.~L. and {Lindegren}, L. and {Luri}, X. and {Mignard}, F. and {Panem}, C. and {Pourbaix}, D. and {Randich}, S. and {Sartoretti}, P. and {Soubiran}, C. and {Tanga}, P. and {Walton}, N.~A. and {Bailer-Jones}, C.~A.~L. and {Bastian}, U. and {Drimmel}, R. and {Jansen}, F. and {Katz}, D. and {Lattanzi}, M.~G. and {van Leeuwen}, F. and {Bakker}, J. and {Cacciari}, C. and {Casta{\~n}eda}, J. and {De Angeli}, F. and {Fabricius}, C. and {Fouesneau}, M. and {Fr{\'e}mat}, Y. and {Galluccio}, L. and {Guerrier}, A. and {Heiter}, U. and {Masana}, E. and {Messineo}, R. and {Mowlavi}, N. and {Nicolas}, C. and {Nienartowicz}, K. and {Pailler}, F. and {Panuzzo}, P. and {Riclet}, F. and {Roux}, W. and {Seabroke}, G.~M. and {Sordo}, R. and {Th{\'e}venin}, F. and {Gracia-Abril}, G. and {Portell}, J. and {Teyssier}, D. and {Altmann}, M. and {Andrae}, R. and {Audard}, M. and {Bellas-Velidis}, I. and {Benson}, K. and {Berthier}, J. and {Blomme}, R. and {Burgess}, P.~W. and {Busonero}, D. and {Busso}, G. and {C{\'a}novas}, H. and {Carry}, B. and {Cellino}, A. and {Cheek}, N. and {Clementini}, G. and {Damerdji}, Y. and {Davidson}, M. and {de Teodoro}, P. and {Nu{\~n}ez Campos}, M. and {Delchambre}, L. and {Dell'Oro}, A. and {Esquej}, P. and {Fern{\'a}ndez-Hern{\'a}ndez}, J. and {Fraile}, E. and {Garabato}, D. and {Garc{\'\i}a-Lario}, P. and {Gosset}, E. and {Haigron}, R. and {Halbwachs}, J.-L. and {Hambly}, N.~C. and {Harrison}, D.~L. and {Hern{\'a}ndez}, J. and {Hestroffer}, D. and {Hodgkin}, S.~T. and {Holl}, B. and {Jan{\ss}en}, K. and {Jevardat de Fombelle}, G. and {Jordan}, S. and {Krone-Martins}, A. and {Lanzafame}, A.~C. and {L{\"o}ffler}, W. and {Marchal}, O. and {Marrese}, P.~M. and {Moitinho}, A. and {Muinonen}, K. and {Osborne}, P. and {Pancino}, E. and {Pauwels}, T. and {Recio-Blanco}, A. and {Reyl{\'e}}, C. and {Riello}, M. and {Rimoldini}, L. and {Roegiers}, T. and {Rybizki}, J. and {Sarro}, L.~M. and {Siopis}, C. and {Smith}, M. and {Sozzetti}, A. and {Utrilla}, E. and {van Leeuwen}, M. and {Abbas}, U. and {{\'A}brah{\'a}m}, P. and {Abreu Aramburu}, A. and {Aerts}, C. and {Aguado}, J.~J. and {Ajaj}, M. and {Aldea-Montero}, F. and {Altavilla}, G. and {{\'A}lvarez}, M.~A. and {Alves}, J. and {Anders}, F. and {Anderson}, R.~I. and {Anglada Varela}, E. and {Antoja}, T. and {Baines}, D. and {Baker}, S.~G. and {Balaguer-N{\'u}{\~n}ez}, L. and {Balbinot}, E. and {Balog}, Z. and {Barache}, C. and {Barbato}, D. and {Barros}, M. and {Barstow}, M.~A. and {Bartolom{\'e}}, S. and {Bassilana}, J.-L. and {Bauchet}, N. and {Becciani}, U. and {Bellazzini}, M. and {Berihuete}, A. and {Bernet}, M. and {Bertone}, S. and {Bianchi}, L. and {Binnenfeld}, A. and {Blanco-Cuaresma}, S. and {Blazere}, A. and {Boch}, T. and {Bombrun}, A. and {Bossini}, D. and {Bouquillon}, S. and {Bragaglia}, A. and {Bramante}, L. and {Breedt}, E. and {Bressan}, A. and {Brouillet}, N. and {Brugaletta}, E. and {Bucciarelli}, B. and {Burlacu}, A. and {Butkevich}, A.~G. and {Buzzi}, R. and {Caffau}, E. and {Cancelliere}, R. and {Cantat-Gaudin}, T. and {Carballo}, R. and {Carlucci}, T. and {Carnerero}, M.~I. and {Carrasco}, J.~M. and {Casamiquela}, L. and {Castellani}, M. and {Castro-Ginard}, A. and {Chaoul}, L. and {Charlot}, P. and {Chemin}, L. and {Chiaramida}, V. and {Chiavassa}, A. and {Chornay}, N. and {Comoretto}, G. and {Contursi}, G. and {Cooper}, W.~J. and {Cornez}, T. and {Cowell}, S. and {Crifo}, F. and {Cropper}, M. and {Crosta}, M. and {Crowley}, C. and {Dafonte}, C. and {Dapergolas}, A. and {David}, M. and {David}, P. and {de Laverny}, P. and {De Luise}, F. and {De March}, R.},
        title = "{Gaia Data Release 3. Summary of the content and survey properties}",
      journal = {\aap},
         year = 2023,
        month = jun,
       volume = {674},
          eid = {A1},
        pages = {A1},
          doi = {10.1051/0004-6361/202243940},
archivePrefix = {arXiv},
       eprint = {2208.00211},
 primaryClass = {astro-ph.GA},
       adsurl = {https://ui.adsabs.harvard.edu/abs/2023A&A...674A...1G}
}

@ARTICLE{Xuan2024b,
       author = {{Xuan}, Jerry W. and {Hsu}, Chih-Chun and {Finnerty}, Luke and {Wang}, Jason and {Ruffio}, Jean-Baptiste and {Zhang}, Yapeng and {Knutson}, Heather A. and {Mawet}, Dimitri and {Mamajek}, Eric E. and {Inglis}, Julie and {Wallack}, Nicole L. and {Bryan}, Marta L. and {Blake}, Geoffrey A. and {Molli{\`e}re}, Paul and {Hejazi}, Neda and {Baker}, Ashley and {Bartos}, Randall and {Calvin}, Benjamin and {Cetre}, Sylvain and {Delorme}, Jacques-Robert and {Doppmann}, Greg and {Echeverri}, Daniel and {Fitzgerald}, Michael P. and {Jovanovic}, Nemanja and {Liberman}, Joshua and {L{\'o}pez}, Ronald A. and {Morris}, Evan and {Pezzato}, Jacklyn and {Sappey}, Ben and {Schofield}, Tobias and {Skemer}, Andrew and {Wallace}, J. Kent and {Wang}, Ji and {Agrawal}, Shubh and {Horstman}, Katelyn},
        title = "{Are These Planets or Brown Dwarfs? Broadly Solar Compositions from High-resolution Atmospheric Retrievals of {\ensuremath{\sim}}10{\textendash}30 M $_{Jup}$ Companions}",
      journal = {\apj},
         year = 2024,
        month = jul,
       volume = {970},
       number = {1},
          eid = {71},
        pages = {71},
          doi = {10.3847/1538-4357/ad4796},
archivePrefix = {arXiv},
       eprint = {2405.13128},
 primaryClass = {astro-ph.EP},
       adsurl = {https://ui.adsabs.harvard.edu/abs/2024ApJ...970...71X}
}

@article{Thorngren2016,
   author = {Daniel P. Thorngren and Jonathan J. Fortney and Ruth A. Murray-Clay and Eric D. Lopez},
   doi = {10.3847/0004-637x/831/1/64},
   issn = {15384357},
   issue = {1},
   journal = {\apj},
   month = {10},
   pages = {64},
   publisher = {American Astronomical Society},
   title = {THE MASS–METALLICITY RELATION FOR GIANT PLANETS},
   volume = {831},
   year = {2016},
}

@article{Turrini2021,
   author = {D. Turrini and E. Schisano and S. Fonte and S. Molinari and R. Politi and D. Fedele and O. Panić and M. Kama and Q. Changeat and G. Tinetti},
   doi = {10.3847/1538-4357/abd6e5},
   issn = {0004-637X},
   issue = {1},
   journal = {\apj},
   month = {3},
   pages = {40},
   publisher = {American Astronomical Society},
   title = {Tracing the Formation History of Giant Planets in Protoplanetary Disks with Carbon, Oxygen, Nitrogen, and Sulfur},
   volume = {909},
   year = {2021},
}

@ARTICLE{Reggiani2024,
       author = {{Reggiani}, Henrique and {Galarza}, Jhon Yana and {Schlaufman}, Kevin C. and {Sing}, David K. and {Healy}, Brian F. and {McWilliam}, Andrew and {Lothringer}, Joshua D. and {Pueyo}, Laurent},
        title = "{Insight into the Formation of {\ensuremath{\beta}} Pic b through the Composition of Its Parent Protoplanetary Disk as Revealed by the {\ensuremath{\beta}} Pic Moving Group Member HD 181327}",
      journal = {\aj},
         year = 2024,
        month = jan,
       volume = {167},
       number = {1},
          eid = {45},
        pages = {45},
          doi = {10.3847/1538-3881/ad0f93},
archivePrefix = {arXiv},
       eprint = {2311.12210},
 primaryClass = {astro-ph.SR},
       adsurl = {https://ui.adsabs.harvard.edu/abs/2024AJ....167...45R}
}
\bibliographystyle{aasjournalv7}

\end{document}